\documentclass[twocolumn]{aastex631}

\newcommand{\pkg}[1]{\texttt{#1}}

\usepackage{amssymb}
\usepackage{amsmath}
\usepackage{float}
\usepackage{hyperref}
\usepackage{subfigure}
\usepackage{rotating}
\usepackage{xcolor}
\hypersetup{
    colorlinks,
    linkcolor={blue!50!black},
    citecolor={blue!50!black},
    urlcolor={blue!80!black}
}

\usepackage{graphicx}
\usepackage{verbatim}
\usepackage{placeins}
\usepackage{multirow}
\usepackage{tabularx}
\usepackage{url}
\usepackage{enumitem}

\begin{document}

\title{Revealing the {magnetization of the} intracluster medium of Abell 3581 using background Faraday rotation measures from the POSSUM survey}

\author[0009-0001-2196-8251]{Affan Khadir}
\affil{David A. Dunlap Department of Astronomy and Astrophysics, University of Toronto, 50 St. George Street, Toronto, M5S 3H4, ON, Canada}
\affil{Dunlap Institute for Astronomy and Astrophysics, University of Toronto, 50 St. George Street, Toronto, M5S 3H4, ON, Canada}
\affil{Department of Physics and Trottier Space Institute, McGill University, 3600 Rue University, Montréal, QC H3A 2T8, Canada}
\correspondingauthor{Affan Khadir}
\email{affan.khadir@mail.mcgill.ca}

\author[0000-0002-5815-8965]{Erik Osinga}
\affil{Dunlap Institute for Astronomy and Astrophysics, University of Toronto, 50 St. George Street, Toronto, M5S 3H4, ON, Canada}

\author[0000-0002-1566-5094]{Wonki Lee}
\affil{Yonsei University, Department of Astronomy, Seoul, Republic of Korea}

\author[0000-0002-2819-9977]{David McConnell}\affil{CSIRO Astronomy and Space Science, PO Box 76, Epping, NSW 1710 Australia}

\author[0000-0002-3382-9558]{B. M. Gaensler}\affil{Department of Astronomy and Astrophysics, University of California Santa Cruz, 1156 High Street, Santa Cruz, CA 95064, USA
}\affil{Dunlap Institute for Astronomy and Astrophysics, University of Toronto, 50 St. George Street, Toronto, M5S 3H4, ON, Canada}\affil{David A. Dunlap Department of Astronomy and Astrophysics, University of Toronto, 50 St. George Street, Toronto, M5S 3H4, ON, Canada}

\author[0000-0003-1619-3479]{Chiara Stuardi}\affil{INAF - Istituto di Radioastronomia, Via Gobetti 101, 40129
Bologna, Italy
}

\author[0000-0002-6243-7879]{Craig Anderson}\affil{Research School of Astronomy \& Astrophysics, The Australian National University, Canberra, ACT 2611, Australia}

\author[0000-0002-3973-8403]
{Ettore Carretti}\affil{INAF - Istituto di Radioastronomia, Via Gobetti 101, 40129
Bologna, Italy
}
\author[0000-0001-9399-5331]{Takuya Akahori}\affil{Mizusawa VLBI Observatory, National Astronomical Observatory of Japan, 2-21-1, Osawa, Mitaka, Tokyo 181-8588, Japan
}

\author[0000-0002-3968-3051]{Shane P. O'Sullivan}\affil{Departamento de Física de la Tierra y Astrofísica \& IPARCOS-UCM, Universidad Complutense de Madrid, 28040 Madrid, Spain
}
\author[0000-0003-0520-0696]{Lerato Baidoo}\affil{Dunlap Institute for Astronomy and Astrophysics, University of Toronto, 50 St. George Street, Toronto, M5S 3H4, ON, Canada}

\author[0000-0001-7722-8458]{Jennifer West}\affil{National Research Council Canada, Herzberg Research Centre for Astronomy and Astrophysics, Dominion Radio Astrophysical Observatory, PO Box 248, Penticton, BC V2A 6J9, Canada}

\author[0000-0002-7641-9946]{Cameron Van Eck}
\affil{Research School of Astronomy \& Astrophysics, The Australian National University, Canberra, ACT 2611, Australia}

\author[0000-0001-5636-7213]{Lawrence Rudnick}\affil{Minnesota Institute for Astrophysics, University of Minnesota, 116 Church Street SE, Minneapolis, MN 55455, USA}

\author[0000-0003-2730-957X]{Naomi McClure-Griffiths}\affil{Research School of Astronomy \& Astrophysics, The Australian National University, Canberra, ACT 2611, Australia}

\author[0000-0003-0742-2006]{Yik Ki Ma}\affil{Max-Planck-Institut f\"ur Radioastronomie, Auf dem H\"ugel 69, 53121 Bonn, Germany}

\author[0000-0001-9006-0725]{David Alonso-L\'opez}\affil{Departamento de Física de la Tierra y Astrofísica \& IPARCOS-UCM, Universidad Complutense de Madrid, 28040 Madrid, Spain
}
\author[0009-0003-9392-4711]{Paris Gordon-Hall}\affil{Research School of Astronomy \& Astrophysics, The Australian National University, Canberra, ACT 2611, Australia}



\begin{abstract}

The line-of-sight magnetic field of galaxy clusters can be probed using Faraday rotation measure (RM) data. However, our understanding of cluster magnetism is limited due to the scarcity of polarized background radio sources, with most previous studies being constrained to $\sim 10$ sources per cluster. Leveraging the increased source density of the POlarisation Sky Survey of the Universe's Magnetism (POSSUM), we probe the magnetic field properties of the galaxy cluster Abell 3581 {(A3581)} with 111 RMs. We find that the standard deviation in the RM declines monotonically with increasing radius up to 0.75 Mpc{, agreeing with a radially declining magnetic field and electron density profile modeled as Gaussian and lognormal random fields, respectively. {We compare our observations of the inner 0.75 Mpc of A3581 to various semi-analytic models of the magnetic field and electron density, and obtain several best-fit models.} For the first time, we compare the observed RMs in a cluster to full magnetohydrodynamic simulated clusters from TNG-Cluster and find that the non-monotonic trend in RM standard deviation past 0.75 Mpc in A3581 is likely caused by past or present merger activity. We identify a possible candidate for a merger to be the galaxy group [DZ2015b] 276, which would be the first group detected in RMs that is not strongly emitting in X-rays. We find a possible merger axis of A3581 with this group at a position angle of $\theta = 52\pm 4$ deg.}

\end{abstract}



\section{Introduction} 
\label{sec:intro}
Most of the baryonic universe is composed of magnetoionic plasma that resides in the cosmic web {\citep{2020Natur.581..391M}}. In the densest regions of the cosmic web, gravity causes the formation of galaxy clusters {\citep[e.g.,][]{2022MNRAS.510..581K}}. The vast majority of the baryonic mass inside the characteristic gravitational radii of galaxy clusters is contained in the intracluster medium  (ICM), which is known to be magnetized {\citep[e.g.,][]{2018SSRv..214..122D}}. The magnetic field strengths of galaxy clusters are on the levels of $\mu$G {\citep[e.g.,][]{2004IJMPD..13.1549G, 2025A&A...694A..44O}}, and these fields are crucially involved in the non-thermal processes that occur in clusters, including the acceleration of cosmic rays {\citep[e.g.,][]{2015ASSL..407..557B}} and the turbulent motions in the ICM {\citep[e.g.,][]{2006MNRAS.366.1437S}}. 

The exact {structure and origins} of the magnetic fields of clusters remain unknown. Still, it is believed that the magnetic field strength, $B$, is likely correlated with the thermal electron density, $n_e$, {both observationally \citep[e.g.,][]{2010A&A...513A..30B, 2012A&A...540A..38V} and from simulations \citep[e.g.,][]{2005MNRAS.364..753D, 2018MNRAS.474.1672V}}. Both of these quantities appear to decrease with radius from the cluster center {\citep[e.g.,][]{1976A&A....49..137C, 2004A&A...424..429M, 2008MNRAS.391..521L}}. The magnetic field strength is {often modeled as a function of thermal electron density} as: 

\begin{align}
    B(r) = B_0 \left(\frac{n_e(r)}{n_e(0)}\right)^\eta, \label{eq:1}
\end{align}
where $ B_0 $ is the magnetic field strength at the center of the cluster, $n_e(0)$ is the thermal electron density at the center of the cluster, $r$ is the distance from the cluster center, and $\eta$ is a power-law index with typical values of 0.5 {\citep[e.g.,][]{2004A&A...424..429M}}.

A method of probing the line-of-sight (LOS) magnetic field is through the use of Faraday rotation, which is the change in polarization angle as polarized light travels through a magnetoionic medium. This change in polarization angle is given by: 
\begin{align}
    \psi_{\mathrm{obs}} - \psi_0 = \mathrm{RM} \lambda^2
\end{align}
where $\psi_0$ is the intrinsic polarization angle at the source, $\psi_{\mathrm{obs}}$ is the observed polarization angle (in radians), and $\lambda$ is the wavelength (in meters).
The observed polarization angle is determined as: 
\begin{align}
    \psi_{\mathrm{obs}} = \frac{1}{2}\arctan\left(\frac{U}{Q}\right),
\end{align}
where $Q$ and $U$ are the two linear polarization Stokes parameters. Faraday rotation is quantified using the rotation measure (RM). The RM is defined to be: 

\begin{equation}
    \mathrm{RM}= {0.812} \mathrm{\ rad\ m}^{-2} \int_{z_s}^0 \frac{1}{(1+z)^2}\frac{n_e(z)}{\mathrm{cm}^{-3}} \frac{B_{\parallel}(z)}{\mu\rm{G}} \frac{dl}{dz ~\rm{pc}}dz,
    \label{eq:2}
\end{equation}
where $n_e$ is the thermal electron density, $B_\parallel$ is the LOS magnetic field strength, $dl$ is the infinitesimal path length along the LOS, $z_s$ is the redshift of the polarised background radio source, and $z$ is the redshift \citep[e.g.][]{2021MNRAS.507.4968F, 2014MNRAS.442.3329X}; RM is taken to be positive for LOS magnetic fields pointing towards the observer. The RM sources are polarised background or embedded radio sources (usually radio galaxies).

The largest catalog of RMs from a single survey {to date was conducted by the Very Large Array  \citep[VLA;][]{1980ApJS...44..151T}:} the NRAO VLA Sky Survey \citep[NVSS;][]{1998AJ....115.1693C, 2009ApJ...702.1230T}. NVSS has an RM grid density of $\sim $1 source deg$^{-2}$ covering $\delta > -40$ deg. In contrast to this, the POlarisation Sky Survey of the Universe's Magnetism \citep[POSSUM;][]{2010AAS...21547013G, BMG} is producing an RM catalog with grid densities of $\sim 40$ polarized sources deg$^{-2}$ \citep{2024AJ....167..226V} and will eventually cover the entire southern sky. The greater sky density of POSSUM RMs, as well as their measurement via the more robust RM-synthesis technique \citep{2005A&A...441.1217B}, allow us to probe individual clusters at much greater precision than before.

Cluster magnetic fields have been studied using Faraday rotation in both single nearby clusters \citep[e.g.,][]{2006A&A...460..425G, 2008A&A...483..699G, 2010A&A...513A..30B, 2012A&A...540A..38V, 2017A&A...603A.122G} and stacked samples of higher redshift clusters \citep{2001ApJ...547L.111C,2011A&A...530A..24B,2016A&A...596A..22B,2019MNRAS.487.4768S,2022A&A...665A..71O, 2025A&A...694A..44O}. Stacking experiments constrain the average magnetic field strength of clusters to the $1-10$ $\mu$G range, with possible differences between merging and non-merging clusters \citep{2019MNRAS.487.4768S}. \citet{2022A&A...665A..71O, 2025A&A...694A..44O} for the first time combined both depolarization and Faraday rotation in a stacking study, and found mean magnetic field strengths of a few $\mu$G with central magnetic field strengths of $5-10$ $\mu$G. However, they found that Gaussian random field models could not fully explain the data. The greatest caveat of stacking studies is that they are unable to discern specific features of the magnetic field of individual clusters. 

Studies of single clusters also generally find magnetic fields in the $1-10 \ \mu$G range \citep[e.g.,][]{1990ApJ...355...29K, 1995A&A...302..680F}. Notably, \citet{2010A&A...513A..30B} constrained the magnetic field profile of the Coma cluster to have $ B_0  = 4.7 \ \mu$G and $\eta = 0.5$ with high statistical confidence, albeit using only 7 resolved radio galaxies; $\eta = 0.5$ implies that the magnetic field energy density scales with the thermal energy density. Most single cluster studies have compared observations to simple models of Gaussian random fields for the magnetic field \citep[often assuming $\eta = 0.5$, e.g.,][]{2011A&A...530A..24B, 2024A&A...691A..23D}, without considering fluctuations in the electron density, and based on small samples of polarized radio sources (only using five to ten), while generally underestimating uncertainties \citep{2020ApJ...888..101J}. In a more detailed study, \citet{2021MNRAS.502.2518S} allowed the exponent to vary and found $\eta \sim 0.9 - 1$ for the ICM of the merging galaxy cluster Abell 2345; furthermore, they obtained the power spectrum of the magnetic field from magnetohydrodynamic (MHD) simulations of clusters, rather than assuming a Kolmogorov power spectrum as is often done. In a recent work, \citet{2024A&A...691A..23D} compared the depolarization trend of radio relics in the galaxy cluster PSZ2 G096.88+24.18 to model magnetic fields imposed on density cubes obtained from MHD simulations of clusters, and they found that the MHD simulation does not produce the same depolarization as the observations, attributing this to a lower magnetic field strength in the simulation. However, no one-to-one comparison of MHD simulations with observed RM grids of clusters has been made so far.


Precursors and pathfinders to the Square Kilometre Array \citep[{SKA;}][]{2009IEEEP..97.1482D} such as MeerKAT \citep{2009IEEEP..97.1522J} and the Australian Square Kilometre Array Pathfinder \citep[ASKAP;][]{2021PASA...38....9H} have been enabling a much more detailed look at cluster magnetism with high-density RM grids. { Using early data from POSSUM, \citet{2021PASA...38...20A} conducted a study of the magnetized plasma in the Fornax cluster. They demonstrated that RM grids can reveal reservoirs of ionized gas not observable using X-rays. Additionally, they noted that mergers of subclusters and galaxies in Fornax are likely the cause of substructures of RM enhancement. More recently, \citet{2025arXiv250105519L} conducted the highest density RM survey of a single cluster, obtaining $\sim 80$ RMs deg$^{-2}$. They found a significant RM enhancement along an RM `stripe', which they attribute to possible inflow of matter into the cluster along a cosmic filament.}

{Given the low number of polarized radio sources in most previous studies of single clusters, and the difficulties associated with stacking experiments, it is clear that the next step in the field is detailed high density RM grid studies of single clusters.} In this work, we conduct a study of the magnetic field properties of Abell 3581 (hereafter A3581) using radio data from ASKAP. We use polarization data from {POSSUM}, with total intensity data from the Evolutionary Map of the Universe \citep[EMU;][]{2011PASA...28..215N, 2022PASA...39...55N, 2025arXiv250508271H}. The aim of this study is to constrain the LOS magnetic field parameters of Abell 3581 from Equation \ref{eq:1} by comparing the RM grid to various magnetic field models and full MHD clusters from the TNG-Cluster simulation \citep{2024A&A...686A.157N}.
 
The remainder of this paper is structured as follows: Section \ref{sec:data} describes our criteria for selecting the target galaxy cluster used, Section \ref{sec:methods} explains the methodology used to analyze the data, Section \ref{sec:results} presents the results of the study, and Section \ref{sec:discussion} provides discussion on the results of this work. Throughout our work, we assume a flat $\Lambda$CDM cosmology with the following cosmological parameters: $H_0 = 70 \text{ km}\text{~s}^{-1}\text{Mpc}^{-1},  ~\Omega_{m,0} = 0.3, ~\Omega_{\Lambda, 0} = 0.7$ \citep{2020A&A...641A...6P}.

\section{Target Selection}
\label{sec:data}

{POSSUM is ideal for observing clusters due to the excellent widefield leakage correction \citep[$\sim 0.1\%$ of Stokes $I$;][Thomson et al. in prep, Anderson et al. in prep]{2023PASA...40...40T}, its angular resolution of $20^{\prime\prime}$ and its typical root-mean-square sensitivity of $18~\mu$Jy beam$^{-1}$ \citep{BMG}. We can find targets for which POSSUM is ideal based on the combination of redshift and $M_{500}$ of the cluster, where $M_{500}$ is the mass contained within the radius, $R_{500}$, where the density of the cluster is 500 times the critical density of matter at that redshift. Furthermore, the large field-of-view of 30 deg$^2$ that ASKAP provides makes POSSUM an ideal survey for nearby clusters that cover large areas of the sky, particularly for clusters that cannot be covered by single observations with more sensitive telescopes such as MeerKAT or the {Jansky VLA \citep[JVLA;][]{2011ApJ...739L...1P}} (i.e. apparent $R_\mathrm{500} > 0.5$ deg). Using this angular size criterion, the best targets are found at $z<0.033$ for $M_\mathrm{500} \sim 5\times 10^{14 }M_\odot$ (and $z<0.024$ for $M_\mathrm{500} \sim  2\times 10^{14} M_\odot$). }

{To find candidate clusters, we cross-matched the Planck Sunyeav Zel'dovich \citep[PSZ2;][]{2016A&A...594A..27P} and the SRG/eROSITA All-Sky Survey DR1 \citep[eRASS1;][]{2024A&A...682A..34M} cluster catalogs to the POSSUM survey coverage as of June 2024. At the time of the start of this work, only two massive clusters in this redshift range were covered by the processed POSSUM fields: Abell 3627 and A3581. While Abell 3627 covers a larger area on the sky, it is also located near the Galactic plane (at Galactic latitude $b = -7.13$ deg) and contains the bright radio galaxy ESO137-006 \citep{2020A&A...636L...1R}, which is not accounted for properly in the automatic POSSUM pipelines and significantly affects the field. For these reasons, we have chosen to focus this study on A3581. }

\subsection{Properties of Abell 3581}
\label{sec:abell_prop}
{A3581 is a cool core (CC) cluster \citep{2005MNRAS.356..237J}  and is covered by the POSSUM field ``1412-28" which spans the area } $ 209.5 \text{ deg }\leq \alpha  \text{ (J2000) }\leq 216.3 \text{ deg}$ and $-30.5 \text{ deg } \leq \delta \text{ (J2000)}\leq -25.3 \text{ deg}$. The field has been observed as the ASKAP SBID 50413 on June 07, 2023 as part of the POSSUM band 1 survey, which has an observing frequency range of $800 - 1088$ MHz. We have considered only analyzing polarized background radio sources that are within 2$R_{500}$ of the center of the cluster. We will analyze the properties of the RMs outside the cluster in this SBID and a neighboring SBID in an upcoming paper. 

For A3581, there are various values of $R_{500}$ in the literature. A study by \citet{2018JCAP...10..010R} used $R_{500} = 0.719$ Mpc, which was { obtained from X-ray observations of the cluster.} A cluster catalog produced by \citet{2024ApJS..272...39W} {identified cluster properties from optical galaxies} and found that $R_{500} = 0.656$ Mpc for A3581 
In contrast to these studies, the eRASS1 cluster catalog \citep{2024A&A...685A.106B} found a larger value of $R_{500} = 0.925$ Mpc; hereafter, all references to $R_{500}$ will be to this value unless explicitly specified. {The eRASS1 catalog infers the cluster mass (and therefore $R_{500}$) using an X-ray mass-relation that has been calibrated with multiple clusters. Because of this calibration, we determined this to be a more accurate radius estimate and will henceforth use it for the remainder of our analysis. Important properties of A3581 are reported in Table \ref{tab:1}.} 

\begin{table}[!htb]
    \centering
    \caption{Basic properties of A3581}
    \begin{tabular}{|c|c|}

        \hline
        {\textbf{Property}} & \textbf{{Measurement}} \\
        \hline
        \hline 
        X-ray Centroid (ICRS) & (14h 07m 29.8s, $-27^\circ$ 01$^{\prime}$ 04$^{\prime\prime}$)\\
        Cluster redshift & $0.0221 \pm 0.0050$\\
        Angular to physical scale & 1 arcsec = 0.447 kpc\\
        $R_{500}$ (Mpc) & {0.925}\\
        $M_{500}$ ($M_\odot$) & $2.15 \times 10^{14}$\\
        \hline
    \end{tabular}
    \tablecomments{Measurements of the X-ray centroid and redshift were taken from the ROSAT All-Sky Survey \citep{2022A&A...658A..59X}. The $M_{500}$ and $R_{500}$ values were taken from the eRASS1 cluster catalog \citep{2024A&A...685A.106B}.}
    \label{tab:1}
\end{table}

\section{Methods}
In this section, we describe the methods we carried out for obtaining the RMs from the Stokes $I, Q, U$ cubes from ASKAP, for analyzing the statistical properties of these RMs, and for modeling the cluster magnetic fields. 
\label{sec:methods}
\subsection{{The POSSUM Single Scheduling Block Pipeline}}
\label{sec:dave}

{To process early POSSUM survey data where sky coverage was disjoint, the POSSUM collaboration developed a single scheduling block (SB) pipeline, which modifies the pipelines described by \citet{BMG} to operate on single observations. We note that a full description of the POSSUM pipeline will appear in an upcoming paper (Van Eck et al. in prep); here, we only give a description of the single SB pipeline.

The single SB pipeline takes image cubes in Stokes parameters $I, Q$ and $U$ from an ASKAP observation and, for a set of source positions, extracts spectra for each parameter, and performs RM-synthesis \citep{2005A&A...441.1217B} using those spectra.  The pipeline products are three files containing results for each source position:}
\begin{enumerate}[label=(\roman*), topsep=0pt,itemsep=-1ex,partopsep=1ex,parsep=1ex]
    \item {the $I, Q$ and $U$ spectra (\pkg{FITS});}
    \item {the complex Faraday depth spectra (\pkg{FITS});}
    \item {some derived quantities characterizing the source (\pkg{csv astropy} table).}
\end{enumerate}

{The pipeline is a python script that is adapted to run on the Australian National University's Research School of Astronomy \& Astrophysics server \textit{avatar}, which has 21 nodes with 128 GB of memory.  The design of the pipeline is predicated on the fact that the extraction of source spectra from the three input cubes is much faster if cubes can be held entirely in node memory.  Each of the three input cubes occupy 183 GB, so a piecewise approach is needed.  We partition each cube into a number of sub-cubes  along the two directional axes and execute the spectra extraction for each in a separate node.}

{The pipeline performs the following steps:}
\begin{enumerate}[topsep=0pt,itemsep=-1ex,partopsep=1ex,parsep=1ex]
    \item {From the CSIRO ASKAP Science Data Archive (CASDA), download the $I, Q, U$ cubes and a source catalog that is generated using Selavy \citep{2012PASA...29..371W, 2017ASPC..512..431W} by the Observatory from the Stokes $I$ cube.}
    \item {Acquire estimates of the free electron content in the ionosphere over the observatory at observation time. The application \pkg{frion\_predict}\footnote{\url{https://frion.readthedocs.io/en/latest/}} is used to do this. It uses total electron content (TEC) maps obtainable from the Jet Propulsion Laboratory within several days of the observation \citep[see][]{2019MNRAS.483.4100P}. }
    \item {Form a subset of the source catalog. The input catalog, generated from the Stokes $I$ cube as above, lists all sources with peak brightness, $B_{\mathrm{peak}}$, above five times the root-mean-square brightness ($B_{\mathrm{peak}} > 5\sigma$).  Since the typical polarized fraction is typically less than 10 per cent, and sources with more than 30\% polarized emission are very rare, we remove from the catalog sources with $B_{\mathrm{peak}} < 15\sigma$.  This step reduces the number of spectra to extract per field from over 20,000 to around 8,000.}
    \item {Divide cubes and the filtered catalog into sub-fields. To match the sub-fields to the memory available on the compute nodes, we divide the approximately square initial field into nine parts. The sub-fields are defined with a bordering guard zone so that each field overlaps its neighbor, ensuring that no sources are missed from laying too close to a sub-field edge.  The catalog is also split into nine parts corresponding to each sub-field.}
    
\end{enumerate}
{The next three steps are performed on nine compute nodes, each dealing with a separate sub-field.}
\begin{enumerate}[topsep=0pt,itemsep=-1ex,partopsep=1ex,parsep=1ex]
    \setcounter{enumi}{4}
    \item {Convolve each image plane to ensure that all spectral channels have the same point-spread-function.}
    \item {Multiply the $Q$ and $U$ cubes by factors that remove rotation of the polarization angle induced by the ionosphere.}
    \item {Perform the main part of the processing in a number of steps executed within the `1d-pipeline' (Vane Eck et al. in prep):}
    \begin{enumerate}[label=(\roman*), topsep=-1ex,itemsep=-1ex,partopsep=0pt,parsep=1ex]
        \item  {Read the input source list;}
        \item {Extract $I, Q, U$ spectra for each source;}
        \item {Diffuse subtraction: use a guard zone around the source to determine the spectrum of diffuse emission and subtract that from the source spectrum \citep[][]{oberhelman24};}
        \item {Perform RM-synthesis on the spectra using \pkg{RM-Tools} \citep{2020ascl.soft05003P} to derive the Faraday dispersion function (FDF) and the RM from the highest amplitude peak of the FDF;}
        \item {Create a catalog that adds polarimeteric parameters to the input source list.}
    \end{enumerate}
    \item  {On a single compute node, merge the products from each sub-field to form the three final data products for the field.}
    \item {Generate a summary plot suitable for a quick assessment of the results.}
    \item {Upload the processing products to the Canadian Advanced Network for Astronomical Research data server.}
\end{enumerate} 

{After running the single SB pipeline, we removed all RMs that have a signal-to-noise ratio (SNR) in polarization of less than 8, following the threshold that has been used in previous POSSUM studies {\citep[e.g.,][]{2024AJ....167..226V}}. Additionally, we removed RMs that had a fractional polarization of less than 1\% because for POSSUM fields that were observed before October 5, 2023, the on-axis polarization leakage correction was applied twice in error {\citep[][Anderson et al. in prep]{BMG}}, resulting in a substantial fraction of leakage-dominated RMs below a polarization fraction of 1\%. Additionally, there were {10} RMs that were incorrectly detected more than once by the Selavy source-finder program. For this reason, we only decided to retain the version of each duplicate that had the highest SNR in our catalog. We obtained 115 RMs within $2R_{500}$ of the cluster once these restrictions were applied, which is an order of magnitude better than most previous studies of single clusters. The most important columns to our analysis in this table are included in Appendix \ref{app:catalog}. The full catalog will be made available on the CDS.

\subsection{QU-fitting and Faraday complexity}
The Stokes $Q$ and $U$ spectra have different levels of complexity, with the most `simple' Stokes $Q$ and $U$ spectra being modeled by single component sinusoidal functions of $\lambda^2$ (used to model the rotation of polarization angle with $\lambda^2$) and more complex spectra having multiple sinusoidal or exponential components (used to model the reduction in polarized intensity as a function of $\lambda^2$ due to depolarization). Sources that exhibit multiple polarized components can be representing multiple physical components \citep[e.g. two distinct radio lobes;][]{2017MNRAS.469.4034O, 2019MNRAS.487.3432M}, making it challenging to decide which polarized component has the RM value best representing the ICM magnetism. Thus, it is important to classify the Faraday complexity of RRMs. 


To quantify if our sources are Faraday simple or Faraday complex, we use $QU$-fitting, which fits various models to the $Q$ and $U$ spectra; for details regarding $QU$ models, we refer to \citet{1966MNRAS.133...67B, 1998MNRAS.299..189S, 2012MNRAS.421.3300O}. We emphasize that $QU$-fitting was not done to obtain the RMs but only to classify complexity; the RMs were obtained from the main peak of the FDF, using RM-synthesis as described in Section \ref{sec:dave}. 

We define a Faraday simple model to model an external Faraday screen that is purely sinusoidal in $Q$ and $U$:
\begin{align}
    P(\lambda) = p_0Ie^{2i(\psi_0 + \mathrm{RM} \lambda^2)},
    \label{eq:5}
\end{align}
where $p_0$ and $\psi_0$ are the intrinsic polarization fraction and the intrinsic polarization angle, respectively,  and $P(\lambda)$ is the complex polarization vector given by: 
\begin{align}
    P(\lambda) = Q(\lambda) + iU(\lambda).
\end{align}

The second model introduces an exponential depolarization term into the Stokes $Q$ and $U$ as: 
\begin{align}
    P(\lambda) = p_0 I e^{2i (\psi_0 + \mathrm{RM} \lambda^2)} e^{-2\Sigma_{\mathrm{RM}}^2 \lambda^4},
\end{align}
where $\Sigma_{\mathrm{RM}}$ is the RM dispersion. 

The next model that we use contains two separate Faraday rotation components for the complex polarization vector, but does not have any depolarization terms: 
\begin{align}
    P(\lambda) = I( p_{0, 1} e^{2i (\psi_{0, 1}+\mathrm{RM_1} \lambda^2)} + p_{0, 2} e^{2i (\psi_{0, 2}+\mathrm{RM_2} \lambda^2)}). \label{eq:m11}
\end{align}

Next, we consider a model with both components having a single depolarization term: 

\begin{align}
    P(\lambda) = I e^{-2\Sigma_{\mathrm{RM}}^2 \lambda^4} (&p_{0, 1} e^{2i (\psi_{0, 1}+\mathrm{RM_1} \lambda^2)} \\ \nonumber &+ p_{0, 2} e^{2i (\psi_{0, 2}+\mathrm{RM_2} \lambda^2)}).
    \label{eq:m3}
\end{align}

Finally, we consider a two-component source that has separate depolarization parameters for its components: 
\begin{align}
    P = I( &p_{0, 1} e^{-2 \Sigma_{\mathrm{RM}, 1}^2 \lambda^4} e^{2i (\psi_{0, 1}+\mathrm{RM_1} \lambda^2)} \\ \nonumber &+ p_{0, 2} e^{-2\Sigma_{\mathrm{RM}, 2}^2 \lambda^4} e^{2i (\psi_{0, 2}+\mathrm{RM_2} \lambda^2)}).
\end{align}
We consider the model given by Equation \ref{eq:5} to be ``simple" and the others to be ``complex". 

For fitting the Stokes $Q$, $U$ spectra with the models outlined above, we use \pkg{RM-Tools} \citep{2020ascl.soft05003P}, which outputs the natural logarithm of the Bayesian evidence for each of the models. When comparing two models ($i$ and $j$), we compute the natural logarithm of the Bayes factor, $B_{j,i}$, defined as: 

\begin{align}
    \ln(B_{j,i}) = \ln(\mathrm{pr} (D|M_j)) - \ln(\mathrm{pr} (D|M_i)),
\end{align}
where $\ln(\mathrm{pr} (D|M_i))$ is the natural logarithm of the Bayesian evidence for the $i$-th model. Following \citet{Kass:1995loi}, we only consider the second (more complicated) model to be a better fit than the first model if $\ln(B_{j, i}) > 5$. In addition to this, if the reduced chi-squared, $\bar{\chi}^2$, of the best-fit model is not in the range $0.5\leq \bar{\chi}^2  \leq 1$, we designate that there was no best-fit $QU$ model found. 

{We found that the distribution of the $\chi^2$ values for the best-fit $QU$ models are modeled well by the theoretical $\chi^2$ probability distribution function, indicating that our models are good fits to the data}. The theoretical $\chi^2$ probability distribution function is completely determined by the degrees of freedom, which is given by $N  = v - k$, where $v$ is the number of data points and $k$ is the number of parameters in the model (which is at most 10 for the models given here). {We note} that it is not possible to rigorously choose a single $N$ as the number of parameters varies between models{, and all parameters are not necessarily linearly independent}. Therefore, we have chosen $N$ to be the number of frequency channels, which is 288. This is a reasonable assumption as $v >>k$. 

In addition to $QU$-fitting, we use the second moment of the cleaned peaks (obtained from RM-synthesis and RM-cleaning) and the $\sigma_{\mathrm{add}}$ (obtained from $QU$-fitting) complexity metrics, following \citet{2024AJ....167..226V}. Further details regarding these complexity metrics and about the classification of complexity of RRM sources can be found in Appendix \ref{app:complexity}. In all, we found {99} Faraday simple RMs, and {16} Faraday complex RMs. We note here that we will conduct our analysis both with the Faraday simple and Faraday complex RMs to gauge the effect that Faraday complexity has on our results. }

\subsection{Galactic RM correction}
\label{sec:gal}

Since RM probes the entire LOS to the background RM grid sources, any medium between the background source and the observing telescope will affect the measured RM. The largest source of contamination in extragalactic RMs comes from Galactic RM (GRM) contributions. Once the GRM has been estimated the residual RM (RRM) of the object of interest is calculated as: 
\begin{equation}
    \mathrm{RRM} = \mathrm{RM}_{\mathrm{obs}}  - \mathrm{GRM},   
\end{equation}
where $\mathrm{RM}_{\mathrm{obs}}$ is the observed RM.

{There have been various different approaches that have been used to remove GRM contributions. In recent years, the most widespread method has been to use the GRM map created by \citet{2022A&A...657A..43H}, who modeled the GRM as a product of a sign and an amplitude field and inferred the hyperparameters of the model from RM measurements. Hereafter, we refer to the inference technique used in this work as the Bayesian Rotation Measure Sky (BRMS), and the \citet{2022A&A...657A..43H} GRM map as H22. \citet{2024ApJ...977..276K} tested BRMS, along with other spatial and geometric interpolation techniques to reconstruct GRM maps; they found that natural neighbor interpolation (NNI), which is a geometric interpolation technique, produces results that are comparable to BRMS.}

{In contrast to these works}, \citet{2024MNRAS.533.4068A} remove GRM contributions using statistical properties of the RMs. They aimed to estimate the GRM contribution at each RM source {by} defining an exclusion zone {(a circle of some radius $r$)} around it so that the GRM model does not erroneously include coherent RM signal from the extragalactic RM structure that is being studied. The GRM is then taken as the median of the 40 closest RMs outside this exclusion radius; the choice of this number is motivated because the outer radius of these 40 sources is typically on the order of 1 deg (around 1.5 Mpc at A3581's redshift) and therefore any local RM contribution due to the cluster in our GRM estimate is minimized. In our work, we used an exclusion radius of 1 Mpc {$\approx R_{500}$}, so that RM structure from the cluster is not removed\footnote{We note here that for the purposes of the GRM correction we use all sources in the POSSUM tile (not just sources within $2R_{500}$). This is done to prevent the correction for sources near the edge from being dominated by internal cluster sources.}. {Hereafter, we refer to this method as the exclusion radius GRM subtraction (ERGS).}

{In the subsequent analysis, we use ERGS to obtain the RRMs, use bootstrapping (of the median of the 40 closest RMs outside the exclusion zone) to obtain errors on the correction, and calculate the total RRM error by adding the error in the correction and the error in the observed RM from RM-synthesis in quadrature. We decided against using the GRM map produced by \citet{2022A&A...657A..43H} as in this particular region, they were limited to using $\sim 1.7$ RMs deg$^{-2}$ for the inference; therefore, the map might be unreliable for smaller scales. We avoided using NNI for the reconstruction of the GRM map as it required too many assumptions about the spatial distribution of the RMs on the sky (see Appendix \ref{app:a} for further details). Although we believe the EGRS method is best suited for this field given the reasons above, the RRM scatter profiles (the standard deviation in the RRMs as a function of distance from the cluster center; see Section \ref{sec:rrm_scatter}) after all three correction methods are very similar (see Figure \ref{fig:grm_scatters}), and comparable to what is found in previous studies of other clusters \citep[e.g.,][]{2025A&A...694A..44O}.

Figure \ref{fig:skyRM} displays the observed RMs (without any Galactic correction) on the sky. There are a total of 888 RMs in the POSSUM field 1412-28, with a mean RM of $-27.2$ rad m$^{-2}$, a standard deviation of $12.4$ rad m$^{-2}$, and a root-mean-square of $29.9$ rad m$^{-2}$. The full data for the POSSUM field 1412-28 will be released as part of POSSUM's Data Release 1. Additionally, in this figure, we have plotted circles indicating $2R_{500}$ of A3581 and of three nearby clusters identified by \citet{2024ApJS..272...39W} to give a sense of the large-scale structure in the neighborhood of A3581; the properties of these additional clusters are listed in Table \ref{tab:clust}. To identify the closest clusters in redshift to A3581, we used a fixed velocity gap (the maximum allowed difference in the recession velocity of objects) of 6000 km s$^{-1}$. Figure \ref{fig:a3581rrm} displays the values of the RRMs observed within 2$R_{500}$ of A3581 (see Figure \ref{fig:GRM_grms} for the GRM values determined using ERGS). Notably, the RRM values do not appear to be completely randomly distributed, with positive values preferentially in the north-west and negative values in the south-east. 

\begin{table*}[]
    \centering
    \caption{Properties of the three closest clusters to A3581 identified from the \citet{2024ApJS..272...39W} galaxy cluster catalog}
    \begin{tabular}{|c|c|c|c|}
    \hline 
        \textbf{Cluster Name} & \textbf{Cluster center (ICRS)} & $\mathbf{R_{500}}$ \textbf{(Mpc)} &  $\mathbf{z}$\\
    \hline
    \hline
        WH-J135418.5-265338 & (13h 54m 18.5s, $-26^\circ~53^\prime~38^{\prime\prime}$) & 0.625 & 0.0200\\
       WH-J141826.6-272244 & (14h 18m 26.6s, $-27^{\circ}~22^\prime~44^{\prime\prime}$) & 0.825 & 0.0257\\
        WH-J142949.1-294455 & (14h 29m 49.1s, $-29^{\circ}~44^\prime~55^{\prime\prime}$) & 0.522 & 0.0230\\
        \hline
    \end{tabular}
    \label{tab:clust}
\end{table*}

\begin{figure}[!htb]
    \centering
    \subfigure[]{\includegraphics[width=0.48\textwidth]{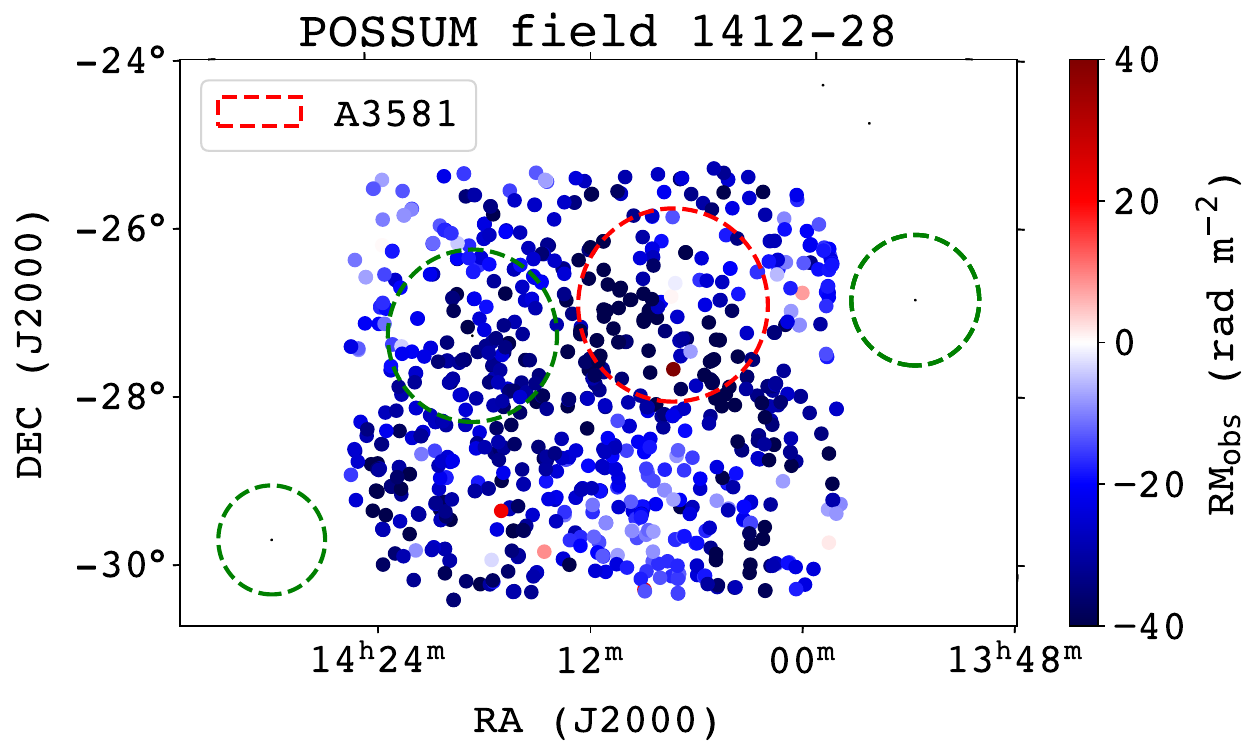}\label{fig:skyRM}}    
    \subfigure[]{\includegraphics[width = 0.48\textwidth]{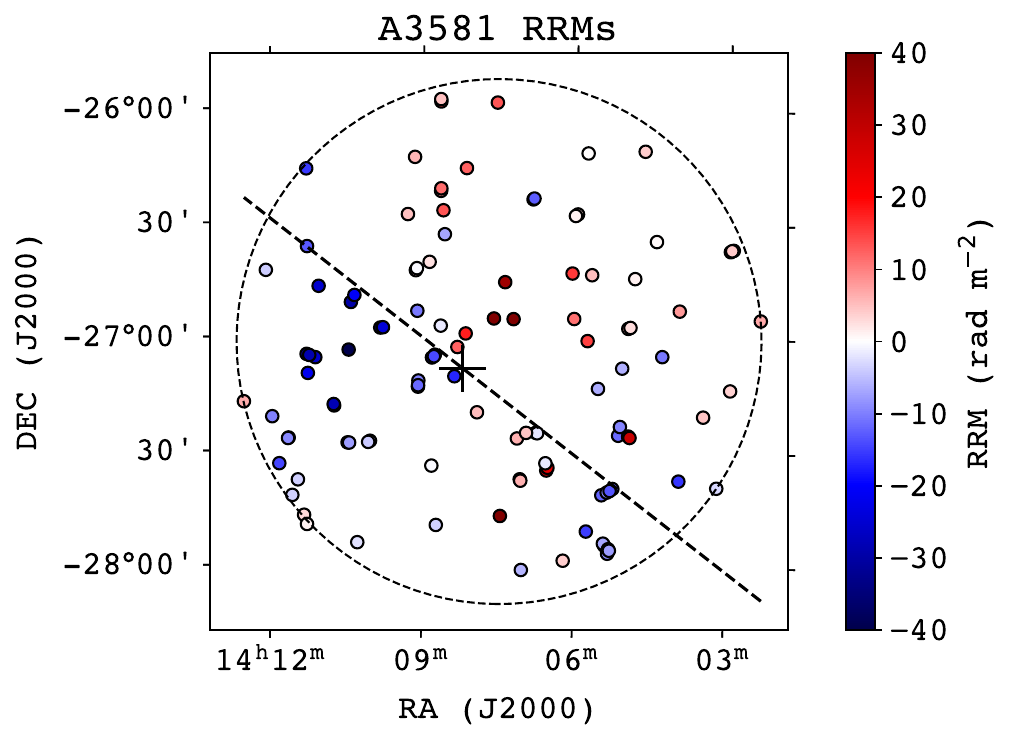}\label{fig:a3581rrm}}
    \caption{(a) The locations and values of the observed RMs across the whole POSSUM field, along with identified nearby clusters. The black circle indicates $2R_{500}$ for A3581 and the green circles indicate $2R_{500}$ for nearby clusters. (b) The locations and values of the RRMs that are within $2R_{500}$ (marked with a dashed circle) of A3581. The color bars represent the RRMs saturated from $-40$ rad m$^{-2}$ to $+40$ rad m$^{-2}$. The plus sign shows the location of the center of RM, and the dashed line portrays the axis of symmetry (see Section \ref{sec:asymmetry} for further details).}
\end{figure}

\subsection{Cluster membership of sources}
{Since Faraday rotation is an integrated effect along the line of sight, it is important to know the location of each RM source with respect to the medium that we are probing. However, given the significant velocity dispersion of cluster members, it is impossible to determine where they are located with respect to the ICM. Background RMs do not suffer this uncertainty as they are located fully behind the cluster}. For this reason, we only retain background radio sources for our analysis. 
{We determine cluster membership of sources using the photometric and spectroscopic redshift of sources (see Appendix \ref{app:clustermember}). In all, we found that only 4 RMs are inside the cluster, leaving us with 111 RRMs projected within $2R_{500}$ of the X-ray centroid.}

\subsection{Magnetic field modeling}
\label{sec:mag_modeling}
{In the simplified picture of Kolmogorov turbulence with scale-by-scale equipartition between the energy density of magnetic fields and turbulent motions, the magnetic field is expected to behave as a Gaussian random field with a single power-law power spectrum \citep[e.g.,][]{2004ApJ...612..276S}:}
\begin{equation}
    |B_k| \sim k^{-5/3},
\end{equation}
where $|B_k|$ is the Fourier amplitude of the magnetic field and $k$ is the magnitude of the wave vector given by $ k = \frac{\pi}{\Lambda}$, where $\Lambda$ is the physical fluctuation scale. {We note that this is the 1D power spectrum; the 3D power spectrum has an index of $-11/3$.} In our models, we use a box size of 2048$^3$, with each pixel representing 2 kpc. We set the maximum fluctuation scale to be $\Lambda_{\mathrm{max}} = $ 100 kpc \citep[this matches well with the $\sim 10^2$ kpc maximum fluctuation scale found in polarized emission observations and simulations of clusters, e.g.,][]{2004A&A...424..429M, 2005A&A...430L...5G} and the minimum fluctuation scale to be $\Lambda_{\mathrm{min}} = 4$ kpc (this corresponds to a field reversal between adjacent pixels). We note that we do not test different fluctuation scales, which can also affect the RM scatter profiles, but are partially degenerate with other parameters such as the magnetic field strength. To keep the number of free parameters limited, we model the magnetic field as a Gaussian random field, following the Kolmogorov power spectrum{, and scaling with the electron density as in Equation \ref{eq:1}}. {We explore the parameter space $\eta = {0, 0.25, 0.5}$ and treat $B_0$ as a free parameter, which we infer from MCMC sampling. We do not explore $\eta$ as a free parameter as it is computationally expensive to do so (we would need to create multiple realizations of the random fields for every $\eta$); however, it is possible to explore the $B_0$ parameter space because $\sigma_{\mathrm{RM}}\propto B_0$. Our choice for $\eta$ is motivated by the following: $\eta = 0$ corresponds to the extreme case where the magnetic field has no dependence on the electron density, $\eta  = 0.5$ corresponds to the standard power-law index observed in clusters like Coma \citep[e.g.][]{2010A&A...513A..30B}, and $\eta = 0.25$ corresponds to the intermediate case. We do not explore higher power-law indices as our observed RRM scatter profile is relatively flat and going for higher $\eta$ will produce models with scatter profiles far steeper than our observations.} To calculate the magnetic field models and RM observables, we use the \pkg{GRAMPA}\footnote{\url{https://pypi.org/project/grampa/}} Python module.

{Although the assumption of a Gaussian random magnetic field with a Kolmogorov power spectrum is an idealized case, this has been the standard assumption in cluster magnetic field studies. However recent works have expanded on this \citep[e.g.,][]{2021MNRAS.502.2518S} or have shown that more advanced modeling is needed \citep{2025A&A...694A..44O}. Still, for simplicity and consistency with previous works, we initially compare the observations with Gaussian random field models of the magnetic field. }
 {Previous studies \citep[e.g.,][]{2004A&A...424..429M} have modeled the electron density profile as a simple single $\beta$ model}: 
\begin{align}
    n_e(r) = n_e(0) \left[ 1 + \left(\frac{r}{r_c}\right)^2\right]^{-3\beta/2},\label{eq:ne}
\end{align}
 where $n_e(0)$ is the thermal electron density at the cluster center, $r_c$ is the radius of the X-ray core and $\beta$ is the power-law index. {The only available parameters in the literature for the A3581 $\beta-$model are from \citet{2004PASJ...56..965F}, who found $n_e(0) = (33.60_ {- 14.02}^{+0.00}) \times 10^{-3}~\mathrm{cm}^{-3},~ r_c = 10.18_{-0.00}^{+6.70}~\mathrm{kpc},~ \beta = 0.47_{-0.00}^{+0.05} $, using archival X-ray data from the Advanced Satellite for Cosmology and Astrophysics \citep{1994PASJ...46L..37T}. However, X-ray emission was only detected out to a radius of $\sim$ 10 kpc, so the radial profile beyond this radius is strongly unconstrained.} Because clusters are relatively self-similar \citep{2010A&A...517A..92A}, we {model the radial profile of the electron density distribution using} the mean $n_e$ profile {from \citet{2022A&A...665A..71O}, determined from X-ray observations of 102 clusters from the Chandra-Planck Legacy Program for Massive Clusters of Galaxies \footnote{\url{https://hea-www.cfa.harvard.edu/CHANDRA_PLANCK_CLUSTERS/}}}. {We rescale this {mean} $n_e$ profile to be consistent with the measurements of the central electron density made by \citet{2004PASJ...56..965F}{, while following the scaling with radius as determined by \citet{2022A&A...665A..71O}}.}
 
 In addition to fluctuations in the magnetic field strength, fluctuations in the electron density might also contribute to the RM scatter. Multiple studies \citep[e.g.,][]{2007ApJ...659..257K, 2014A&A...569A..67G, 2024ApJ...976...67M} of simulations of clusters have demonstrated that the electron density in the ICM has lognormal fluctuations. We model the power spectrum to be {Kolmogorov as a first-order approximation which broadly agrees with simulations \citep[e.g.,][]{2014A&A...569A..67G}, although thermal conduction could flatten this spectrum in reality. We note that while the large-scale magnetic field amplitude is normalized to the radial profile of $n_e$, we treat the magnetic field and electron density fluctuations to be uncorrelated and statistically independent in our models. \pkg{GRAMPA} allows fluctuations in the electron density model that are generated with the \pkg{pyFC}\footnote{\url{https://www2.ccs.tsukuba.ac.jp/Astro/Members/ayw/code/pyFC/}} module. We limited the fluctuations to be within 10\% of the mean $n_e$ profile, as found in the Coma cluster by \citet{2012MNRAS.421.1123C}. Together, these assumptions allow us to construct a simplified but tractable model of Faraday rotation in a turbulent ICM.}

For all the models, we sample the modeled RM maps at the same locations (with respect to the cluster center) as the RRM observations in A3581 to fully address any spatial correlation between RRMs. Furthermore, since complex RMs might be experiencing beam depolarization, we attempt to imitate the effects of depolarization when we samle our models by averaging Stokes $Q$ and $U$ separately (across all POSSUM frequency channels) for all pixels within one ASKAP telescope beam, which has a size (full width at half maximum) of  $20^{\prime\prime}$ corresponding to a circle with diameter $\sim8.93$ kpc at A3581's redshift, and this results in 13 pixels within a beam. For each pixel within one telescope beam around the complex RM, we assume a simple model for the Stokes $Q$ and $U$ parameters (given by Equation \ref{eq:5}), where we assume the RM to be the RM of the pixel, $p_0 = 0.0654$ (which we obtain from the median polarization fraction for our sources) and assume a power law for Stokes I:
\begin{align}
    I = I_0\left(\frac{\nu}{\nu_0}\right)^\alpha, 
\end{align}
where we found that the median $I_0 = 7.45$ mJy for our polarized sources at a reference frequency of $\nu_0 = 800$ MHz, and we found the median spectral index to be $\alpha = -0.768$. This is similar to the weighted mean spectral index of $\langle \alpha \rangle = 
 -0.7870 \pm 0.0003$ found by \citet{2018MNRAS.474.5008D} for radio sources in the TIFR GMRT Sky Survey and the NVSS. Once we have produced an average Stokes $Q$ and average Stokes $U$ for the complex RMs, we conduct RM-synthesis and RM-cleaning to obtain the RMs for the complex RMs.
 
{We note here that we have made several simplifying assumptions in our model, chief of which is that the fluctuations in the magnetic field are independent of the fluctuations in the electron density field. A proper treatment of this would require a full MHD simulation. Thus, we also compare our} results with {simulated galaxy clusters} from the TNG-Cluster project \citep{2024A&A...686A.157N} in Section \ref{sec:TNG}.

\section{Results}
\label{sec:results}

\subsection{RRM scatter profile}
\label{sec:rrm_scatter}
{Cluster magnetic fields are theorized to have been amplified from random seed magnetic fields by a turbulent dynamo process \citep{2018SSRv..214..122D}. In this process, random velocity fields stretch and fold pre-existing field lines to amplify the magnetic field to a saturation level. Given the random nature of this process, the magnetic field orientations should be random and the average RRM will thus be zero.} Therefore, traditionally, the magnetic field of clusters is probed by studying the scatter in RRM as a function of radius from the cluster center; a larger scatter in RRM generally indicates a stronger $B_\parallel$ or larger $n_e$.

We computed the standard deviation in the RRMs in annuli over the sky as a function of the projected distance to the cluster center. We used a moving bin (with the radius for each bin being determined by its left edge) with 20 points (corresponding to a median bin width of {0.27} Mpc) and computed the scatter in the RRM, denoted as $\sigma_{\mathrm{RRM}}$, to be the interquartile range (IQR) divided by 1.349 in each bin. {We note that to test how the number of points per bin affect the results, we conducted the analysis with 10 points per bin too and found no changes in the results.} Furthermore, we also corrected for the extrinsic scatter ({the RM scatter due to the intergalactic medium, the local environment of a radio source and the ionosphere}), denoted as $\sigma_{\mathrm{RRM, ext}}$. {Initially, we computed $\sigma_{\rm{RRM, ext}}$ as the mean of running standard deviation from 2$R_{500}$ to 4$R_{500}$, as we expect RM enhancement due to the ICM to be relatively low in this region and it is also local to the cluster, therefore giving a good representation of the extrinsic scatter in the cluster's neighborhood. This approach resulted in $\sigma_{\mathrm{RRM, ext}} = 5.4 \pm 1.9$ rad m$^{-2}$, where the error is taken to be the standard deviation in the running RRM scatter from 2$R_{500}$ to 4$R_{500}$. However, we decided against this as this range will inadvertently encroach into the neighboring clusters and also possible bridge regions between the clusters, and therefore not give a reliable estimate of the extrinsic scatter. For this reason, we define a region with possible extragalactic plasma to be a collection of cylinders that connects (and contains) all the clusters with a radius of 1 Mpc (see Figure \ref{fig:masked}), which is the typical radius of short filaments between clusters as found in cosmological simulations \citep{2021A&A...649A.117G}. Thus, we define the extrinsic scatter as the mean running scatter in {two control regions (see Appendix \ref{app:control}) where there is no large extragalactic structure} (like galaxy clusters and possible bridges between clusters)}. {When the extragalactic regions} are masked, we found $\sigma_{\mathrm{RRM, ext}} = {7.0 \pm 1.3 }$ rad m$^{-2}$. This agrees within error with the extragalactic RM scatter of $6.5 \pm 0.1$ rad m$^{-2}$ found by \citet{2010MNRAS.409L..99S}, and also agrees with the values found by \citet{2024MNRAS.528.2511T} of $5.9 \pm 2.7$ rad m$^{-2}$ and $6.3 \pm 2.2$ rad m$^{-2}$ for the COSMOS and \textit{XMM}-LSS fields, respectively.

Then, we calculate the corrected RRM scatter as:
\begin{equation}
\label{eq:sigma_corr}
    \sigma_{\mathrm{RRM, corr}} = \sqrt{\sigma_{\mathrm{RRM}}^2 -  {\frac{\sum_{i = 1}^N \delta\mathrm{RRM}_i^2}{N-1}} - \sigma_{\mathrm{RRM, ext}}^2},
\end{equation}
where $\delta \mathrm{RRM}_i$ is the {uncertainty in the RRMs and the sum is taken over all the RRMs in the bin}. We also note here that we calculate $\sigma_{\mathrm{RRM, ext}}$ by removing measurement uncertainties (as in Equation \ref{eq:sigma_corr}) and that the value of $\sigma_{\mathrm{RRM, ext}}$ we obtain is highly dependent on the signal-to-noise threshold used for retaining sources in the RM grid as shown by \citet{2024AJ....167..226V}.

Figure \ref{fig:rrm} displays the RRMs as a function of the projected distance to the center of the cluster. As expected, most of the RRMs {scatter around zero}. The only clear outlier in these plots is the RRM that is around $\sim 100$ rad m$^{-2}$, and it is likely due to a local increase in the magnetic field strength or the electron density around the emitting source; since we use statistics that are robust against outliers (e.g. IQR), this outlier will not affect our results.

\begin{figure*}
    \centering
    \subfigure[]{\includegraphics[width=0.45\textwidth]{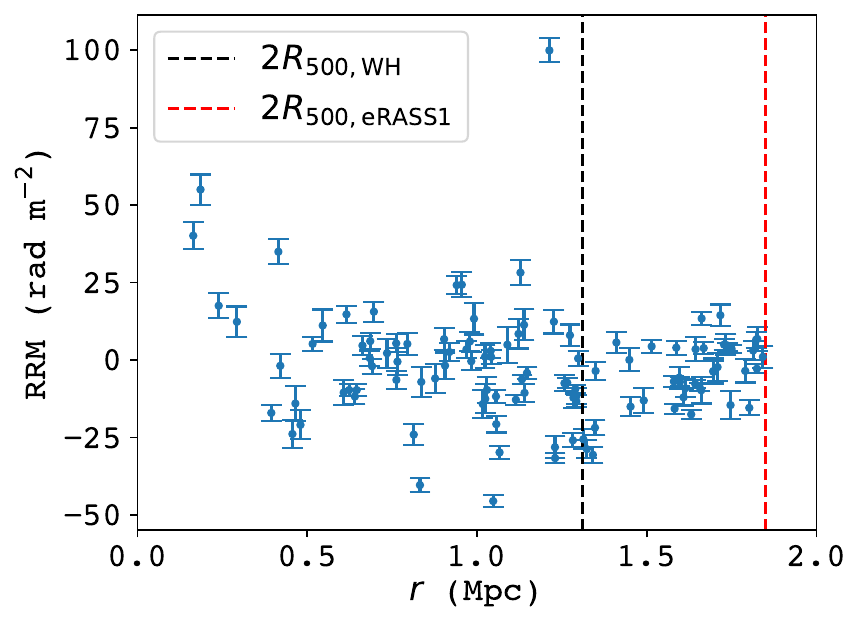} \label{fig:rrm}}
    \subfigure[]{\includegraphics[width = 0.45 \textwidth]{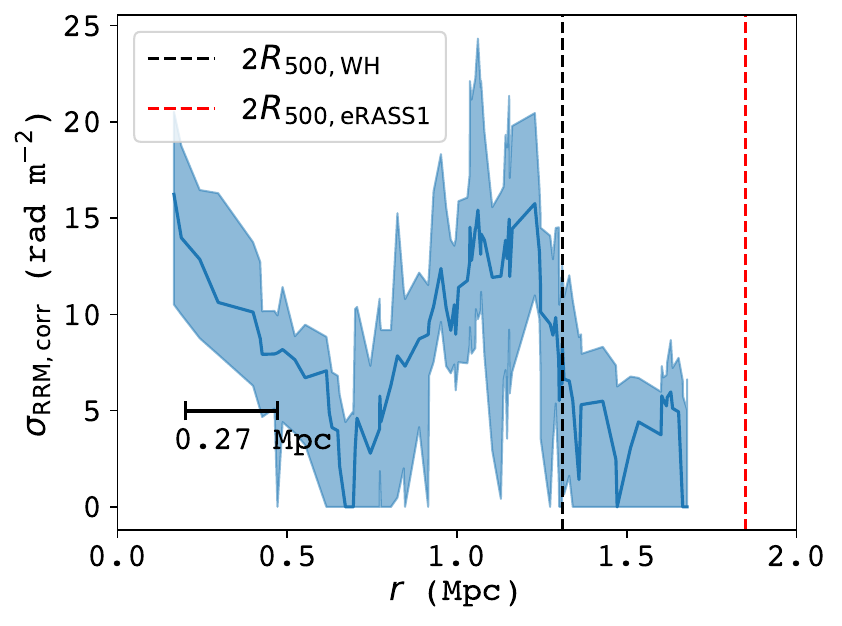}  \label{fig:rrm_scatter}}
    \caption{(a) The RRMs as a function of the projected distance to the X-ray centroid. The black line indicates the $2R_{500}$ radius using the value from \citet{2024ApJS..272...39W}, and the red line indicates the $2R_{500}$ radius using the value from the SRG/eROSITA all-sky survey \citep{2024A&A...685A.106B}. (b) {The RRM scatter as a function of the projected distance from the center of the cluster. To calculate the scatter we used a running bin and fixed the number of points per bin to be 20. The median bin width is 0.27 Mpc.}}
\end{figure*}

The blue line in Figure \ref{fig:rrm_scatter} displays the scatter profile of the RRMs within 2$R_{500}$ of the cluster. This profile was produced by including all complex RMs; the profile created after excluding the 15 background complex RMs (one of the complex RMs was identified to be embedded in the cluster) was similar (within uncertainties) to this profile. Based on Equation \ref{eq:1} and the typical electron density profile of a galaxy cluster, we expect the scatter in the RRM to decrease monotonically as a function of the distance from the cluster center. This is the case in the interior of the cluster (at $r < 0.75$ Mpc). However, for $r > 0.75$ Mpc, the cluster's scatter does not decay monotonically, contrary to what is expected.
{We note} that there is still {measurable non-zero} scatter between $2R_{500, \mathrm{WH}}$ and $2R_{500, \mathrm{eRASS1}}$. This {means that the ICM extends significantly out to $\sim 1.75$ Mpc, being more consistent with the eRASS1 estimate of $R_{500}$}. {We explore if the presence of the cluster is what contributes to the enhanced in the RRM scatter by comparing the scatter profile of A3581 to two control regions of the same size on this POSSUM tile in Appendix \ref{app:control}.} 

\subsection{Magnetic field modeling in the interior of A3581}
\label{sec:mod_res}
In this section, we compare the observed RM grid to semi-analytic magnetic field models of increasing complexity, as has been done in previous studies \citep[e.g.,][]{2004A&A...424..429M, 2010A&A...513A..30B, 2025A&A...694A..44O}. In particular, we only attempt to model the interior of the cluster ($r < 0.75$ Mpc); this is the region over which the observed RRM scatter is monotonically decreasing. The behavior of the RRM scatter outside this radius is more complicated and will not be well-described by a simple radially declining magnetic field and electron density model. {This will be addressed in the following sections.}

{In order to determine which model most accurately represents the observed RRM, we use the Bhattacharya coefficient \citep[BC;][]{lee2012separability}, which is a bounded, symmetric similarity measure for two Gaussian distributions that accounts for both differences in mean and variance. Here, we model the RRM scatter at each radius to be a normal distribution, with mean given by $\mu_ {\sigma_{\mathrm{x}}}$ (the solid lines in Figure \ref{fig:lognormal_model_scatters}) and standard deviations given by the error in the RRM scatter $\delta_{ \sigma_{\mathrm{x}}}$ (the filled regions in Figure \ref{fig:lognormal_model_scatters}); here, $x$ is either the model or the observation. }

{Then, the BC of the model and the observation for a particular radius represents the overlap of the two scatters (for a fixed radius) and is given by: }

\begin{multline}\label{eq:BC}
    {\mathrm{BC}(\sigma_{\mathrm{obs}}, \sigma_{\mathrm{mod}}, r) =}\\ {\sqrt{\frac{2\delta_{\sigma_{\mathrm{obs}}}(r) \delta_{\sigma_{\mathrm{mod}}}(r)}{\delta_{\sigma_{\mathrm{obs}}}^2(r) +\delta_{\sigma_{\mathrm{mod}}}^2(r)}} \exp\left(-\frac{\{\mu_{\sigma_{\mathrm{obs}}}(r) - \mu_{{\sigma_\mathrm{mod}}}(r)\}^2}{4\{\delta_{\sigma_{\mathrm{obs}}}^2(r) +\delta_{\sigma_{\mathrm{mod}}}^2(r)\}}\right),}
\end{multline}

{Then, we define the normalized overlap metric, $\Phi$ as follows:}

\begin{multline}\label{eq:overlap}
{{\Phi(\sigma_{\mathrm{obs}}, \sigma_{\mathrm{mod}})}} \\  {{ = 1 -  \dfrac{\int \mathrm{BC}(\sigma_{\mathrm{obs}}, \sigma_{\mathrm{mod}}, r) dr}{\int \mathrm{BC}(\sigma_{\mathrm{obs}}, \sigma_{\mathrm{obs}}, r) dr},}} 
\end{multline}
{where $r$ is the distance from the cluster center, $\sigma_{\mathrm{obs}}$, $ \sigma_{\mathrm{mod}}$, are the scatter in the RRM for the observation and the model, respectively. From our definition of $\Phi$, models that have scatter profiles that are more similar to that of A3581 will produce a $\Phi$ that is closer to zero. We also note that, we are only modeling the interior of A3581.}

First, we compare our observation to a model with uncorrelated lognormal fluctuations in the electron density content and normal fluctuations in the magnetic field. {For each value of $\eta,$ we calculated the scatter profile (which is measured from the X-ray centroid in Table \ref{tab:1} because the X-ray centroid probes the peak of the gas density profile) for the various $B_0$; we used the loss function in Equation \ref{eq:overlap} to define a likelihood and used Markov Chain Monte Carlo sampling to explore the posterior distribution of $B_0$, from which we infer the best-fit $B_0$ values for each $\eta$. In doing this, we found the following best-fit values of $B_0$, for $\eta = 0, B_0 = 0.16^{+0.03}_{-0.02}~\mu\mathrm{G}$; for $\eta = 0.25, B_0 = 0.59^{+0.09}_{-0.08}~\mu$G; for $\eta = 0.5, B_0= 2.20^{+0.36}_{-0.32}~\mu$G.}
{Figure \ref{fig:lognormal_model_scatters} displays the comparison plots of the observed and modeled RRM scatter for the best-fit $B_0$ for each $\eta$. All the individual scatter profiles for the models follow the expected trend of a monotonically decaying RRM scatter, and appear to fit the observed scatter profile well for $r < 0.75$ Mpc.} However, none of the models are able to reproduce the full complexity of the observed RRM scatter profile of A3581; in particular, none show the non-monotonic behavior at $r > 0.75$ Mpc.
\begin{figure*}
    \centering
    \gridline{\includegraphics[width = 0.33\textwidth]{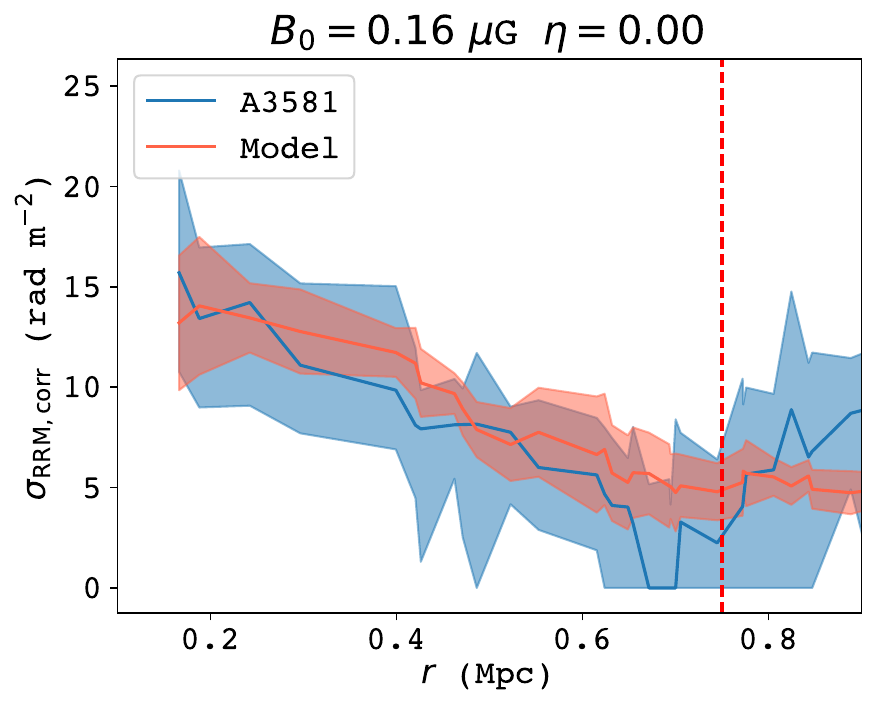} \includegraphics[width = 0.33\textwidth]{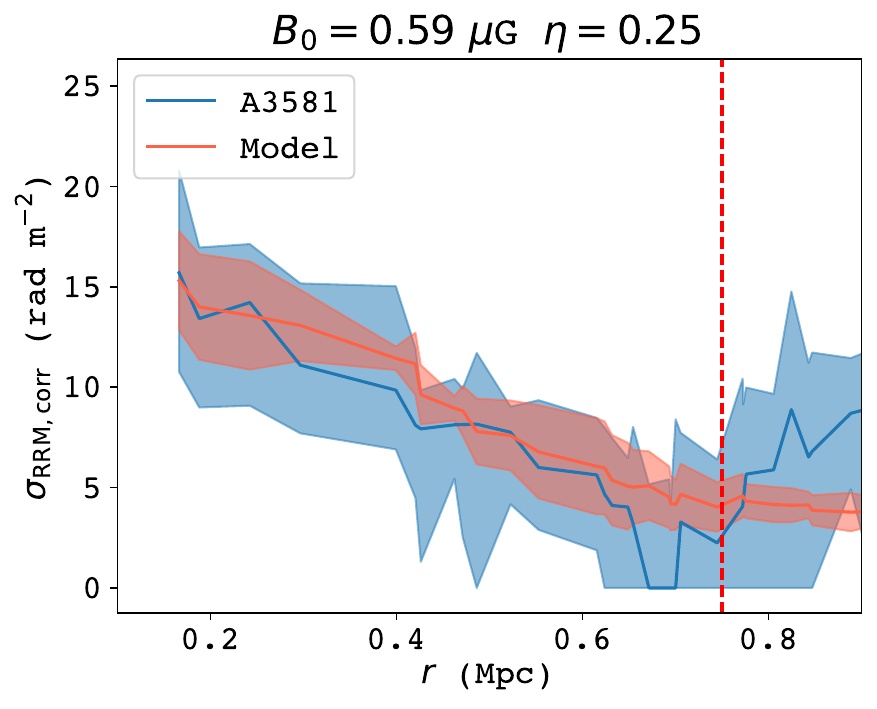} \includegraphics[width = 0.33\textwidth]{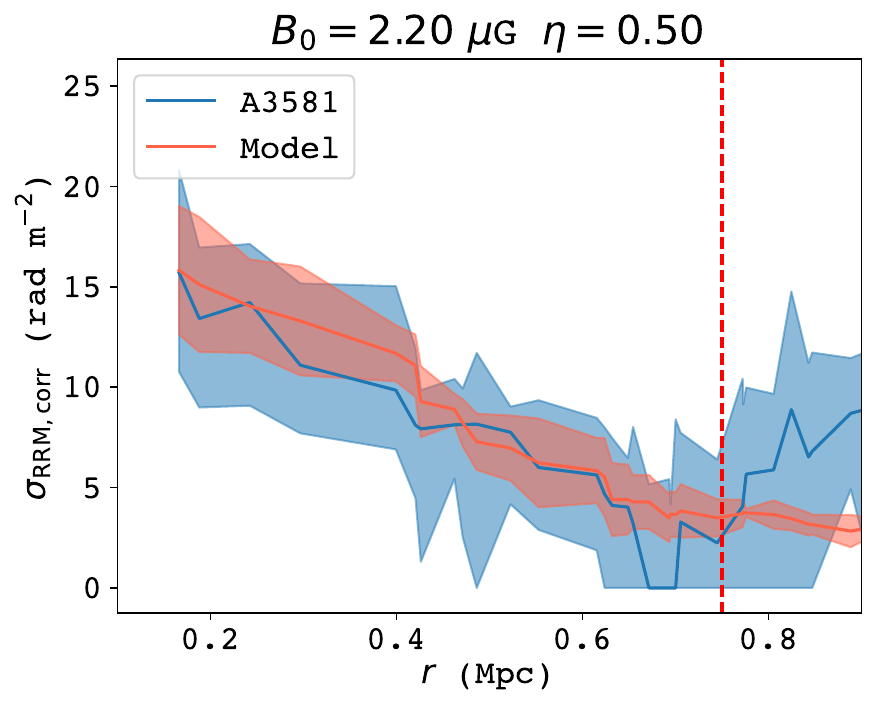}}
    \caption{{RRM scatter plots as a function of radius for {the best-fit $B_0$ cluster models for $\eta = {0, 0.25, 0.5}$}. For each model, we ran ten iterations; the solid red line indicates the median RRM scatter and the shaded region indicates the 1$\sigma$ scatter. The solid blue line is the observed RRM scatter in A3581 as in Figure \ref{fig:rrm_scatter}, with the shaded region indicating the uncertainty. {The red dotted line indicates $r = 0.75$ Mpc, beyond which we do not compute the overlap metric (Eq. \ref{eq:overlap}).} }}
    \label{fig:lognormal_model_scatters}
\end{figure*}

 {We computed the values for $\Phi_{r < 0.75\rm{~Mpc}}$ (the overlap metric in the interior) for the models with fluctuations in both electron density and the magnetic fields, {centered on the X-ray peak with values of $n_e(0)$ fixed at the best value from the literature of $33.6\times 10^{-3}$ cm$^{-3}$ \citep{2004PASJ...56..965F} and the radial profile of the electron density determined from a self-similar scaling. The values of $\Phi_{r < 0.75\rm{~Mpc}}$ for the best-fit models are: for $\eta = 0, \Phi_{r < 0.75\rm{~Mpc}} = 0.124$; for $\eta = 0.25, \Phi_{r < 0.75\rm{~Mpc}} = 0.112$; and for $\eta = 0.5, \Phi_{r < 0.75\rm{~Mpc}} = 0.126$}}. We found that the model without any fluctuations in the electron density also results in similar scatter profiles to the model with fluctuations in the electron density for the vast majority of magnetic field strengths and values of $\eta$. This also results in these models predicting similar best-fit models.

\subsection{{RM scatter profiles in TNG-Cluster}}
\label{sec:TNG}

The analytic models are able to estimate the best-fit mean magnetic strength and scaling with electron density of A3581 from a set of assumed values. However, the models fail to reproduce the non-monotonic nature of the RRM scatter profile for $r > 0.75$ Mpc. To better understand the origin of this behavior, we investigate the RRM scatter profiles of the simulated clusters in the MHD cosmological zoom-in simulation TNG-Cluster\footnote{https://www.tng-project.org/cluster/} \citep{2024A&A...686A.157N}.


TNG-Cluster re-simulated 352 massive clusters sampled from a 1 Gpc$^3$ size cosmological box with a high baryonic mass resolution $\sim10^{7}M_{\odot}$. The simulations were performed using the moving-mesh code \pkg{AREPO} \citep{2010MNRAS.401..791S}, which implements state-of-the-art astrophysics models that successfully reproduce a broad range of observed properties across different scales \citep[e.g.,][]{2018MNRAS.475..648P,2018MNRAS.474.2073V,2018MNRAS.481.1809B,2018MNRAS.480.5113M}. Unlike the analytic models, TNG-Cluster provides a direct estimate of RM by solving the ideal continuum MHD equations, allowing for the self-consistent evolution and amplification of intracluster magnetic fields \citep{2011MNRAS.418.1392P}. From an initial homogeneous magnetic field strength of $10^{-14}$ comoving Gauss, the field is amplified through compression, turbulence, and shear flow, reaching $\mu$G-scale strengths in the cluster environment \citep{2018MNRAS.480.5113M,2024A&A...686A.157N}.  


We estimate RM in the simulated clusters by mimicking the observation. We begin by selecting 121 galaxy clusters from the simulation at redshift $z=0$, with masses in the range $M_{\rm 500} = [1.4,~3.4] \times 10^{14}~M_{\odot}$. 
For each simulated cluster, the RMs are placed at the observed positions in A3581 and scaled by $R_{500}$ to preserve their spatial distribution relative to the cluster center. 
The RM contribution from the simulated ICM is computed using all gas particles within a projected depth of $\pm 2 R_{\rm 200}$ from the cluster center along the LOS. 
The size of each gas particle is estimated from its mass and density, assuming a spherical geometry. 
We identify particles whose radial size is larger than their shortest distance to the line of sight to an RM, such that they intersect the LOS and contribute to the RM.
Then, the contribution to the RM from each intersecting particle is calculated using its LOS magnetic field component, electron density, and the chord length of the LOS path through its spherical volume.
For RMs identified as Faraday complex,
we follow the same procedure of averaging Stokes $Q$ and $U$ as in Section \ref{sec:mag_modeling}.

Finally, to improve the statistics and incorporate projection effects, we repeat the procedure along the $x$, $y$, and $z$ projection axes. For each projection, we generate 18 different realizations by rotating the RM positions around the cluster center in 20-degree increments, while preserving their relative positions in units of $R_{500}$. This results in 54 realizations for each cluster, producing $\sim$6500 RRM profiles in total. 

Figure \ref{fig:TNG_rms} presents the three simulated analogues of separate clusters from TNG-Cluster whose RM scatter profiles most closely resemble that of A3581. These clusters were identified by searching for cluster configurations that minimize the overlap metric $\Phi$ (given in Equation \ref{eq:overlap}) over all radii. The overlap metric for these clusters was found to be ${0.228},~ {0.217},~ {0.241}$. As shown in Figure \ref{fig:TNG_rms}, TNG-Cluster exhibits analogues where the simulated clusters show a comparable enhancement in the RRM scatter at $\sim 1.1 R_{\rm 500}$, with the elevated scatter {extending over a radial width of $\sim 0.4R_{\rm 500}$.} This trend is highly sensitive to the spatial distribution of RMs, as the scatter profile becomes monotonically declining when RMs are projected along a different axis or under a different rotation. One of the simulated analogues is a CC cluster {(Halo ID 250)}, one is a weak cool core (WCC) cluster {Halo ID 231}, and the other is a non-cool core (NCC) cluster {Halo ID 255} based on the central entropy \citep{2024A&A...687A.129L}. {We display the sampled RMs from these simulated clusters in Figure \ref{fig:TNG_rm_dots}.}

\begin{figure*}
    \centering
    \subfigure[]{\includegraphics[width=0.49\textwidth]{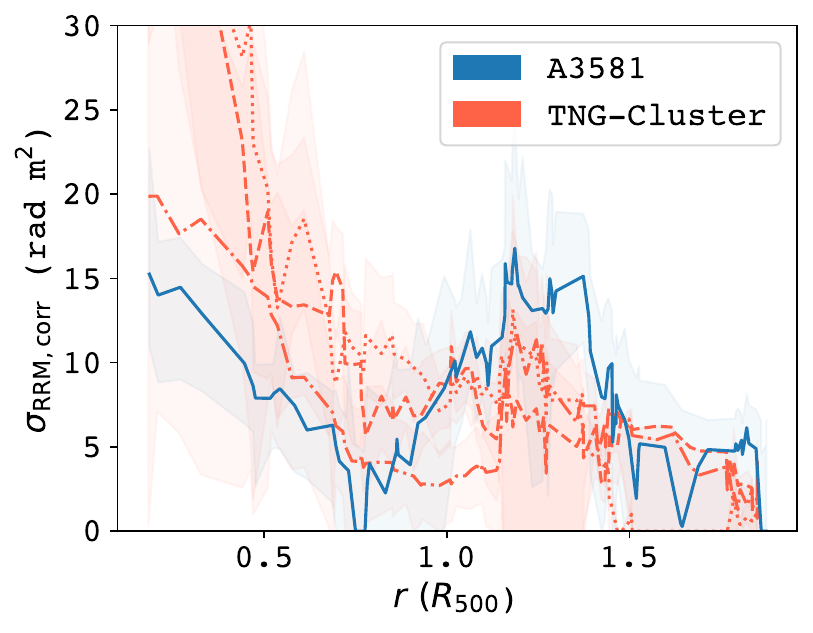}\label{fig:TNG_rms}}
    \subfigure[]{\includegraphics[width = 0.49\textwidth]{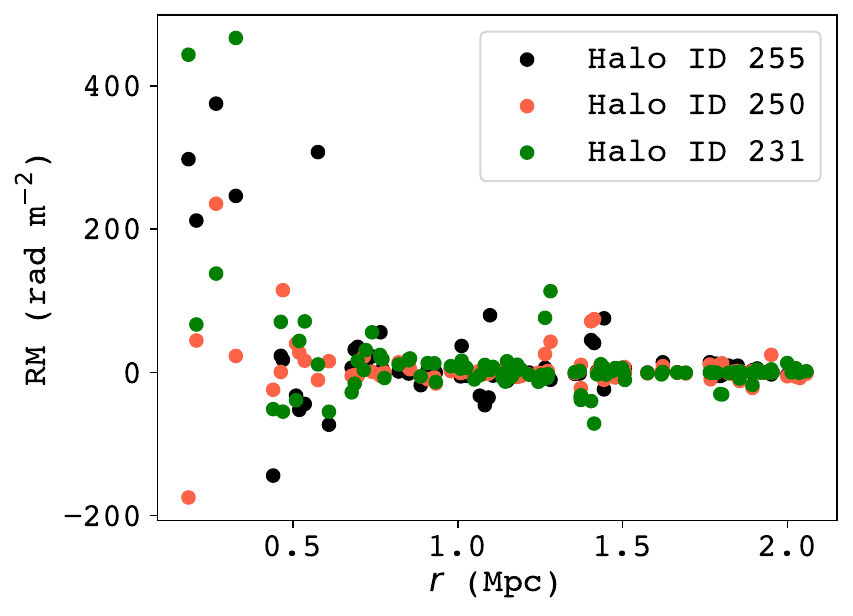}\label{fig:TNG_rm_dots}}
    \caption{{(a)} Comparison of the RM scatter in A3581 (in blue) and three simulated analogues (in red) of separate clusters that were found in TNG-Cluster. The different simulated clusters are indicated by different line styles. {(b) The RMs of the closest analogs to A3581 from TNG-Cluster.}}
    
\end{figure*}

Figure \ref{fig:sim_analaog} presents the LOS magnetic field of the CC analog to A3581 in TNG-Cluster; this analog has the Halo ID 250, and was found to have an overlap metric value $\Phi = {0.217}$ and is shown as the dash-dot line in Figure \ref{fig:TNG_rms}. This system appears to be interacting with a neighboring cluster to the north through accreting mass and also in the process of merging with a subcluster to the east. As presented in the RM map of this cluster, this activity has resulted in the enhancement of RM scatter in the outskirts of the cluster, and has stretched the magnetized ICM along the axis of collision with the subcluster beyond $R_{500}$. This suggests that the non-monotonic nature of the RRM scatter at the cluster outskirts is tracing the in-falling subcluster and that we are observing a complex scatter profile that cannot be fully described by a single halo profile.


\begin{figure}
    \centering
    \includegraphics[width=0.49\textwidth]{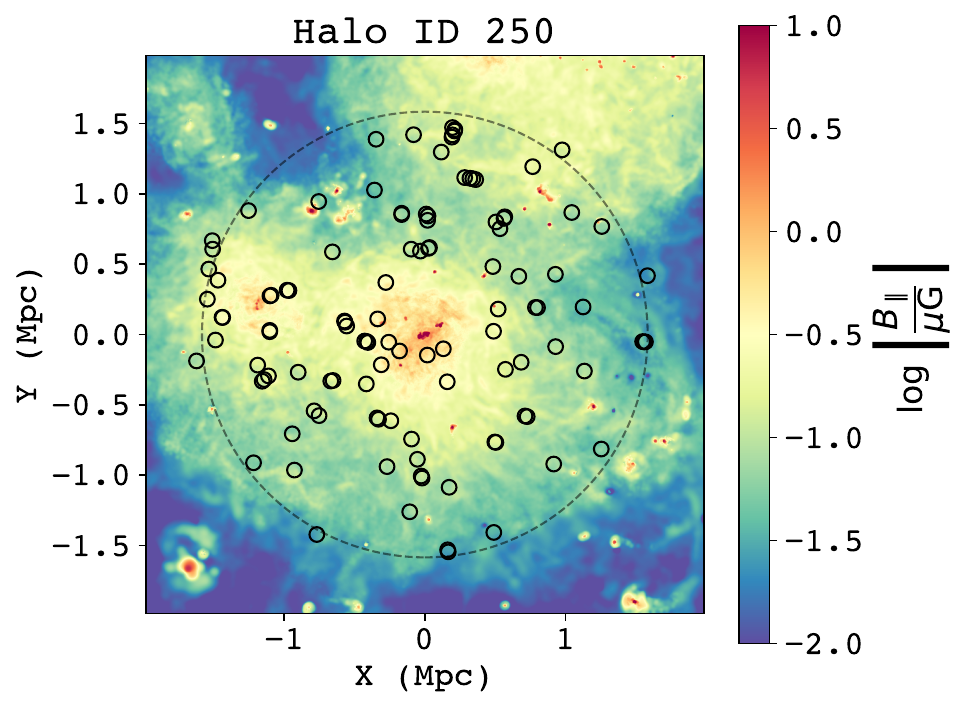}
    \caption{Logarithm of the LOS magnetic field strength for the CC simulated analogue cluster from the TNG-Cluster simulation. The dotted circle represents $2R_{500}$ and the solid circles represent the positions of the sampled RMs (obtained from the observation of A3581).}
    \label{fig:sim_analaog}
\end{figure}

\subsection{{The RRM clump in Abell 3581}}
From MHD simulations, it is known that CC clusters often undergo sloshing motions that create cold fronts that lead to amplification of the magnetic field and large-scale asymmetry in the magnetic field strength and structure \citep{2016JPlPh..82c5301Z, 2018SSRv..214..122D}. Furthermore, the infalling of mass into a cluster also creates cold pockets around the infalling matter, leading to local amplification in the magnetic field strength and to small-scale asymmetry in the magnetic field \citep{2024arXiv241100103T}. Additionally, as noted in Section \ref{sec:TNG}, CC clusters that are currently undergoing a (minor) merger might also portray large-scale asymmetry in the magnetic field and the electron density of the cluster. {This is shown by the enhanced magnetic field at (X,Y) $= (-1, 0.4)$ Mpc in the simulated cluster from TNG-Cluster, presented in Figure \ref{fig:sim_analaog}.}

Figure \ref{fig:bubble} {presents an RRM bubble plot for A3581 overlaid on an X-ray image taken from eRASS1 \citep[][]{2024A&A...682A..34M} in the 0.2$-10$ keV band. Based on this figure, it is likely that A3581 also possesses significant substructures in the ICM that are causing the (radially averaged) scatter to be non-monotonic. In particular, we note the clumping of the high-magnitude RRMs east of the cluster center at a radius of $\sim 1.1$ Mpc, which is the radius at which the RRM scatter profile for A3581 seems to peak. The RRMs in this clump have the opposite sign to the RRMs in the centre of the cluster (which are predominantly positive). Furthermore, we have identified an optical sub-group [DZA2015b] 276 within A3581 as a possible cause for this clump of high magnitude RRMs. This group was identified by \citet{2015A&A...578A..61D} as part of a compact group catalog using velocity-filtered compact groups from the Two Micron All Sky Survey \citep[][]{2006AJ....131.1163S} and the 2M++ galaxy redshift catalog \citep{2011MNRAS.416.2840L}; we list some of the important properties of this group in Table \ref{tab:compact}. As far as we are aware, this is the first single galaxy group that is detected in RMs while not strongly emitting in X-rays. }

{Another possible cause for the clumping of high magnitude RRMs is a clustering of background radio sources. \citet{2024ApJS..272...39W} have identified the galaxy group WH-J140921.4-270516 at a redshift of 0.7416 that has an $R_{500}$ of 0.35 Mpc with central coordinates near [DZA2015b] 276, at $(\alpha, \delta, ~\mathrm{J2000}) = (212.339~\rm{deg}, -27.088~\rm{deg})$. However, none of our RM sightlines intersect WH-J140921.4-270516 within its $R_\mathrm{500}$, so it is unlikely to be contributing to the enhanced RM scatter.}

\begin{figure}
    \centering
    \includegraphics[width=0.49\textwidth]{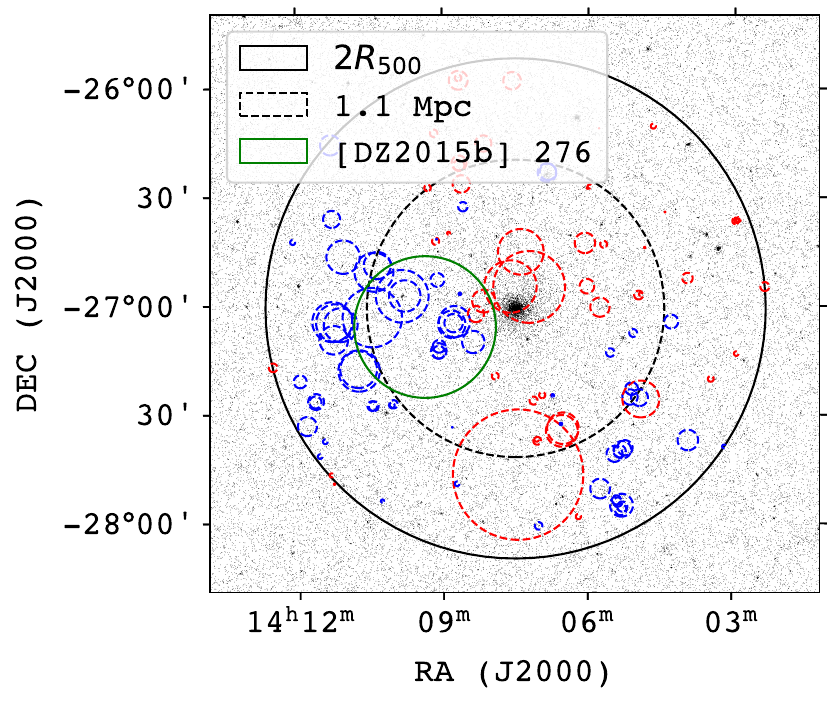}
    \caption{RM bubble plot for the A3581 RRM values {overlaid on an X-ray image taken from eRASS1 in the $0.2-10$ keV band}. The bubbles represent the location of the RRMs. Red bubbles indicate positive RRMs, and blue bubbles indicate negative RRMs. The size of the bubble is linearly proportional to the magnitude of the RRM, with the largest bubble of radius 0.3 deg on the sky representing an RRM of 100 rad m$^{-2}$. The solid black circle indicates $2R_{500}$ for A3581 and the dashed black line indicates a circle of radius 1.1 Mpc. {The solid green circle indicates the virial radius of the galaxy group [DZ2015b] 276.}}
    \label{fig:bubble}
\end{figure}

\begin{table}[!htb]
    \centering
    \caption{{Basic properties of [DZ2015b] 276 taken from \citet{2015A&A...578A..61D}.}}    
    \begin{tabular}{|c|c|}

        \hline
        {\textbf{Property}} & \textbf{{Measurement}} \\
        \hline
        \hline 
        Centre (ICRS) & (14h 09m 22.1s, $-27^\circ$ 06$^{\prime}$ 07$^{\prime\prime}$)\\
        Group redshift & $0.0214$\\
        $R_{\rm{vir}}$ (Mpc) & {0.507}\\
        $M_{\rm{vir}}$ ($M_\odot$) & $3.32\times 10^{13}$\\
        \hline
    \end{tabular}

    \label{tab:compact}
\end{table}

\subsection{Cluster merger axis from RRM grid}
\label{sec:asymmetry}
In the simplest scenario of uniform magnetic field strength $B$ and electron density $n_e$ \citep[e.g.][]{2004A&A...424..429M,2016A&A...596A..22B}, the RRM scatter (or the variance of RRMs) probes the combination of electron density, magnetic field strength and magnetic field coherence scale as: 
\begin{equation}\label{eq:simplesigma}
    \sigma_\mathrm{RRM}^2 \propto {\ell_c \int_0^d [n_e B_\parallel]^2 ~dl},
\end{equation}
where $\ell_c$ is the scale on which the magnetic field direction is coherent. In reality, all of these parameters can vary as a function of location in the cluster. From Equation \ref{eq:simplesigma}, and as illustrated in Fig \ref{fig:sim_analaog}, we expect the RM scatter to be most axially symmetric about the projected axis of a merger, as this is the axis about which the projected electron density and the magnetic field strength are most symmetric. Therefore, we probe possible merger axes using the axis of symmetry of the RRM scatter. We note here that we are unable to discern the full three-dimensional structure of possible merger axes as we are limited to only discerning two-dimensional information of merger axes as the RM is a LOS probe.


In general, RRMs are expected to have higher magnitudes near the center of the cluster (as a greater column depth is probed through the ICM) but the RRM grid of the cluster might also have a separate preferred center (near where the $|\rm{RRM}|$ peaks) than the X-ray centroid. Therefore, we define the  `center of rotation measure' (CORM) as:
\begin{align}
    (\alpha_{\mathrm{CORM}}, \delta_{\mathrm{CORM}}) = \dfrac{\sum_{i=1}^N  (\alpha_i, \delta_i) |\mathrm{RRM}_{i}|}{\sum_{i=1}^N |\mathrm{RRM}_i|}, 
\end{align}
where $(\alpha, \delta)$ are the right ascension and declination at J2000 in ICRS, and the sum is taken over all RRMs. 

The main motivation behind defining this quantity is that we are searching for an axis of symmetry in the RRM grid. Therefore, it is not ideal for us to search for an axis of symmetry about the X-ray centroid. Then, we split the cluster into two halves through the CORM and calculate the scatter in each of the split regions (taken to be IQR/1.349) as a function of the position angle of the splitting axis. The axis of symmetry of the RRM scatter is determined by minimizing the difference in the standard deviations of the two sides. 

We calculated the CORM and the axis of symmetry for the RRM scatter for the simulated analogs from TNG-Cluster, as displayed in Figures \ref{fig:250}, \ref{fig:255} and \ref{fig:231}. Here, we have computed these quantities using both the full RM image, as well as sampling the RM image identical to our observations. The difference in the scatter in the two halves for each of the clusters with the sampled RMs is displayed in Figure \ref{fig:asymm}. Notably, we see that there are only certain position angles along which the scatter in the RMs of the two sides is minimized; these correspond to the axis of symmetry for the RM scatter. However, the axis of symmetry from the sampled RMs might deviate mildly (see Figure \ref{fig:250}) to significantly (see Figure \ref{fig:231}) from the axis of symmetry predicted using the full RM image. This discrepancy is primarily caused by the clustering of RMs and the sparsity of the RM grid. Therefore, we apply the axis of symmetry for the RM scatter to our observations with caution, noting that it is possible that we predict a merger axis in A3581 within error or that we are significantly deviating from any real merger axis. 

{Halo IDs 255 and 231 are post-merger systems. For such systems, the TNG-Cluster simulation has defined the merger axis using the relative displacement between the clusters before and after the pericenter passage. For Halo ID 255, we see that the merger axis of the cluster aligns very well with the axis of symmetry obtained from using the full RM image, as is expected based on Equation \ref{eq:simplesigma}. This indicates that the full RM image is a good tracer of the hot plasma of the system and that we are able to accurately predict merger axes from it. On the other hand, for Halo ID 231, we see that the merger axis aligns well with the axis obtained from the sampled RMs (and not with the full RM image). This indicates that Halo ID 231 likely is not a single cluster-cluster merger but has undergone mergers with multiple clusters along different axes, and that the axis of symmetry of the sparse RM grid happens to align with one of these merger axes by chance, while the full RM image is tracing the merger axis of a different merger. Since Halo ID 250 is a pre-merger system, the merger axis could not be calculated as defined above; instead, for this system, we fit an ellipsoid to the X-ray contour encompassing the two merging clusters and report the position angle of this ellipsoid as the potential merger axis. In this system, all three of the axes computed (the full RM image, the sample RMs and the position angle of the X-ray ellipsoid) appear to be very well-aligned with one another. From these three systems, we can conclude that the axis obtained from the sampled RMs is likely to trace potential merger axes, while the full RM image will provide more information about the tenuous plasma in the outskirts.}

\begin{figure*}[!htb]
    \centering
    \subfigure[]{\includegraphics[width=0.47\textwidth]{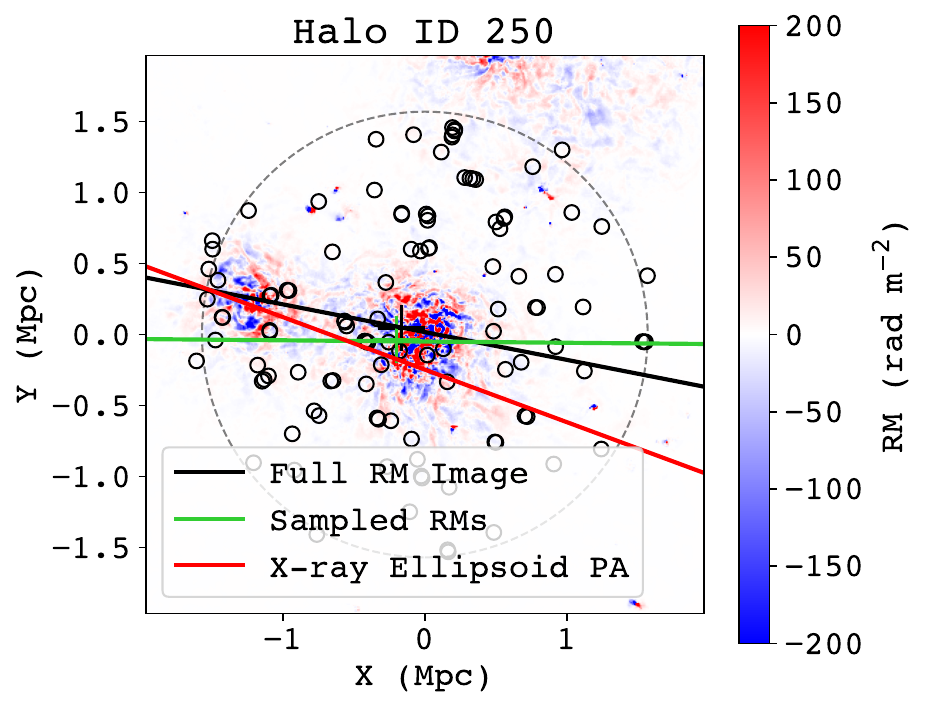} \label{fig:250}} 
    \subfigure[]{\includegraphics[width = 0.47\textwidth]{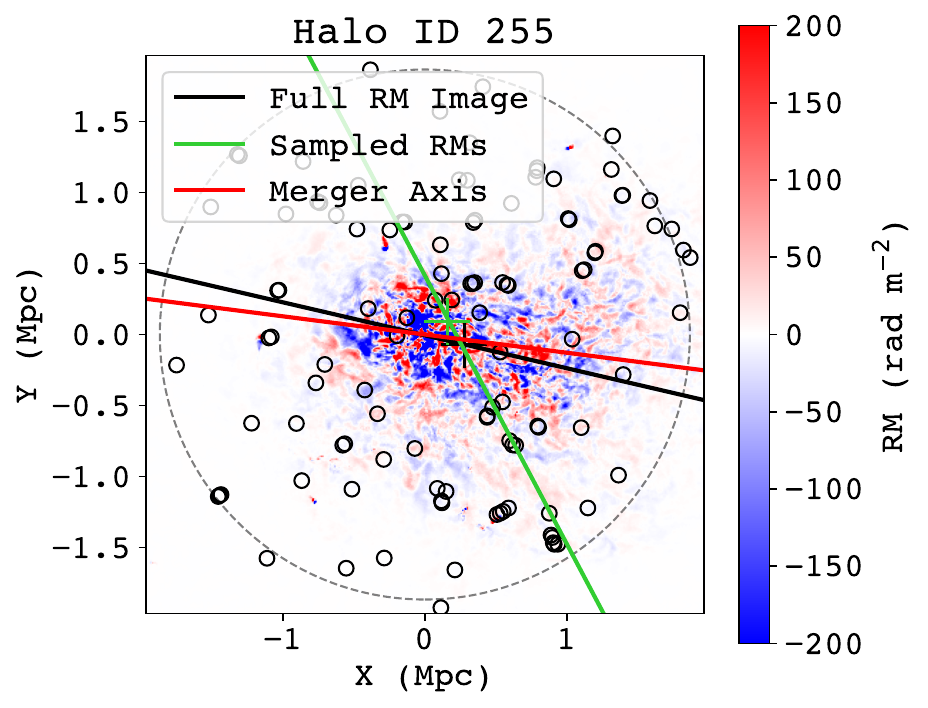} \label{fig:255}}\\
    \subfigure[]{\includegraphics[width = 0.47\textwidth]{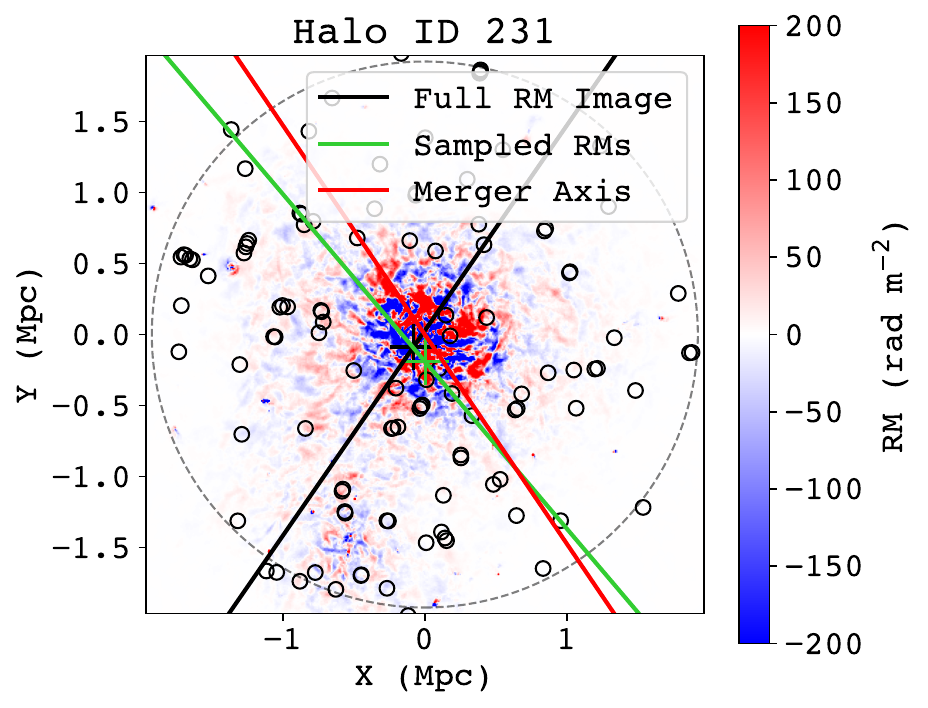}\label{fig:231}}
    \subfigure[]{\includegraphics[width = 0.47\textwidth]{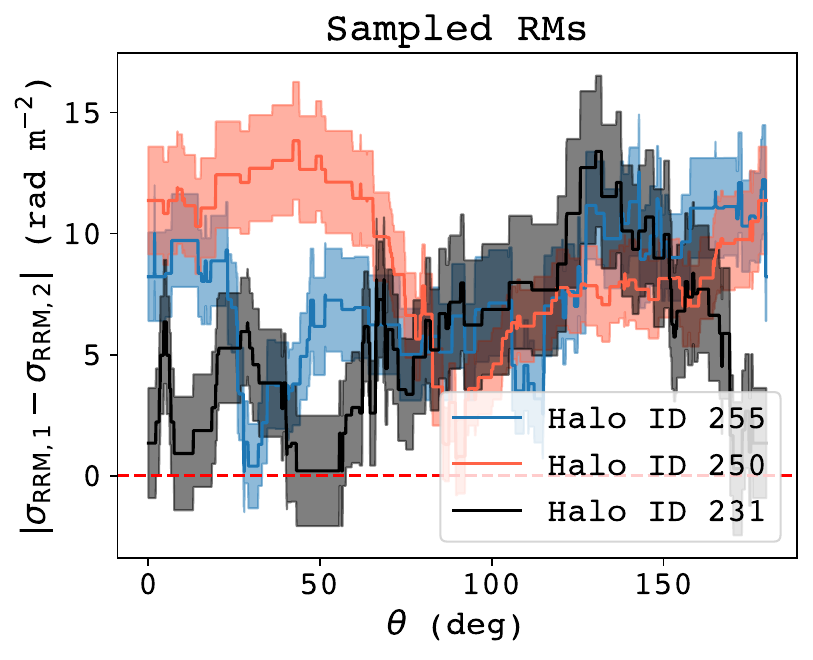}\label{fig:asymm}}

    \caption{(a), (b), (c) RM images of the closest matches to A3581 in TNG-Cluster (Halo IDs 250, 255 and 231, respectively). The dotted circles indicate $2R_{500}$ of the clusters, the solid circles indicate the positions of the sampled RMs from the full image. We note here that the positions of the samples are different for each of the clusters because these are different rotations. The black and green lines (and plus sign) indicate the  axis of symmetry of the RM scatter (and CORM) computed using the full RM image and just the sampled RMs, respectively. {Halo IDs 255 and 231 are post-merger systems; therefore, we have also included the merger axis computed (as red lines). For Halo ID 250, we display the position angle of the X-ray ellipsoid (in red).} (d) The magnitude of the difference in RM scatter for the two split sides of the closest matches in TNG-Cluster as a function of the position angle (displayed with solid lines), along with the 1$\sigma$ error (displayed with the shaded regions). The red dotted line indicates identical scatter in both sides. }
\end{figure*}

{Therefore, we apply this technique to A3581 to investigate potential merger axes.} For A3581, we found that $(\alpha_{\mathrm{CORM}}, \delta_{\mathrm{CORM}}, ~\mathrm{J2000}) = (212.052 \pm 0.014 \ \mathrm{deg}, -27.137\pm 0.013 \ \mathrm{deg})$ (indicated as a plus sign in Figure \ref{fig:a3581rrm}), which is $\sim314\pm29$  kpc to the south-east of the X-ray centroid. Figure {\ref{fig:136} portrays the axis of symmetry for A3581.} Figure \ref{fig:asymmetry} displays the difference in the standard deviation of the RRMs between side 1 and 2. The difference in scatter crosses zero at angles of  $48\ \mathrm{deg}$ and $56\ \mathrm{deg}$. This allows us to determine that the axis of symmetry of the cluster lies at an angle of $\theta = 52 \pm 4$ deg (indicated as a dashed line in Figure \ref{fig:a3581rrm}). This indicates that the combination of properties noted in Equation \ref{eq:simplesigma} are similar, on average, on either sides of this axis. Interestingly, the RRMs have opposite signs on either sides of this axis, indicating a possible preferential direction of the magnetic field on either side of this axis. {Additionally, the merger axis we have calculated also traces the position of the optically identified sub-group [DZ2015b] 276.}



\begin{figure*}
    \centering
    \subfigure[]{\includegraphics[width = 0.48\textwidth]{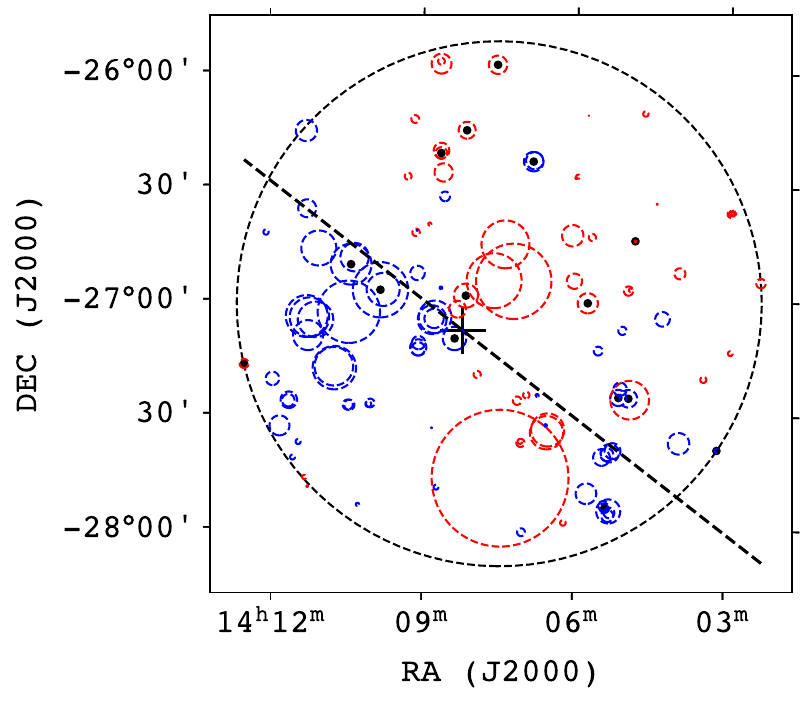} \label{fig:136}}
    \subfigure[]{\includegraphics[width = 0.47 \textwidth]{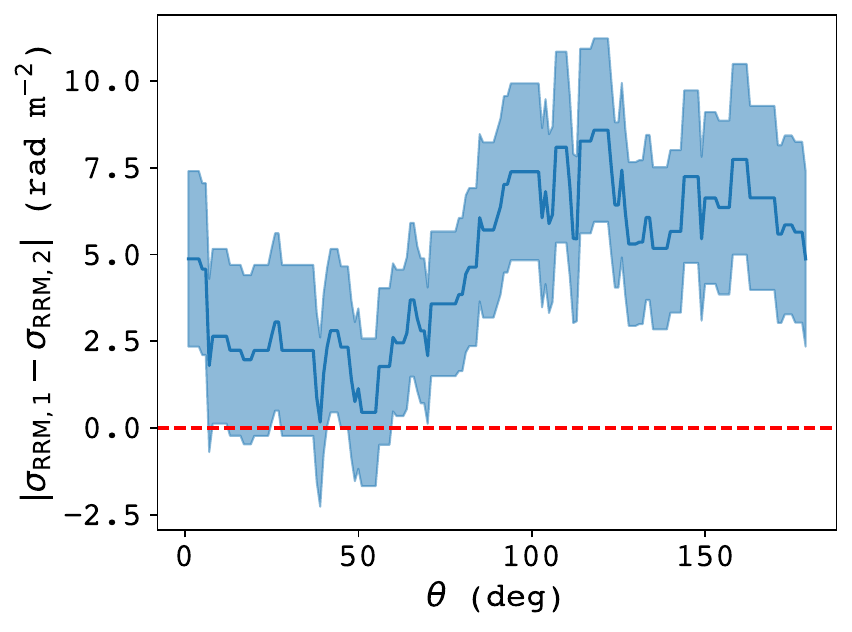} \label{fig:asymmetry}}
    \caption{(a) {An RM bubble plot of A3581. The bubbles represent the location of the RRMs; red bubbles are for positive RRM and blue bubbles are for negative RRMs. The size of the bubble is linearly proportional to the RRM, with 100 rad m$^{-2}$ having a bubble with radius of 0.3 deg on the sky. The CORM is indicated with a plus sign and the axis of symmetry of the RRM scatter is indicated with a straight dashed line. The black dots indicate the location of Faraday complex RMs (see Section \ref{sec:far_comp})} (b) The magnitude of the difference in RM scatter for the two split sides as a function of the position angle. The red dotted line indicates identical scatter in both sides.}
\end{figure*}

\subsection{RRM scatter enhancement due to cluster members}

{For RRMs that have small impact parameters to cluster members in a galaxy cluster, there is likely an enhancement in the RRM scatter due to the circumgalactic medium (CGM) of the member galaxy. To investigate this effect properly, it is best to use spectroscopically confirmed cluster members. However, we are limited to the photometric samples as the sparse availability of optical spectra (only 7 RMs) prevents us from drawing meaningful conclusions with only spectroscopic members.} For this reason, we used galaxies that were observed by Pan-STARRS1 (PS1) survey \citep{2016arXiv161205560C}, calculated the photometric redshifts, and determined cluster membership by using a fixed gap of 1000 km s$^{-1}$. To prevent spurious associations due to large uncertainties in the photometric redshifts, we 
included only galaxies with a fractional uncertainty in the photometric redshift less than 0.4. This left us with 645 potential cluster members within 2$R_{500}$ of the X-ray centroid (and 150 cluster members within $R_{500}$). This number of galaxies is consistent with the richness of similar mass clusters in MHD simulations such as TNG-Cluster; an example of such a simulated cluster in TNG-Cluster is the one with Halo ID 861, which has a richness of 128 galaxies\footnote{\url{https://www.tng-project.org/files/TNG-Cluster_Catalog.txt}} \citep{2024A&A...686A.157N}.

For each RRM, we calculated the impact parameter, $b_{\mathrm{nearest}}$, to the closest member galaxy. {From this, we computed the observed scatter in the RRMs as a function of the impact parameter: $\sigma_{\mathrm{RRM}, \mathrm{CGM}_{\mathrm{obs}}}$. Additionally, to account for the scatter in the RRMs due to the ICM in each bin, we model the statistical distribution of the ICM contribution to the observed RRMs as random variables that are normally distributed around 0 rad m$^{-2}$ with a standard deviation given by $\sigma_{\mathrm{RRM, corr}}$. Then, for each bin, we compute the ICM contribution, $\sigma_{\mathrm{RRM, ICM}}$, to be the interquartile scatter (divided by 1.349) of the resampled RRMs. Finally, we compute the ICM-corrected CGM scatter as:}
\begin{align}
    {\sigma_{\mathrm{RRM, CGM}} = \sqrt{\sigma_{\mathrm{RRM}, \mathrm{CGM}_{\mathrm{obs}}}^2 - \sigma_{\mathrm{RRM, ICM}}^2}.}
\end{align}

Figure \ref{fig:rrm_b} displays the RRMs as a function of the impact parameter to the nearest cluster member, and Figure \ref{fig:impact_scatter} displays the running scatter in the RRM (corrected for ICM contributions) as a function of impact parameter. {We do not observe an increase in RM scatter for sightlines with smaller impact parameters to potential cluster members. }

\begin{figure*}[!htb]
    \centering
    \subfigure[]{\includegraphics[width=0.47\textwidth]{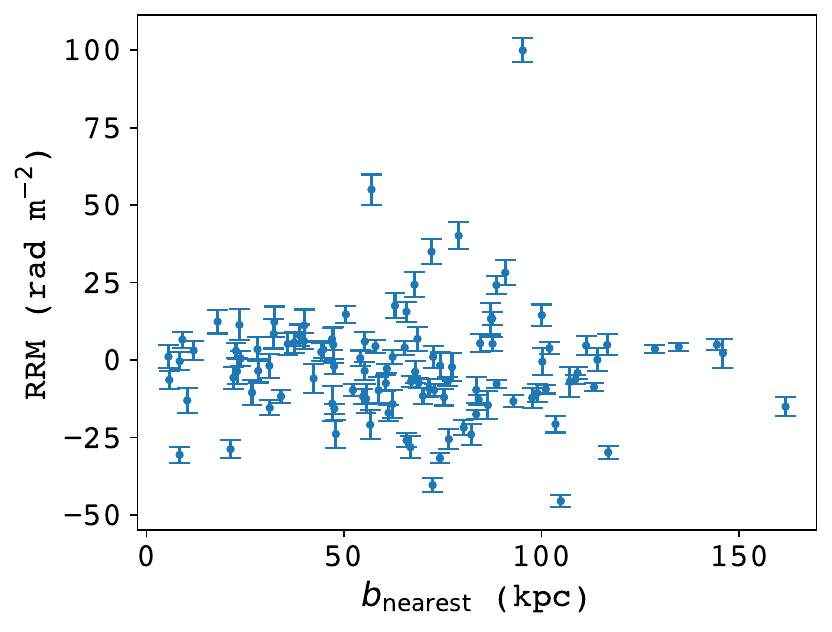} \label{fig:rrm_b}}
    \subfigure[]{\includegraphics[width = 0.47\textwidth]{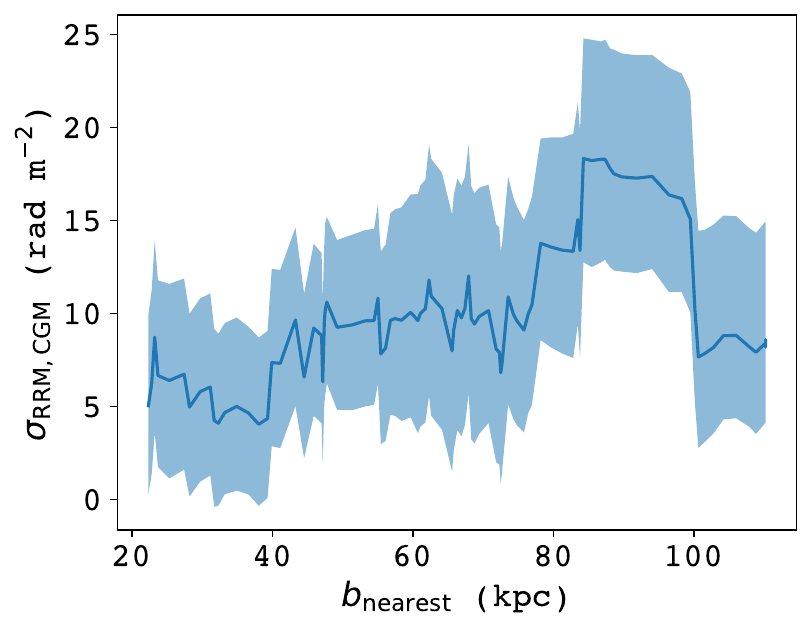} \label{fig:impact_scatter}}
    \caption{(a) The RRMs as a function of the impact parameter to the nearest member galaxies to the RRM. (b) The ICM-corrected running scatter in the RRM as a function of the impact parameter to the closest member galaxy to the RRM. To calculate the scatter, we used a running bin and fixed the number points per bin to be 20. }
\end{figure*}

\subsection{{Faraday complexity}}
\label{sec:far_comp}
As noted previously, most of the RRMs in our sample have been found to be simple using the criteria outlined in Appendix \ref{app:complexity}. This indicates that the cluster (and any other material along the LOS) does not cause significant depolarization observable within the POSSUM band. {The relatively small bandwidth of the POSSUM observations (800-1088 MHz) might be a reason why we fail to detect significant depolarization. However, the fact that we do not observe multiple peaks in most of the FDFs is a good indication that our data are dominated by simple emitting sources that are not associated with a Faraday rotating medium, and that the Faraday rotation is dominated by the ICM and the Milky Way.} 

Furthermore, we explored possible correlations with the distribution of the RRMs on the sky and their Faraday complexity (with complex RMs being indicated by black dots) as displayed in Figure \ref{fig:136}. {In particular, we found that 50\% of the Faraday complex RMs lie along the axis of symmetry. Half of the Faraday complex RMs along the axis have a best-fit $QU$ model given by Equation \ref{eq:m11}, and the other half have the best-fit $QU$ model given by Equation \ref{eq:m3}. In particular, we note that both of these models imply that there are two separate Faraday components within a single telescope beam (and are therefore unresolved)}. This indicates that there are likely multiple different regions that are rotating along the LOS across the axis of symmetry \citep{2005A&A...441.1217B}. 



We also explored if there is any correlation between the magnitude of RRMs and the depolarization parameter ${\Sigma_{\mathrm{RM}}}$. In the case that the model had two depolarization parameters, we took the depolarization of the component that had a higher fractional polarization. When the best-fit model had no depolarization term present, we set $\Sigma_{\mathrm{RM}} = 0$ {rad m$^{-2}$}.  
Figure \ref{fig:depol} displays the magnitude of the RRMs as a function of $\Sigma_{\mathrm{RM}}$. For the RMs for which we were able to detect depolarization, the magnitudes of RRM and $\Sigma_{\mathrm{RM}}$ do not appear to show any correlation. 


\begin{figure} 
    \centering
    \includegraphics[width = 0.47\textwidth]{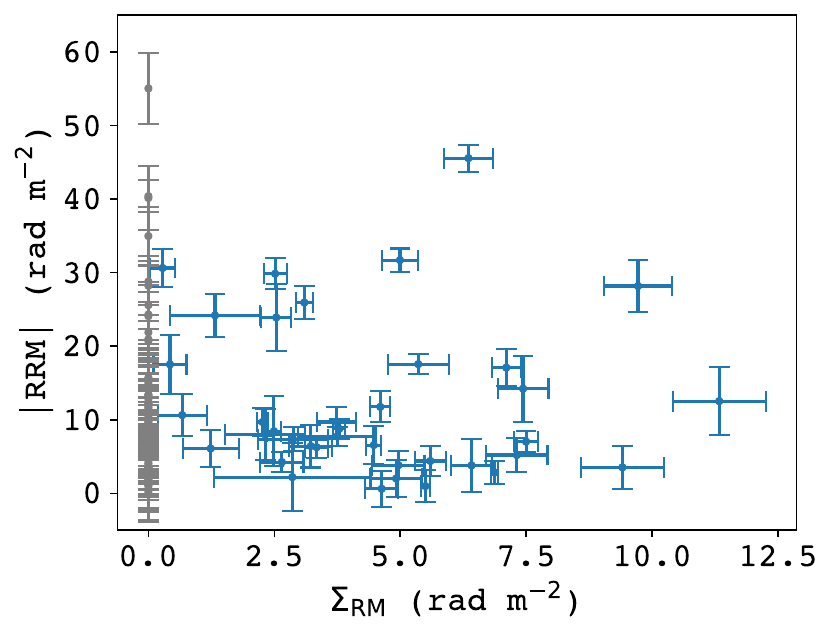}
    \caption{{The RRMs (in rad m$^{-2}$) as a function of the depolarization parameter $\Sigma_{\mathrm{RM}}$ (in {rad m$^{-2}$). The gray points indicate RMs where we were unable to determine $\Sigma_{\mathrm{RM}}$ from $QU$-fitting, and the blue points indicate the ones where we were able to measure $\Sigma_{\mathrm{RM}}$.}}}
    \label{fig:depol}
\end{figure}

\section{Discussion}
\label{sec:discussion}

\subsection{{The RRM scatter profile of Abell 3581}}
\label{sec:scatter_discussion}


{As shown in Figure \ref{fig:bubble}, the non-monotonic nature of A3581's RRM scatter profile for $r > 0.75$ Mpc is most likely caused by the clumping of high magnitude RRMs at $r \sim 1.1$ Mpc. However, another possible explanation for the enhancement at $r \sim 1.1$ Mpc is that there might be more complete depolarization of radio sources near the cluster center due to the increased magnetic field strength and column density of thermal electrons \citep[e.g.,][]{2004A&A...424..429M, 2022A&A...665A..71O}. This increase in complete depolarization would decrease the number of RMs we detect near the cluster center and therefore would lead to a decrease in the scatter of the RRMs near the cluster center that we observe. However, the RRM grid density as shown in Figure \ref{fig:grid_dens} appears to be fairly similar at $r < 1$ Mpc and $r > 1$ Mpc,
indicating that this is not the case\footnote{Here the uncertainties are obtained by assuming that the number count of RMs follows a Poisson process.}. Furthermore, since we sample the models (in Section \ref{sec:mag_modeling}) at equivalent locations to observed RMs, it is unlikely that this is the case as none of the models display this non-monotonic RM scatter profile. }

\begin{figure}
    \centering
    \includegraphics[width=0.47\textwidth]{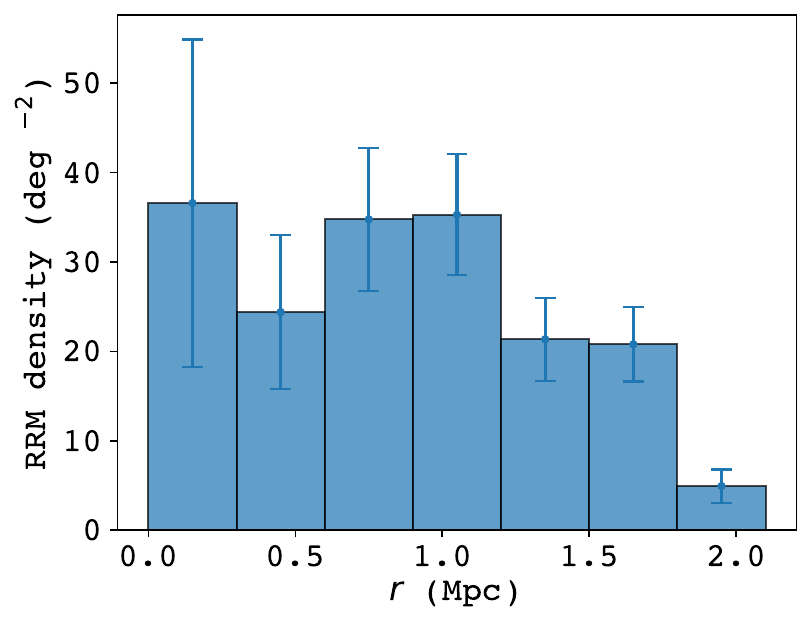}
    \caption{The density of RRMs as a function of radius (in bins of size 0.3 Mpc) from the X-ray centroid of A3581.}
    \label{fig:grid_dens}
\end{figure}

Furthermore, we investigated whether there was any enhancement in RMs due to the CGM of cluster galaxies. However, as shown in Figure \ref{fig:impact_scatter}, there does not seem to be any statistically significant enhancement in scatter close to cluster members (which is what we expect). Because of this, our modeling of the ICM (where we have not considered any contribution due to the CGM) is valid. Nevertheless, the effect of the CGM on studying the surrounding plasma will likely be probed much better for galaxy groups and clusters (and individual galaxies) that are much closer to the observer, where there are multiple RMs to probe the CGM of a single galaxy. This will be investigated further in upcoming POSSUM works.

\subsection{Magnetic field modeling}
All of the magnetic field and electron density models that we have tested produce a monotonic decline in the scatter, indicating that more detailed models are likely needed to include the complexity in real clusters.

{Models of the inner 0.75 Mpc, centered on the X-ray peak with values of $n_e(0)$ fixed at $33.6\times 10^{-3}$ \citep{1994PASJ...46L..37T} and the radial profile of the electron density determined from a self-similar scaling \citep{2025A&A...694A..44O}}, show that A3581's RRM scatter profile is well-modeled by { the following magnetic field profiles: {$B_0 = 0.16 ~\mu \mathrm{G}, \eta = 0; B_0 = 0.59~\mu \mathrm{G}, \eta = 0.25; B_0 = 2.20~\mu \rm{G}, \eta = 0.50$. All of these profiles seem to fit the data equally well, and we cannot distinguish between them based on the current data. The sampled RMs of all these modeled clusters are drastically different; however, their scatter profiles are similar due to the small number of sources near the cluster center. The fixed bin frequency we used in calculating the scatter profile causes some of these differences to be smoothed, reproducing similar RM scatter profiles. Unfortunately, this can only be improved with higher data density in future RM grid surveys of clusters. Therefore, we are unable to conclude anything about which model is favorable; the only one that is unlikely is the one with $\eta = 0$, as we physically expect some coupling between the magnetic field and the electron density.}}

At larger radii, the magnetic field may change its relation with the thermal gas density as mergers or bulk motions have a strong impact on RM scatter. We found that the magnetic field of Abell 3581 cannot be modeled with an analytical profile above $r = 0.75$ Mpc. 
This is consistent with the picture of a relaxed cool core in the interior and enhancement in the outskirts (where the RRM scatter is no longer monotonically decaying) due to enhancements in gas density and magnetic field strength caused by interactions with neighboring systems (as noted in Section \ref{sec:scatter_discussion}). This implies that it is important to revise the simple radial picture of cluster magnetic fields as given by Equation \ref{eq:1} and compare to full MHD simulations of cluster magnetic fields in a cosmological context. 
We note here that one of the biggest caveats in our assumption of modeling the cluster is that we do not know the true electron density profile. {X-ray observations have constrained A3581 to have a cool-core with $n_e(0) = (33.60_ {- 14.02}^{+0.00}) \times 10^{-3}~\mathrm{cm}^{-3}$ \citep[][]{2004PASJ...56..965F,2005MNRAS.356..237J}. Therefore, we assume that its profile is similar to that of other CC clusters, scaled to this central electron density}. 



On a cluster-to-cluster basis, our models (see Section \ref{sec:mag_modeling}) are not accurate enough to predict the complexity of the ICM compared to the MHD simulated clusters from TNG-Cluster, some of which portray similar RM scatter profiles as A3581 as noted in Section \ref{sec:TNG}. However, we note that our cluster models are still applicable when studying the mean properties of clusters, given sufficiently large sample sizes. Figure \ref{fig:mean_TNG_model} displays a comparison of the RM scatter for one of our models and the median RM scatter profile (where the median is taken over all possible realizations) for 11 CC clusters (594 realizations), 86 WCC clusters (4644 realizations), and 24 NCC clusters (1296 realizations) in the mass range $M_{500} = [1.4, 3.4]\times 10^{14} \ M_\odot$ from TNG-Cluster. {In particular, the most significant differences between entropy-based cluster classifications emerge at $r < 0.5~R_{500}$, while the median profiles converge around and beyond $R_{500}$.}

Furthermore, we note that the NCC clusters have the largest RM scatter and CC clusters have the lowest RM scatter at low $R_{500}$. In general, we expect CC clusters to have higher $B$ and $n_e$ at the centers of clusters \citep[e.g.,][]{2004JKAS...37..337C, 2025A&A...694A..44O}. The higher RM scatter in NCC clusters is possibly due to enhanced turbulence, as they are more likely to have recent merger activity \citep[e.g.,][]{2024A&A...686A..55L, 2025arXiv250301969L}.

We found that the model that most closely resembles the RM scatter profile of the MHD clusters has parameter values of $B_0 = 10 \ \mu$G and $\eta = 0.50$. This is consistent with parameter values that have been found in previous studies \citep[e.g.,][]{2010A&A...513A..30B}. The close resemblance of the TNG-Cluster RM scatter profiles to those of our models and to parameter values found in previous observations shows that our models are viable when carrying out stacking studies of clusters, but might fail in capturing the complexity of individual clusters.   

\begin{figure}
    \centering

    \includegraphics[width=0.48\textwidth]{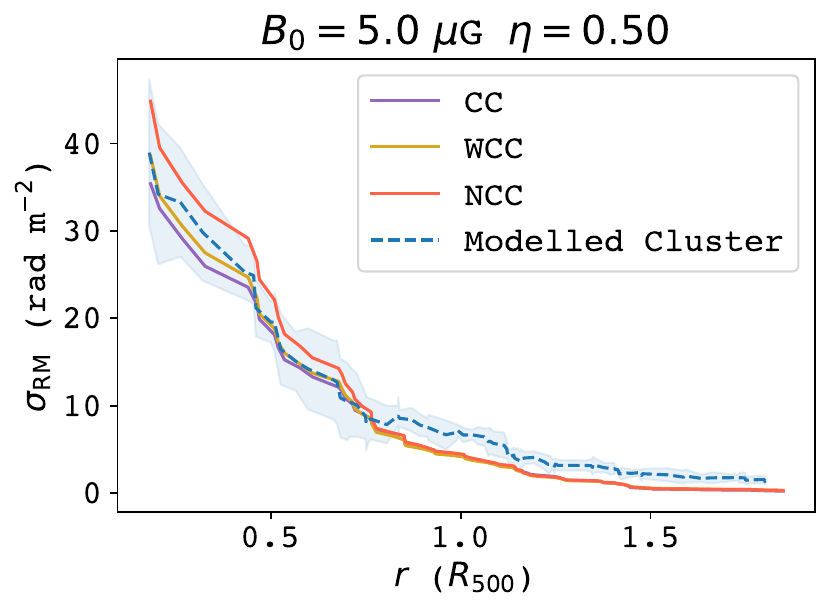}
    \caption{A comparison of the RM scatter profile of our modeled cluster (with uncorrelated fluctuations in both $n_e$ and $B$) with parameters $B_0 = 5 \ \mu$G and $\eta = 0.50$ (in blue) with the median profiles  of the cool core (in purple), weak cool core (in yellow) and non-cool core (in red) clusters from TNG-Cluster.}
    \label{fig:mean_TNG_model}
\end{figure}


\subsection{Mergers in TNG-Cluster and Abell 3581}
\label{sec:RRM_in_merger}

From TNG-Cluster, we have identified clusters with comparable masses to A3581 that also exhibit a similar RRM scatter profile (see Figure \ref{fig:TNG_rms}). To gain further insight into the possible origin of these features, we analyze the cosmic evolution of the simulated analogs. Interestingly, we find that all simulated analogs present elevated magnetic field strengths near the outskirts (regardless of their core's entropy) and they all seem to be currently interacting or have interacted with other nearby clusters and groups (either through cluster mergers or through accretion of gas). Furthermore, we were able to find the merger axis of the clusters using the scatter in the RRMs of the full RM image. We noted that the predicted merger axis may deviate significantly from the true merger axis based on how the RM image is sampled. Finally, given that all the simulated analogues in TNG-Cluster have different entropy cores, we note that the state of the core seems to have no strong implications for the magnetic field strength and electron density content of the ICM at the outskirts of the cluster.




Due to the simulated analogs having undergone past or present merger activity, we explore this explanation for the clumping of high magnitude RRMs at $r \sim 1.1$ Mpc {(which have the opposite sign to the RMs near the X-ray centroid that are predominantly positive)}. {A possible} explanation for the enhancement of RM in the outskirts in A3581 might be present interaction with the neighboring groups or the clusters that we have identified in Figure \ref{fig:skyRM}; this situation is similar to that of the CC simulated analog displayed in Figure \ref{fig:sim_analaog}. {However,} an X-ray analysis of A3581 revealed a sloshing cold front near the X-ray core \citep{2013MNRAS.435.1108C}, which can hint at past merger activity as well. 
Thus, a subcluster that triggered the sloshing motion in A3581 might also be the source of the clumping at $r\sim 1.1$ Mpc and depending on the radiative cooling time at the center, it may be possible to produce the low-entropy core at the center while the disturbances caused by the merger remain at the outskirts. It is also possible that the cool core was never or is not yet destroyed during a recent or current minor off-axis merger \citep[e.g.,][]{2021MNRAS.504.5409V}. {This last explanation seems to be the most likely due to the presence of the optically detected galaxy sub-group [DZ2015b] 276.} For A3581, we have identified a possible merger axis at $\theta = 52 \pm 4$ deg {(which is likely tracing the merger axis of A3581 with the galaxy group [DZ2015b] 276)}, noting that this is highly dependent on the spatial distribution and number of our background RMs. We note here that we are able to begin to quantify the asymmetry in the RM grid due to the increased source density of POSSUM. Previous studies have been constrained to assuming that the RM scatter in clusters is radially symmetric. Future  deep X-ray observations showing the gas density distribution out to and beyond $R_\mathrm{500}$ will shed further light into the merger state of A3581.

To date, the only other single system that is observed to show a similar deviation from spherical symmetry in the RM grid's scatter, as well as having enhancement $\sigma_{\mathrm{RRM}}$ in the outskirts is the Fornax cluster, which was studied first with POSSUM by \citet{2021PASA...38...20A} and then with the MeerKAT telescope by \citet{2025arXiv250105519L}. Both of these studies found a large coherent `RM stripe' in the Fornax cluster. Furthermore, we compare our results with the findings of \citet{2025A&A...694A..44O}, who found a similar non-monotonic RM scatter profile in stacked NCC clusters, albeit at lower projected radii ($\sim 0.3R_\mathrm{500}$). A similar elevated RM scatter at large radii has also been seen in lower mass systems, such as galaxy groups, by \citet{2024MNRAS.533.4068A}. This indicates that the RRM scatter in clusters might be showing a deviation from radial symmetry that can only be observed with increased polarized source densities. The results from these works and the findings in the RRM grid of A3581 indicate that, given a high enough density of polarized background sources, we may detect similar non-monotonic RM scatter profiles due to cluster-specific features. Additionally, we expect our method to probe a cluster's merger axis using the RM scatter to significantly increase in reliability with the SKA RM grid, which is expected to have $\sim 100$ polarized radio sources per square degree \citep[][]{2020Galax...8...53H}. 

Put together, the results of the above studies and our findings in A3581 might indicate that the gas at the outskirts of massive halos also significantly contributes to RM enhancement due to enhancement in gas density and magnetic field strength that are possibly caused by interactions with neighboring clusters and groups. Future POSSUM observations of galaxy groups and galaxy clusters in the local universe will shed further light on this.


Additionally, a notable detail is that the RRMs appear to have opposite signs on either sides of the merger axis; the RRMs to the north of the merger axis are predominantly positive, and those to the south are predominantly negative. This change in the sign could point to the presence of large-scale ordered fields in the ICM, potentially indicating the presence of a tangential or toroidal magnetic field component. However, another plausible explanation is that this large-scale sign change is caused by a residual GRM that we have not corrected for, particularly since we expect the Galactic magnetic field to be ordered on large angular scales, compared to the magnetic field of the ICM. 

\subsection{{Coherency of Magnetic Fields}}
\label{sec:coherency}
{A3581 shows a relatively coherent, weak magnetic field in the center of the cluster as shown by the consistently high magnitude positive RMs within $r \lesssim 0.4$ Mpc of the cluster center in Figure \ref{fig:rrm}. This aligns with what has been found in a population study of simulated clusters in TNG-Cluster by \citet{2025arXiv250301969L}, who found that CC clusters tend to have tangentially oriented magnetic fields near the core. As shown in Figure \ref{fig:TNG_rm_dots}, in the simulated analogs of A3581, we found that two of the clusters (Halo IDs 255 and 231) have strong coherent magnetic fields near the cluster center; this coherency near the cluster center similar to what is seen in A3581. While the merging cluster 250 displays a more random magnetic field. Most notably, the RMs in the simulated analogs are an order of magnitude higher near the cluster center than in A3581. Despite finding a good match between the smoothed RM scatter profiles between the simulated clusters and A3581, the lack of point-to-point agreement between the RMs indicates inconsistencies between the simulations and real clusters.}

{Additionally, we note that the magnetic fields in our semi-analytic models are less coherent than in A3581, as the maximal scale of the fluctuations in the magnetic field has been set to 100 kpc. Because of this, it is important to also optimize over $\Lambda_{\rm{max}}$; however, this is computationally expensive and so we have decided to fix the maximum fluctuation scale. Furthermore, there is some degeneracy between $\Lambda_{\rm{max}}$ and $B_0$ (in terms of the RRM scatter profiles produced); due to this degeneracy and the relative sparsity of the RM grid it is not worthwhile to explore changing $\Lambda_{\rm{max}}$. Therefore, $\Lambda_{\rm{max}}$ as a free parameter can be analyzed in future studies of galaxy clusters with much denser RM grids using telescopes like MeerKAT and the SKA.  }

\section{Conclusion}

We have conducted a detailed study of the magnetic field properties of the nearby massive cool core cluster, Abell 3581 using 111 {rotation} measures (RMs) from the POSSUM survey. {This is the first study focused on a single cluster that uses more than 20 RMs to constrain the properties of the magnetized ICM through comparison with models and MHD simulations.}

{The RMs were obtained using 1D RM-synthesis. We concluded that most of the RMs in our sample were simple (as determined using RM-synthesis and $QU$-fitting), and were then corrected for Galactic contributions to obtain residual RMs (RRMs).} The results of this work are summarized as follows:

\begin{enumerate}[topsep=0pt,itemsep=-1ex,partopsep=1ex,parsep=1ex]
   
    \item The RRM scatter profile of A3581 as a function of radius from the cluster center shows an initial { monotonic} decline but then becomes non-monotonic for $r > 0.8 R_{500}$. 
    \item We compared the observed RRM scatter in A3581 to the scatter in modeled clusters, by modeling the magnetic field  as a Gaussian random field with fluctuations described with a Kolmogorov power spectrum and a universal density profile. Additionally, for the first time, we have also accounted for fluctuations in the electron density content by modeling them as a lognormal field. {The inner 0.75 Mpc, centered on the X-ray peak, is well-modeled {by magnetic fields} with central magnetic field strength{s of $B_0 =B_0 =0.16_{-0.02}^{+0.03},~ 0.59_{-0.09}^{+0.08},~  2.20^{+0.36}_{-0.32}~\mu$G} that scale with the assumed electron density (with $n_e(0) = 33.6\times 10^{-3}$ cm$^{-3}$ and the radial profile of the electron density determined from a self-similar scaling) as {$B \propto n_e^{0},~B \propto n_e^{0.25}$ and $B \propto n_e^{0.5}$, respectively}. }

    \item For the first time, we directly compared the RRM grid of an observed cluster with simulated RM grids from full MHD simulated clusters. We found three simulated analogs in TNG-Cluster that have similar non-monotonic RM scatter profiles to A3581; one of these is a cool core cluster, one of these is a weak cool core cluster, and the other is a non-cool core cluster. {All the analogs display present or past merger activity.}
    \item {We have identified a clump of high magnitude RRMs near $r \sim 1.1$ Mpc that have the opposite sign to the RRMs near the X-ray centroid, and coincide with the position of the optically detected galaxy group [DZ2015b] 276. To our knowledge, this is the first single galaxy group to be detected in RMs while not strongly emitting in X-rays.}
     \item Using the scatter in the RRM grid, we have identified a possible merger axis in A3581 at a position angle of $\theta = 52 \pm 4$ deg{, which traces the positions of the high magnitude RRM clump and the galaxy group [DZ2015b] 276}.  The RRMs have opposite signs on either sides of this axis, indicating a possible preferential large-scale magnetic field direction {or residual galactic rotation measure}.
\end{enumerate}
{In summary, the comparison of the RRM grid to MHD simulations and analytic models paints a picture where Abell 3581 is a dynamically interacting cool-core cluster, with a monotonically declining magnetic field strength out to $r=0.75$ Mpc, consistent with a constant magnetic to thermal energy density ratio, and an enhancement in RRM scatter likely caused by the galaxy group [DZ2015b] 276, which is 1.1 Mpc east of the center.} This paper lays the groundwork for detailed studies of the magnetic field properties of single clusters using upcoming polarization surveys, such as POSSUM and the MeerKAT Large Area Synoptic Survey \citep{2016mks..confE..32S}.

\section*{Acknowledgements}
The University of Toronto operates on the traditional land of the Huron-Wendat, the Seneca, and most recently, the Mississaugas of the Credit River; we are grateful to have the opportunity to work on this land. The Dunlap Institute is funded through an endowment established by the David Dunlap family and the University of Toronto. 

AK would like to acknowledge the support of the Summer Undergraduate Research Program in Astronomy \& Astrophysics. SPO and DAL acknowledge support from the Comunidad de Madrid Atracción de Talento program via grant 2022-T1/TIC-23797, and grant PID2023-146372OB-I00 funded by MICIU/AEI/10.13039/501100011033 and by ERDF, EU. POSSUM is partially funded by the Australian Government through an Australian Research Council Australian Laureate Fellowship (project number FL210100039 awarded to NM-G). DAL also acknowledges support from the Universidad Complutense de Madrid and Banco Santander through the predoctoral grant CT25/24. The authors would like to thank Jennifer Y.H. Chan for meaningful discussion in improving the manuscript.

This scientific work uses data obtained from Inyarrimanha Ilgari Bundara / the Murchison Radio-astronomy Observatory. We acknowledge the Wajarri Yamaji People as the Traditional Owners and native title holders of the Observatory site. The Australian SKA Pathfinder is part of the Australia Telescope National Facility (\url{https://ror.org/05qajvd42}) which is managed by CSIRO. Operation of ASKAP is funded by the Australian Government with support from the National Collaborative Research Infrastructure Strategy. ASKAP uses the resources of the Pawsey Supercomputing center. Establishment of ASKAP, the Murchison Radio-astronomy Observatory and the Pawsey Supercomputing center are initiatives of the Australian Government, with support from the Government of Western Australia and the Science and Industry Endowment Fund. The POSSUM project (\url{https://possum-survey.org}) has been made possible through funding from the Australian Research Council, the Natural Sciences and Engineering Research Council of Canada, the Canada Research Chairs Program, and the Canada Foundation for Innovation. 

The TNG-Cluster simulation suite has been executed on several machines: with compute time awarded under the TNG-Cluster project on the HoreKa supercomputer, funded by the Ministry of Science, Research and the Arts Baden-Württemberg and by the Federal Ministry of Education and Research; the bwForCluster Helix supercomputer, supported by the state of Baden-Württemberg through bwHPC and the German Research Foundation (DFG) through grant INST 35/1597-1 FUGG; the Vera cluster of the Max Planck Institute for Astronomy (MPIA), as well as the Cobra and Raven clusters, all three operated by the Max Planck Computational Data Facility (MPCDF); and the BinAC cluster, supported by the High Performance and Cloud Computing Group at the Zentrum für Datenverarbeitung of the University of Tübingen, the state of Baden-Württemberg through bwHPC and the German Research Foundation (DFG) through grant no INST 37/935-1 FUGG. 
Calculations based on the TNG-Cluster output and used in this paper were performed on the Vera cluster of the MPIA at MPCDF.

The Pan-STARRS1 Surveys (PS1) and the PS1 public science archive have been made possible through contributions by the Institute for Astronomy, the University of Hawaii, the Pan-STARRS Project Office, the Max-Planck Society and its participating institutes, the Max Planck Institute for Astronomy, Heidelberg and the Max Planck Institute for Extraterrestrial Physics, Garching, The Johns Hopkins University, Durham University, the University of Edinburgh, the Queen's University Belfast, the Harvard-Smithsonian Center for Astrophysics, the Las Cumbres Observatory Global Telescope Network Incorporated, the National Central University of Taiwan, the Space Telescope Science Institute, the National Aeronautics and Space Administration under Grant No. NNX08AR22G issued through the Planetary Science Division of the NASA Science Mission Directorate, the National Science Foundation Grant No. AST-1238877, the University of Maryland, Eotvos Lorand University (ELTE), the Los Alamos National Laboratory, and the Gordon and Betty Moore Foundation.

{This work is based on data from eROSITA, the soft X-ray instrument aboard SRG, a joint Russian-German science mission supported by the Russian Space Agency (Roskosmos), in the interests of the Russian Academy of Sciences represented by its Space Research Institute (IKI), and the Deutsches Zentrum für Luft- und Raumfahrt (DLR). The SRG spacecraft was built by Lavochkin Association (NPOL) and its subcontractors, and is operated by NPOL with support from the Max Planck Institute for Extraterrestrial Physics (MPE). The development and construction of the eROSITA X-ray instrument was led by MPE, with contributions from the Dr. Karl Remeis Observatory Bamberg \& ECAP (FAU Erlangen-Nuernberg), the University of Hamburg Observatory, the Leibniz Institute for Astrophysics Potsdam (AIP), and the Institute for Astronomy and Astrophysics of the University of Tübingen, with the support of DLR and the Max Planck Society. The Argelander Institute for Astronomy of the University of Bonn and the Ludwig Maximilians Universität Munich also participated in the science preparation for eROSITA.}

This research has made use of the M2C Galaxy Cluster Database, constructed as part of the ERC project M2C (The Most Massive Clusters across cosmic time, ERC-Adv grant No. 340519).

\FloatBarrier
\appendix
\restartappendixnumbering

\section{Galactic RM correction}
\label{app:a}
\setcounter{figure}{0}

\subsection{Techniques to correct for the Galactic RM}
The Galactic RM map of \citet{2022A&A...657A..43H} was produced by reconstruction from sparse data points (that were compiled from almost all Faraday rotation data sets available at the time) using a Bayesian inference algorithm. They model the Galactic RM sky as the product of a lognormal random amplitude field, $e^\rho$, and a Gaussian random sign field, $\chi$ : 
\begin{align}
    \mathrm{GRM} = \chi e^{\rho}.
\end{align}
They then infer $\chi$ and $\rho$ from sparse data. Their nominal resolution is accurate down to scales of $\lesssim 0.1145$ deg; however, since in some regions of the sky their data density is approximately 1 RM deg$^{-2}$, the resolution ends up being much poorer than the nominal resolution of their GRM map.

{\citet{2024ApJ...977..276K} tested a variety of interpolation techniques (including BRMS) to reconstruct the RM sky. They found that BRMS performs the best, with NNI performing similarly across a variety of RM structures and data properties. To accurately test the use of NNI for producing a GRM map, we ensured that the data being used for the interpolation do not have any extragalactic contribution. For this reason, we removed RMs that probe possible extended extragalactic structure. This includes RMs that lie within $2R_{500}$ of A3581, the other nearby clusters, or any bridges between clusters. The masked RM grid is displayed in Figure \ref{fig:masked}. Further, isolated extragalactic contributions to each of the remaining RMs must also be taken into account. To do this, we followed the method proposed by \citet{2024ApJ...977..276K} and removed any RMs that were not within $3\sigma$ of the mean of the 10 neighboring RMs. This assumes that the fluctuations of the GRM take place at much larger scales than that of isolated extragalctic RMs. Once these RMs were obtained, we ran NNI to reproduce a GRM map.}

{The method used for ERGS is described in Section \ref{sec:gal}, which was used on the un-masked RM grid. Additionally, we also tested the use of ERGS on the masked data; however, there were no marked changes in the results.} {The GRM values for each of the corrections are displayed in Figure \ref{fig:GRM_grms}. The resultant RRM grids from each of the GRM corrections are displayed in Figure \ref{fig:GRM_rrms}. By eye, the RRM grids appear to have the same large-scale structure with many small-scale variations in the RRMs; the difference between H22 and the other techniques is quite stark because H22 uses a nearly constant GRM value across the whole cluster. A possible reason for the difference in RRM grids between NNI and ERGS is the way we have defined extragalactic structure used to produce the NNI GRM map, which indicates that NNI is likely very sensitive to the way the RMs are masked; therefore, we decided against using it in our analysis.}

\begin{figure}
    \centering
    \includegraphics[width=\textwidth]{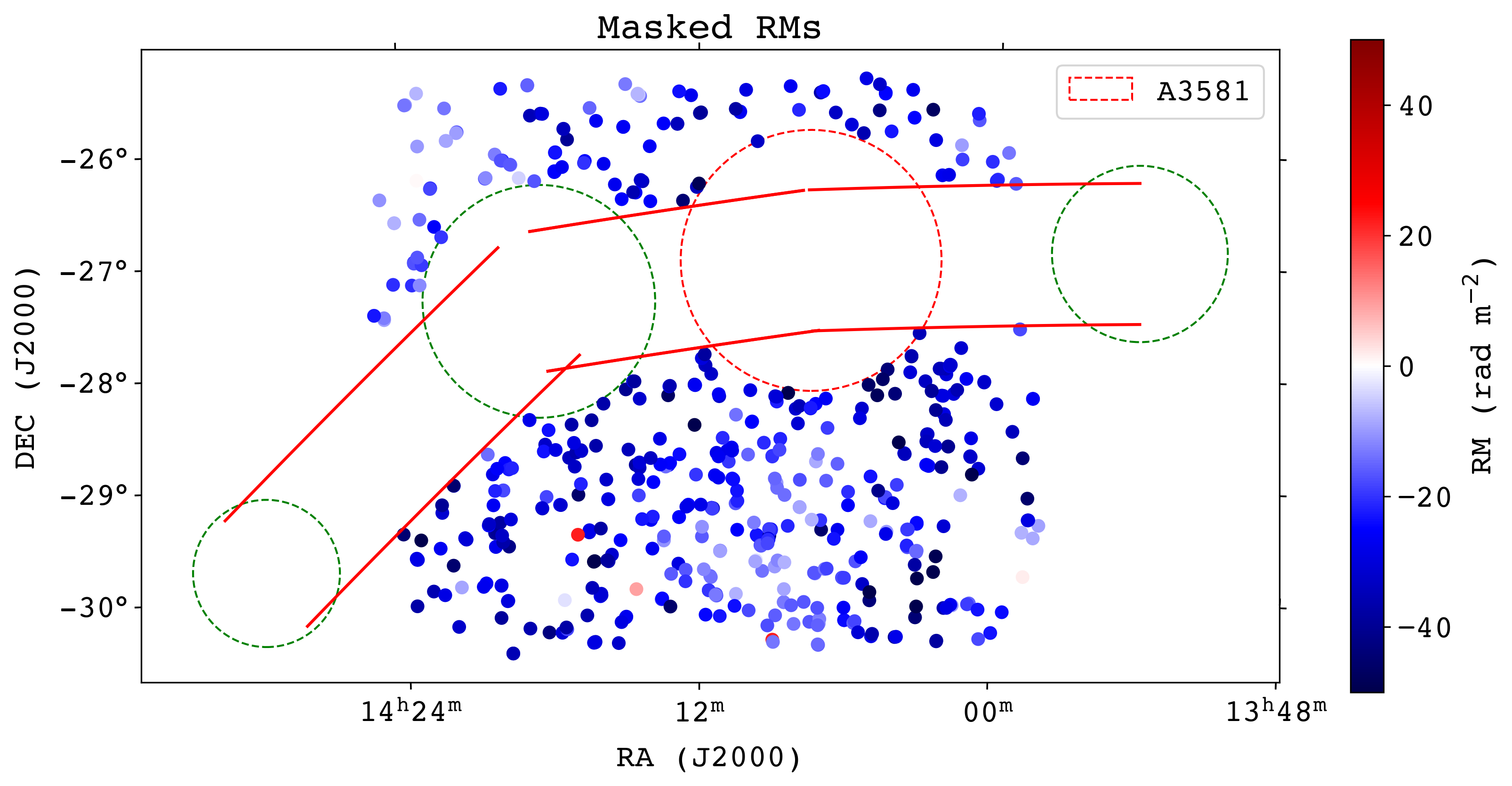}
    \caption{{The masked RM grid (without a GRM correction). The red circle indicates $2R_{500}$ of A3581, and the green circles indicate the same for other nearby clusters identified from the \citet{2024ApJS..272...39W} galaxy cluster catalog. The red lines indicate the boundary of possible bridges between clusters, and we assume a typical bridge radius of $\sim 1 $ Mpc.}}
    \label{fig:masked}
\end{figure}

\begin{figure}
    \centering
\gridline{\includegraphics[width=0.33\textwidth]{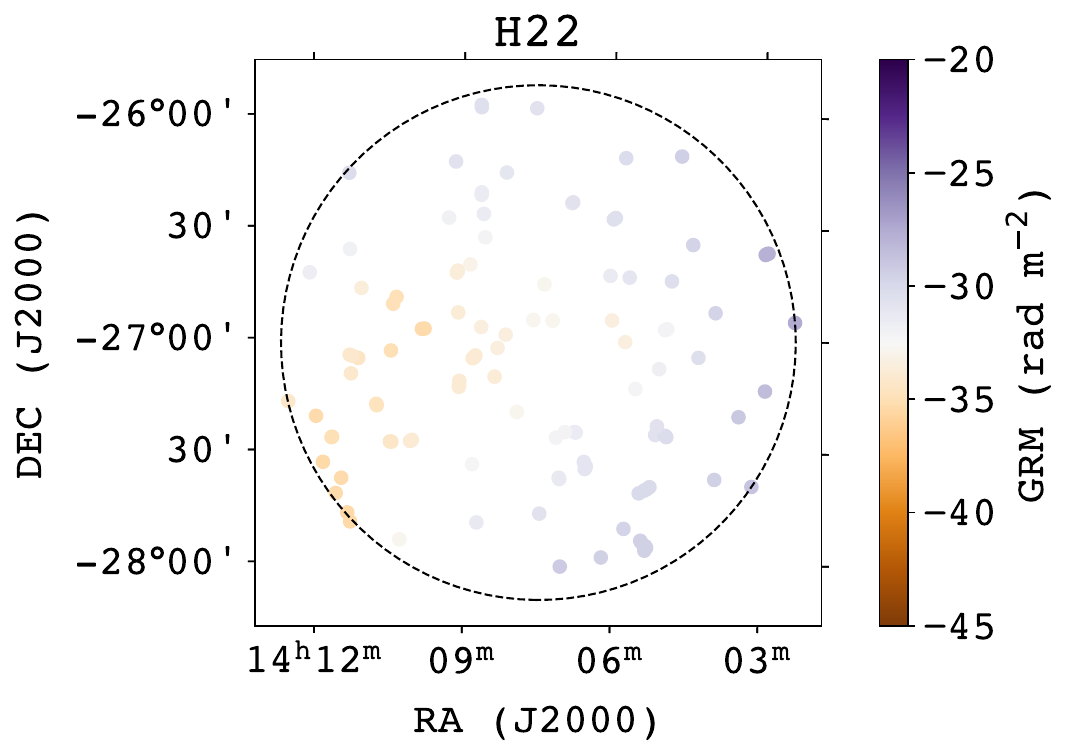}\includegraphics[width = 0.33 \textwidth]{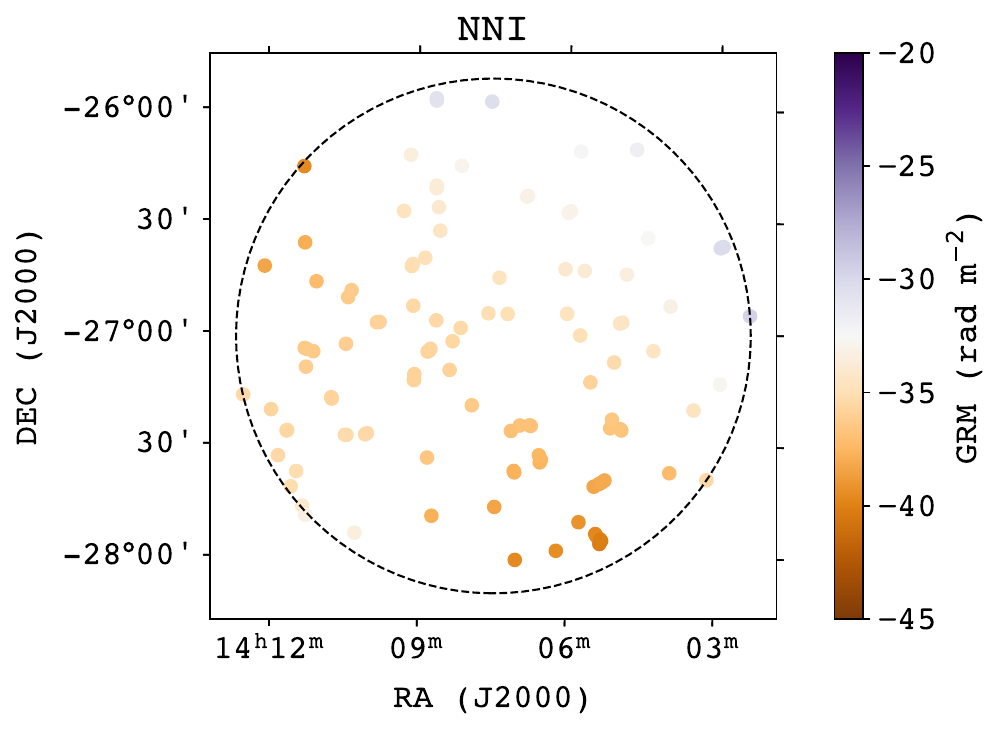}\includegraphics[width = 0.33\textwidth]{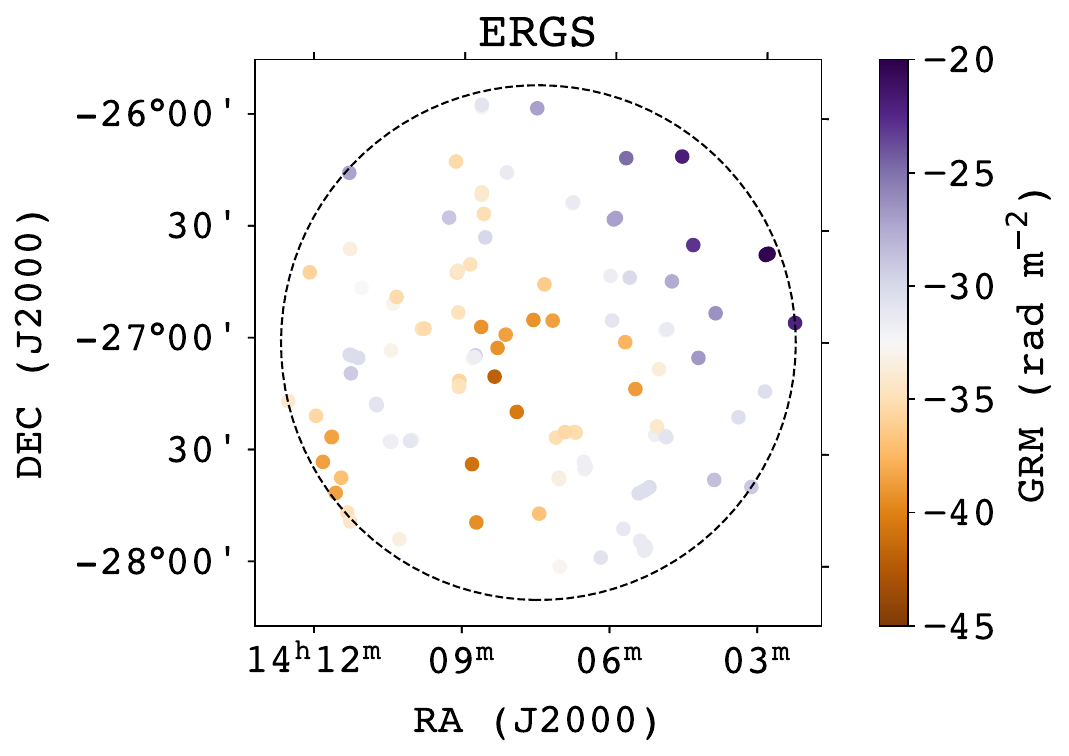}}
    \caption{{The GRM grids obtained from using the H22, NNI and ERGS GRM corrections. The black circles in each case indicate $2R_{500}$ of A3851.}}
    \label{fig:GRM_grms}
\end{figure}

\begin{figure}
    \centering
\gridline{\includegraphics[width=0.33\textwidth]{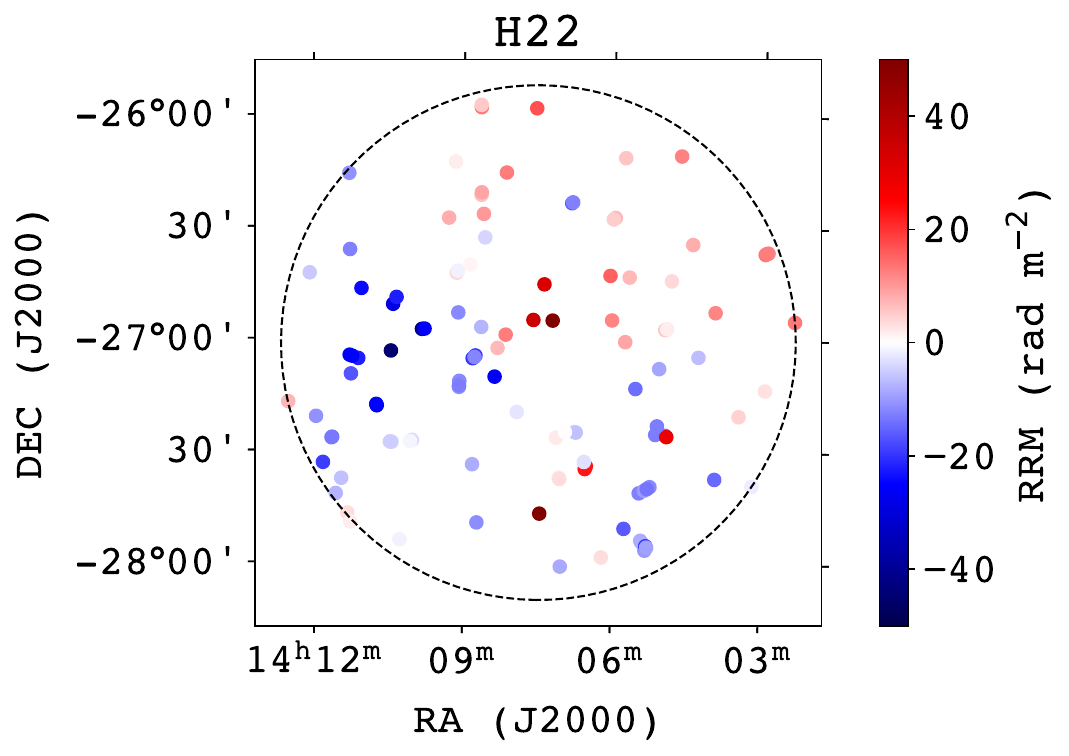}\includegraphics[width = 0.33 \textwidth]{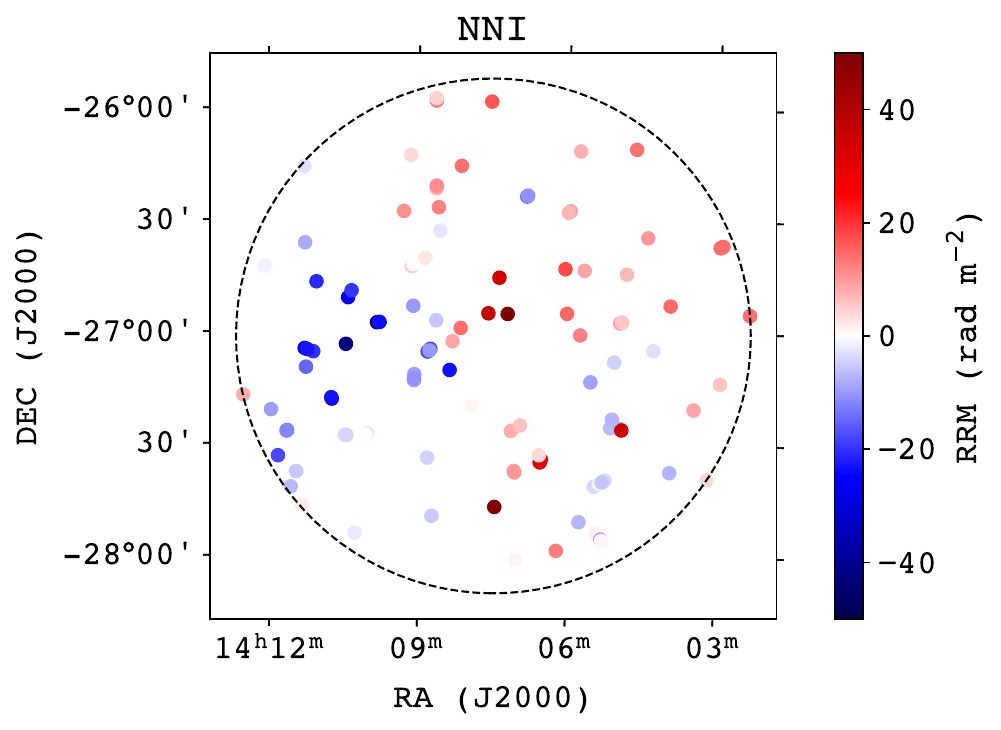}\includegraphics[width = 0.33\textwidth]{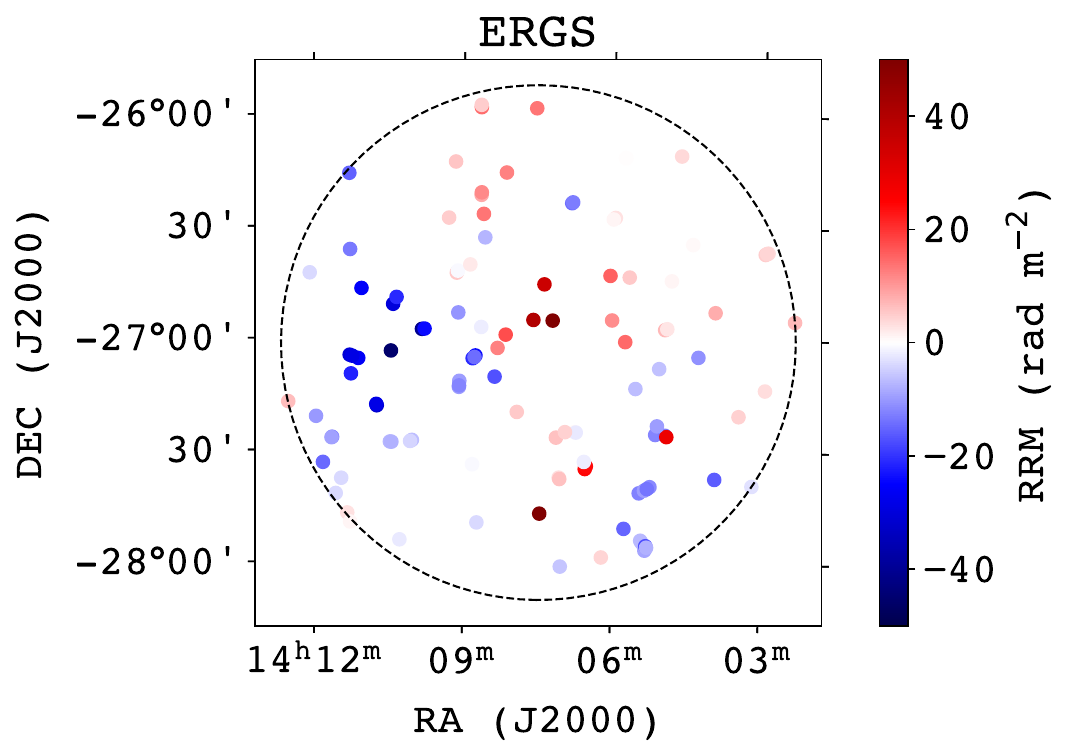}}
    \caption{{The RRM grids obtained from using the H22, NNI and ERGS GRM corrections. The black circles in each case indicate $2R_{500}$ of A3851.}}
    \label{fig:GRM_rrms}
\end{figure}

{\subsection{RRM scatter profiles}}

{We are most interested in the effect that the GRM corrections have on the RRM scatter profile, which is the main observable that we utilize in our analysis. The RRM scatter profiles for each of the GRM corrections is presented in Figure \ref{fig:grm_scatters}. The most notable feature in all the profiles is the nearly zero RRM scatter for the H22 correction, which is incredibly uncharacteristic of cluster scatter profiles. NNI and ERGS both portray the same overall features: a gradual decline until $\sim 0.7$ Mpc, an increase to a peak at around $\sim 1.1$ Mpc, and then a gradual decline until it reaches roughly zero scatter around $\sim 2$ Mpc. This supports our decision to avoid using the H22 correction, and our choice of ERGS. }

\begin{figure}
    \centering
\gridline{\includegraphics[width=0.33\textwidth]{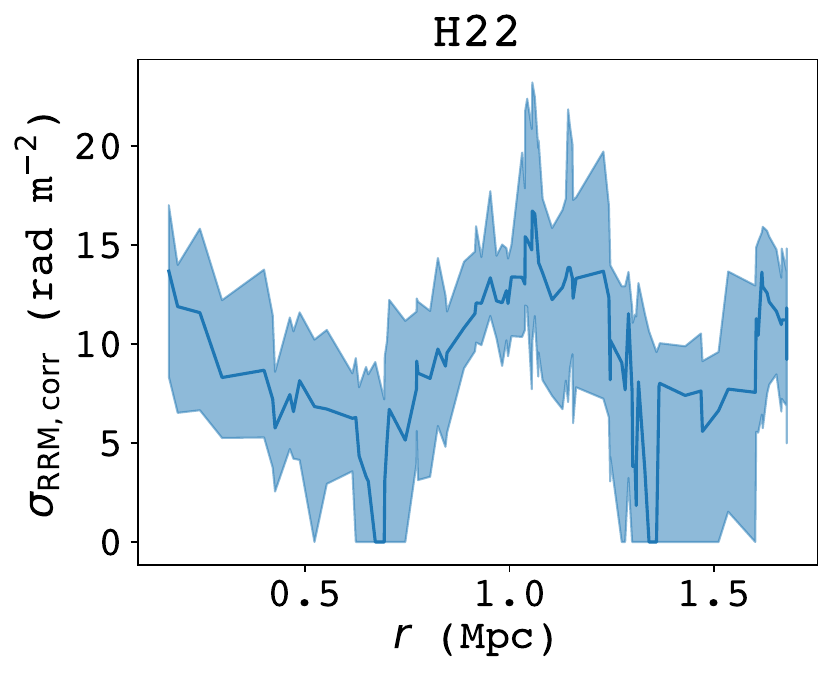}\includegraphics[width = 0.33\textwidth]{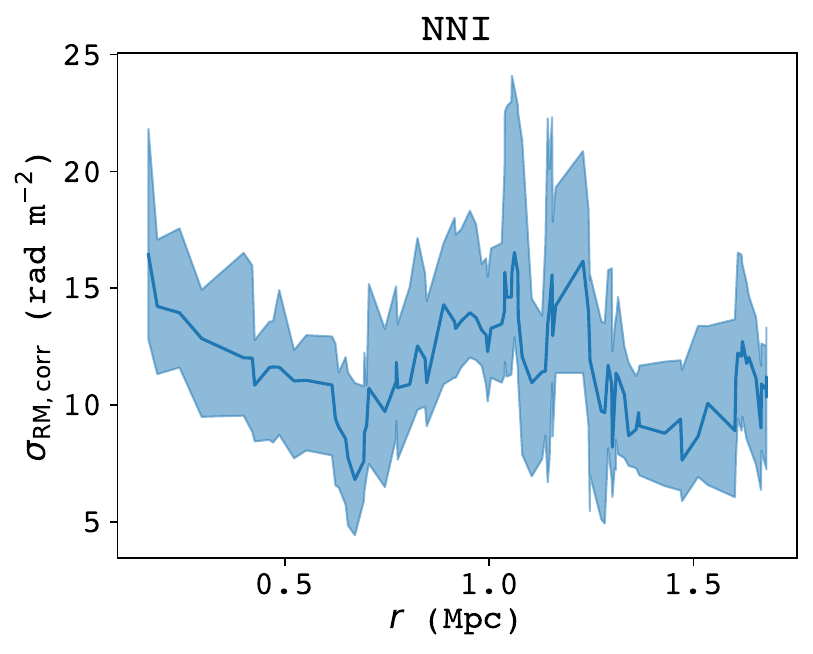} \includegraphics[width = 0.33\textwidth]{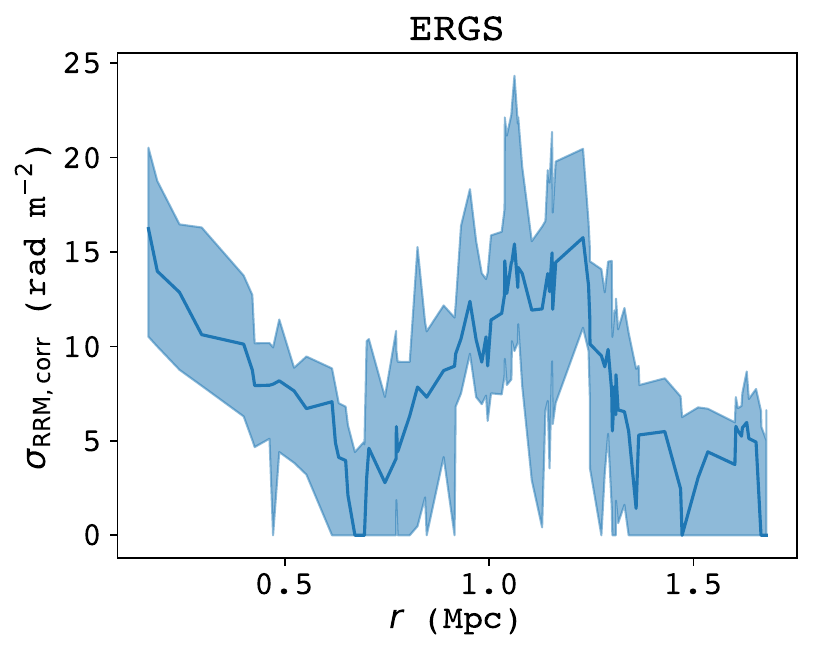}}
    \caption{{The RRM scatter profiles for the H22 (left), NNI (middle), ERGS (right) GRM corrections.} }
    \label{fig:grm_scatters}
\end{figure}

\section{{Control regions}}
\label{app:control}
{To characterize the significance of the $\sigma_{\mathrm{RRM}}$ enhancement in the cluster, we compare the scatter profile of A3581 to the two control regions (of the same size as A3581) displayed in Figure \ref{fig:control}. }
\begin{figure}
    \centering
    \includegraphics[width=0.49\textwidth]{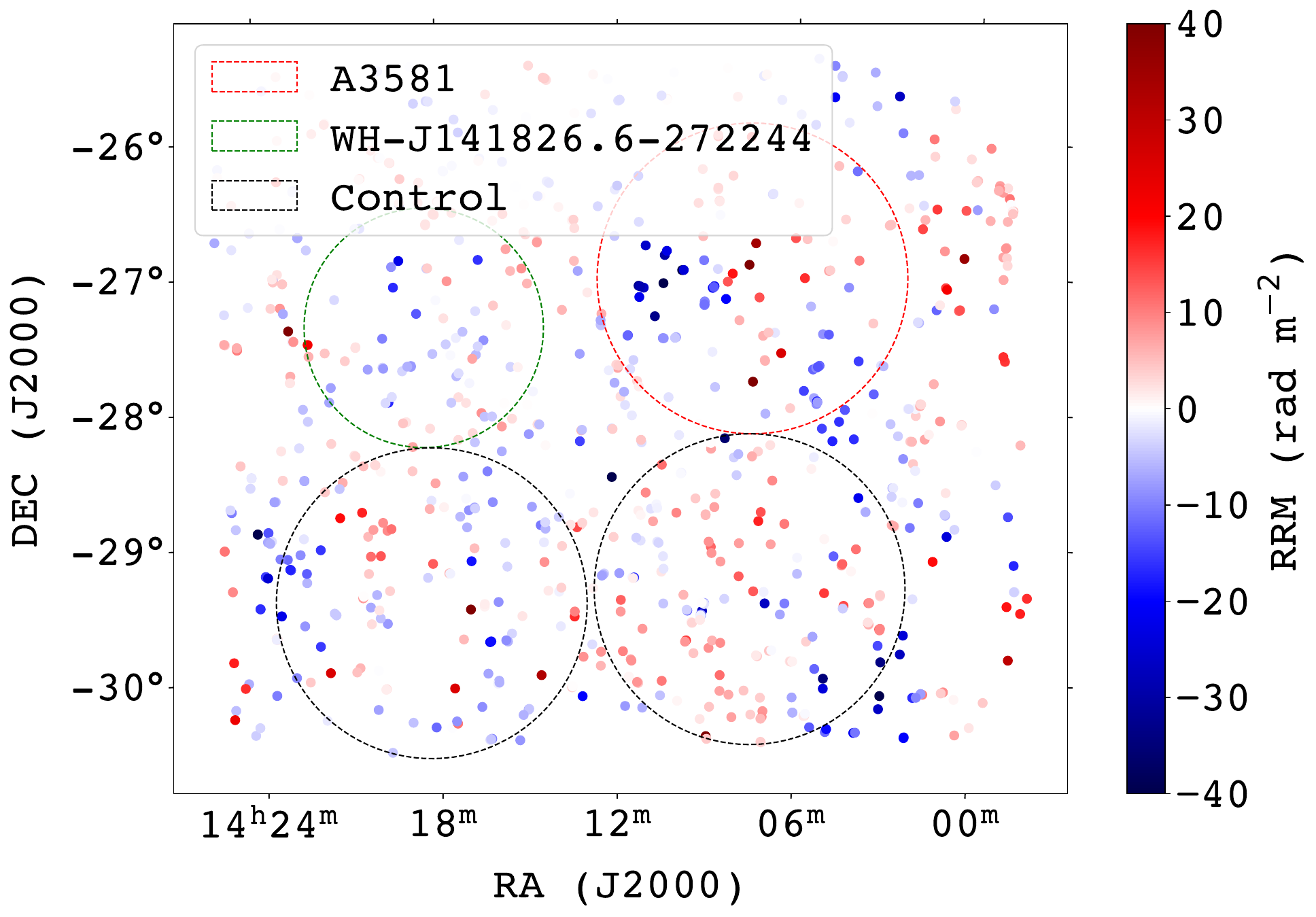}
    \caption{{The control regions used to test the significance of the $\sigma_{\mathrm{RRM}}$ within $2R_{500}$ of A3581.} }
    \label{fig:control}
\end{figure}

{The scatter profile in these control regions is displayed in Figure \ref{fig:control_scatter}. We see that the scatter in the two control regions is identical to zero aside from some random fluctuations, while the scatter in A3581 is several times higher in the center and shows a clear monotonic decline and a pronounced re-enhancement near the outskirts that is not present in the control regions.}

\begin{figure}
    \centering
    \includegraphics[width=0.49\textwidth]{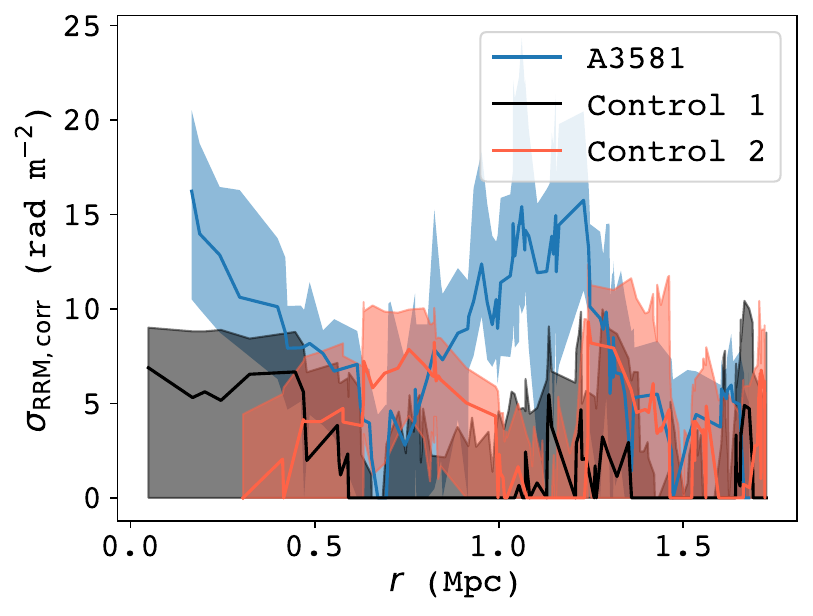}
    \caption{{The scatter profile of the two control regions, in comparison to that of A3581. }}
    \label{fig:control_scatter}
\end{figure}

\section{Cluster membership of sources}
\label{app:clustermember}
\setcounter{figure}{0}

For determining the cluster membership of sources, we used optical data from PS1. The $3\sigma, 6\sigma, 12\sigma, 24\sigma$ radio contours of the radio sources were plotted over optical images from PS1. Then, we determined the most likely optical counterpart for the radio source by eye. {Because A3581 is a low redshift cluster ($z = 0.0221$), we expect to find bright optical counterparts for all radio-emitting sources that are located in the cluster.} {Thus,} if an optical source could not be seen in the image, we determined that the radio source was likely a background source. 

If an optical source was found, we determined the photometric redshift of the source using the code from \citet{2020A&A...642A.102T}. We also cross-matched the photometric redshifts with spectroscopic redshifts for the sources that had spectra available in the literature. If the spectroscopic redshift was significantly different from the calculated photometric {redshift}, we used the spectroscopic redshift. Once we determined a redshift for a source, we determined cluster membership using a fixed gap of 1000 km s$^{-1}$ \citep{1996A&A...310....8K}{, also accounting for uncertainties in the photometric redshift}.

A sample radio-optical overlay plot is displayed in Figure \ref{fig:opt-radio}. {F}or this {optical} source, there are large bent radio lobes and we observed a total of 5 polarized components (marked in red) across these lobes. However, for all the RRMs the optical counterpart was chosen to be the same central source. {For this radio source, we found a photometric redshift of $z = 0.156\pm0.022$ and a spectroscopic redshift of $z = 0.300$ \citep[][]{2023OJAp....6E..49F}, indicating that it is a background source. While Figure \ref{fig:opt-radio} shows an example for a well-resolved radio galaxy, $72$\% of the RRMs in our sample are associated with sources that are unresolved at the POSSUM beam size of 20 arcseconds.}

{Of the 115 polarized RMs in our RM sample, {we were able to visually identify optical counterparts for 51 RMs; of these 51 RMs, we were only able to obtain photometric redshifts for 35 RMs.} Of these 35 RMs, we were able to obtain a spectroscopic redshift for 10 of them. For the remaining 16 RMs (that do not have a photometric redshift but do appear to have an optical counterpart), we obtained a spectroscopic redshift for 7 of them. For the remaining 9 RMs that we identified an optical counterpart for were all incredibly faint so we were able to safely classify them as background sources despite not having photometric or spectroscopic redshifts. Figure \ref{fig:phot_hist} displays a histogram of the redshifts obtained in our sample. Most of the RMs are background to the cluster as they have $z > 0. 1$ and only 4 of the RMs are at A3581's redshift (within error). 

To verify the accuracy of our photometric redshift calculation, we computed photometric redshifts for randomly selected bright unpolarized sources in the POSSUM field and compared this obtained redshift to the spectroscopic redshifts in the literature. Figure \ref{fig:spec_phot} displays a plot of the photometric and spectroscopic redshifts for 7 such sample tests. Our calculated photometric redshifts agree well (within error) of the values found in the literature at low redshifts. The only discrepancy that occurs is at high redshifts ($z > 0.5$). But in this case, the RM is already behind the cluster through either measure so we can safely classify it as a R source.

Using the photometric and spectroscopic redshifts, we found only 4 RRM sources that were cluster members (i.e. within a fixed velocity gap of 1000 km s$^{-1}$ of A3581's recession velocity). These sources were excluded from all our analysis, leaving us with 111 RMs that have sources background to the cluster. }

\begin{figure}[!htb]
    \centering
    \includegraphics[width=0.45\textwidth]{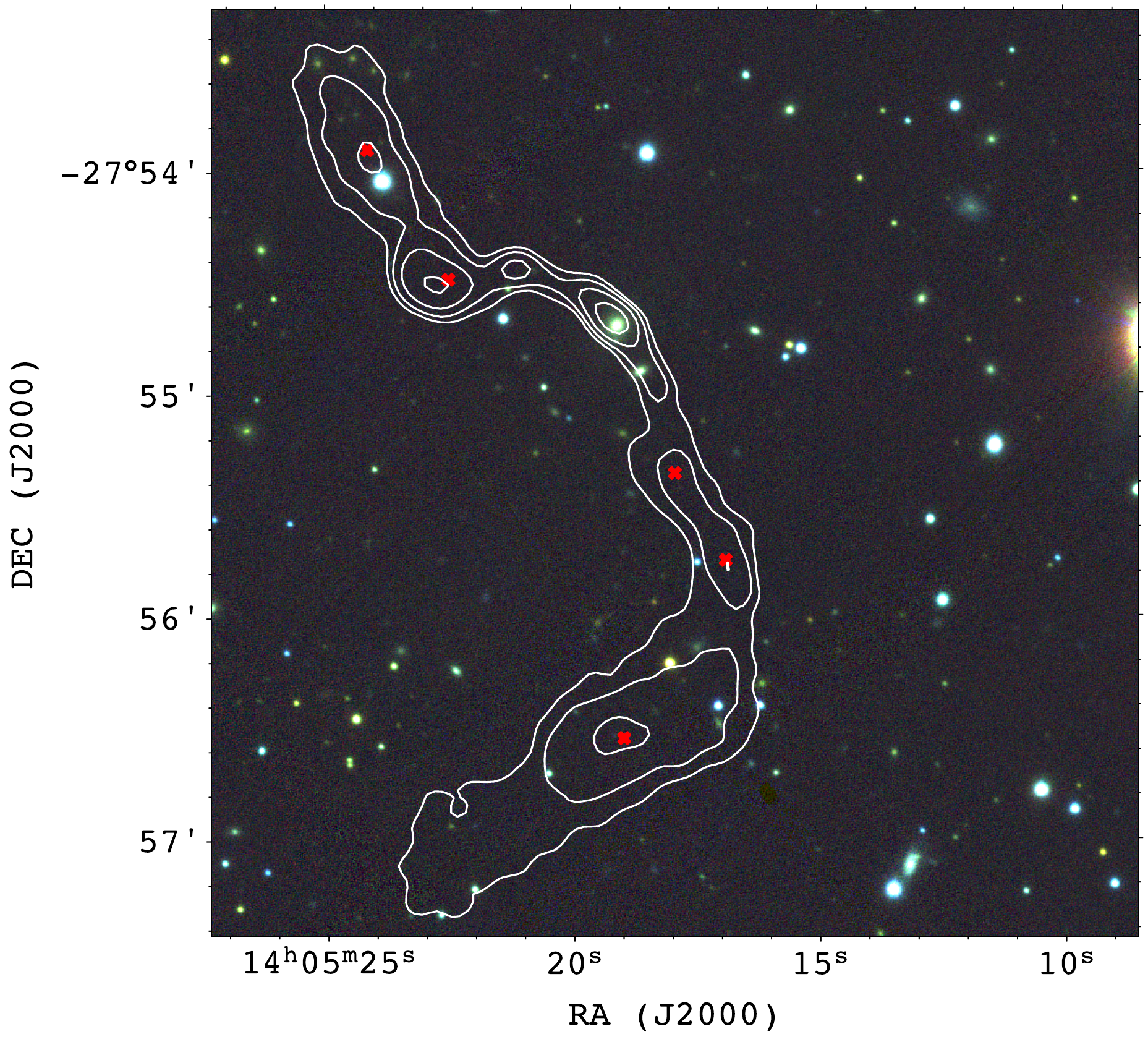}
    \caption{The $3\sigma, 6\sigma, 12\sigma,$ and $ 24\sigma$ radio contours for a radio galaxy in the source catalog plotted on a false-color optical image from PS1. The radio galaxy that hosts the radio lobes is the source in yellow-green is located at the following sky-coordinates ($\alpha, \delta$, J2000) = (14h 05m 19.1s, $-27^\circ$ 54$^\prime$ 42$^{\prime\prime}$). The locations of the RMs obtained from this image are marked in red.}
    \label{fig:opt-radio}
\end{figure}
\begin{figure}[!htb]
    \centering
    \subfigure[]{\includegraphics[width=0.47\textwidth]{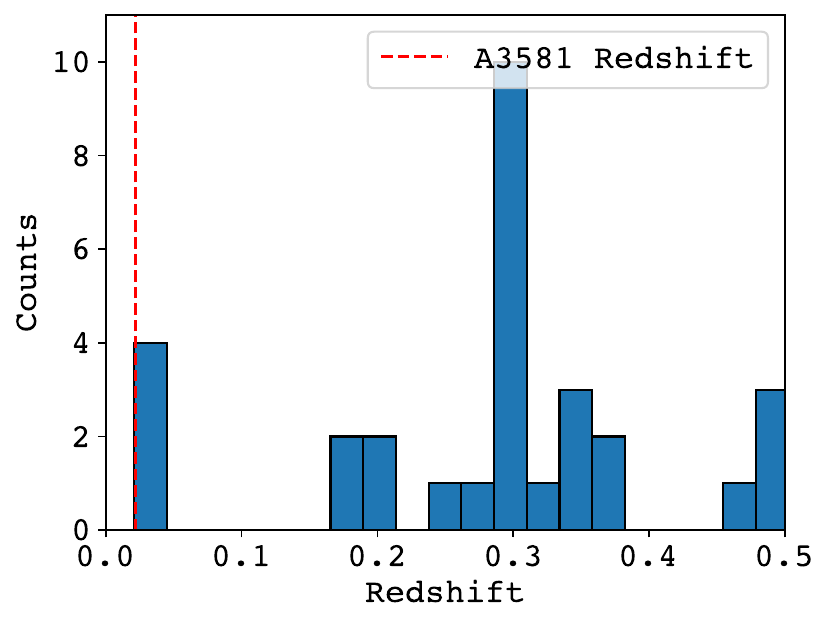}\label{fig:phot_hist}}
    \subfigure[]{\includegraphics[width = 0.47 \textwidth]{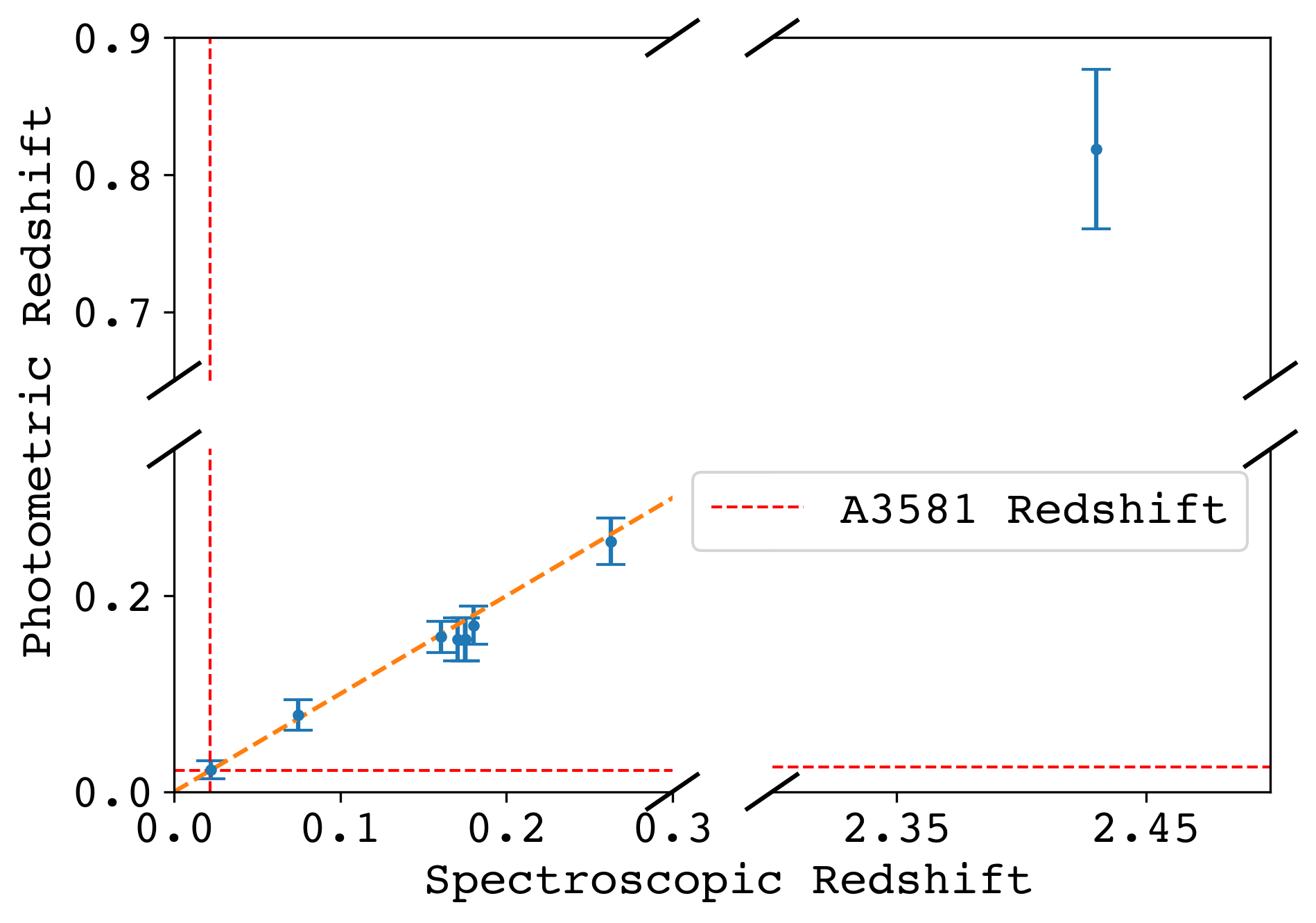}\label{fig:spec_phot}}
    \caption{(a) A histogram of the computed photometric redshifts. The red line indicates the redshift of A3581. Note that we have limited the $x$-axis to a redshift of 0.5 for visibility purposes. Some computed photometric redshifts range to $z = 1$. (b) A scatter plot of the photometric and spectroscopic redshift for 7 randomly selected bright unpolarized sources for which both values were available. The red dotted lines indicate the redshift of A3581. The orange dotted line portrays the $y = x$ line.}
\end{figure}

\section{Faraday complexity metrics}
\label{app:complexity}
\setcounter{figure}{0}
We {used} two metrics to quantify the Faraday complexity. The first metric is referred to as the $\sigma_{\mathrm{add}}$ complexity metric and is based on the $QU$-fitting of the linear Stokes spectra. The second metric is referred to as the second moment of the cleaned peaks metric ($m_2$) and is based on the number and size of peaks in the FDF {\citep[see also][]{2024AJ....167..226V}}.

The $\sigma_{\mathrm{add}}$ complexity metric is obtained from fitting the fractional linear Stokes parameters with a Faraday simple model. The structure in the residuals is then analyzed \citep{purcell2017internal}. If the Faraday simple model is a good fit for the spectra, we expect that the residuals will have a Gaussian distribution with some standard deviation that originates from the noise in the measurements. If the spectra are better fit with a more complex model, we expect there to be some structure in the residuals; $\sigma_{\mathrm{add}}$ quantifies this structure \citep[see][for details]{2024AJ....167..226V}.

The second moment of the cleaned peaks metric, $m_2$, is obtained from performing 1D RM-synthesis  and then RM-cleaning, using \pkg{RM-CLEAN} \citep{2009A&A...503..409H}, {which is implemented in} \pkg{RM-Tools}\footnote{\url{https://github.com/CIRADA-Tools/RM-Tools}} (Van Eck et al., in preparation). \pkg{RM-Tools} deconvolves the Faraday spectrum with the RM transfer function (RMTF) { \citep[analogously to Hogb\"om's \pkg{CLEAN} algorithm for radio imaging;][]{1974A&AS...15..417H}}, which is defined as: 

\begin{align}
    \mathrm{RMTF } = \frac{\sum_j  w_je^{-2i\phi (\lambda_j^2 - \lambda_0^2)}}{\sum_j w_j},
\end{align}
where the sum runs over all frequency channels, $w_j$ are the weights of the channel (which are inversely proportional to the square of the noise in the channel), and $\lambda_0$ is:

\begin{align}
    \lambda_0 = \sqrt{\frac{\sum_j w_j \lambda_j^2}{\sum_j w_j}}.
\end{align}

This results in a `cleaned' Faraday spectrum, $\tilde{F}$. The second moment of the cleaned peaks is then defined as: 
\begin{align}
    m_2 = \left[\frac{\sum_j (\phi_j - \bar{\phi})^2 \tilde{F}_j}{\sum_j \tilde{F}_j}\right]\bigg{/}\delta \phi,
\end{align}
where $\delta \phi$ is the full width half maximum of the RMTF{, the sum is taken over all frequency channels $j$} and $\bar{\phi}$ is:
\begin{align}
    \bar{\phi} = \frac{\sum_j \phi_j \tilde{F}_j}{\sum_j \tilde{F}_j}.
\end{align}

Following \citet{2024AJ....167..226V}, we set a threshold for Faraday complexity of $m_2 = 0.5$ and $\sigma_{\mathrm{add}} = 1$. However, in our sample we found that it was difficult to classify sources near these boundaries. For this reason, we decided to set a buffer region around each of these boundaries. For sources in the buffer region, we determined the complexity using $QU$-fitting: if the best-fit $QU$ model was simple and had a reduced chi-squared that was within {0.5 of 1}, we classified the source as simple, otherwise the source was deemed to be Faraday complex. For $\sigma_{\mathrm{add}}$ we chose the buffer region to be $1- 10^{-0.65} \leq \sigma_{\mathrm{add}}^{\mathrm{buffer}} \leq 1 + 10^{-0.6}$, and for $m_2$ the buffer region is $0.4 \leq m_2^{\mathrm{buffer}} \leq 0.6$. For $\sigma_{\mathrm{RM}}$, we chose different sizes for the lower and upper boundary regions because the sources are distributed logarithmically in $\sigma_{\mathrm{add}}$ space, and to roughly cover the same number of sources on either side of the boundary. 

After computing the complexity metrics using the procedure described {above}, we {investigated} the correlation between the SNR and the complexity metric as shown in Figure \ref{fig:complex}, as a correlation between the two has been observed in previous works \citep[e.g.,][]{2023PASA...40...40T, 2024AJ....167..226V}. Both $m_2$ and $\sigma_{\mathrm{add}}$ generally agree well with regards to the complexity of sources. Most of the sources are observed in the lower left corner of the plot, indicating that most sources in our sample are Faraday simple. Additionally, we also observe a clear increase in the SNR as the complexity of sources increases.
\begin{figure}[!htb]
    \centering
    \includegraphics[width=0.47\textwidth]{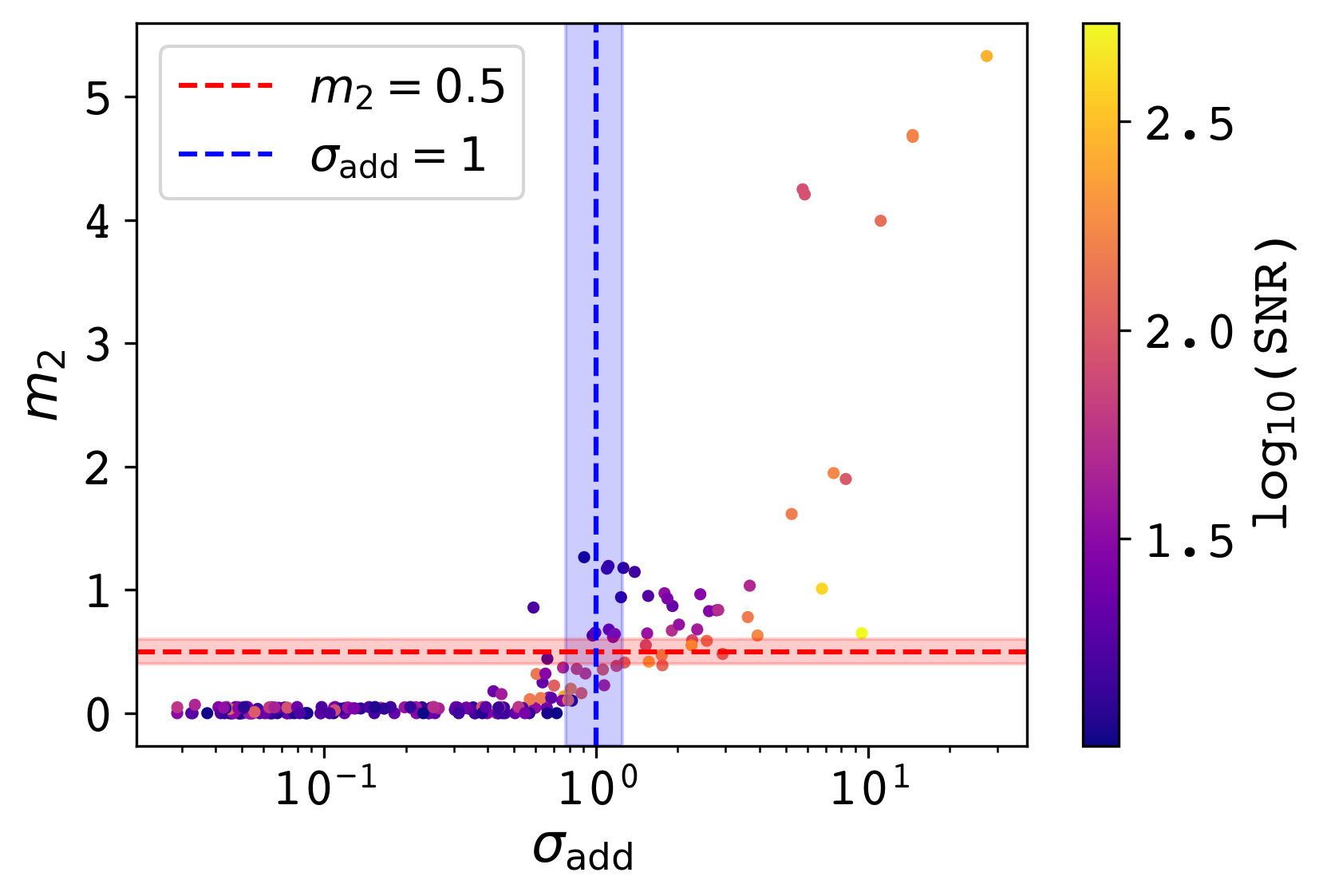}
    \caption{A comparison between the $m_2$ complexity metric, $\sigma_{\mathrm{add}}$, and $\log_{10}(\mathrm{SNR})$. The horizontal axis is in a logarithmic scale, while the vertical axis is in a linear scale. The shaded regions portray the buffer regions for each of the complexity metrics.}
    \label{fig:complex}
\end{figure}

{We used two complexity metrics, rather than one, because there might be sources that are classified as complex by one and not the other (such as the source on the top left quadrant of Figure \ref{fig:complex}). We avoid solely using $QU$-fitting for all sources for this same reason (as $QU$-fitting is used to derive the $\sigma_{\mathrm{add}}$ complexity metric). Additionally, the requirement that the best-fit model has a reduced chi-squared within 0.5 of 1 is necessary because the source might be classified as simple based on the Bayes factors, but the $QU$ spectra might deviate far from the simple model (given by Equation \ref{eq:5}). }

\section{Catalog of sources}
\label{app:catalog}
{The full catalog of polarized components is made available as a machine-readable table, following the RMTable2023 \citep[][]{2023ApJS..267...28V} format. In addition to this, the Stokes spectra for each polarized component are included, following the PolSpectra2023 format \citep[][]{2023ApJS..267...28V}.}

{In Table \ref{tab:catalog}, we describe the columns that are included in the catalog. }
\begin{longrotatetable}  

\begin{deluxetable*}{|c|c|c|}

\tablehead{
\colhead{\textbf{Column Name}} & \colhead{\textbf{Units}} & \colhead{\textbf{Column Description}}}
\tablecomments{}
\tablecaption{Columns of Source Catalog}

\label{tab:catalog}
\startdata
        \pkg{island\_id} & None & Unique identifier for the island of emission as identified by Selavy\\
        \pkg{cat\_ID} & None & Unique identifier per source, as classified by the source-finder(Selavy)\\
        \pkg{component\_name} & None & The IAU-format name taken from the J2000 position of the island's centroid\\
       \pkg{aperture} & px & Size of integrating aperture for spectra extraction\\
        \pkg{ra\_hms\_cont} & hr, min, sec & Right ascension (J2000) in sexagesimal units\\
        \pkg{dec\_dms\_cont} & deg, min, sec & Declination (J2000) in sexagesimal units\\
        \pkg{ra} & deg & Right ascension (J2000)\\
        \pkg{dec} & deg & Declination (J2000) \\
        \pkg{ra\_err} & arcsec & Error in right ascension\\
        \pkg{dec\_err} & arcsec & Error in declination\\
        \pkg{freq\_source-finder} & THz & Reference frequency used by Selavy \\
        \pkg{peak\_flux\_source-finder} & Jy beam$^{-1}$ & Peak flux detected by Selavy at the reference frequency \\
        \pkg{e\_peak\_flux\_source-finder} & Jy beam$^{-1}$ & Error in the peak flux detected by Selavy at the reference frequency\\ 
        \pkg{total\_flux\_source-finder} & Jy & The total flux detected by Selavy across all observing frequencies \\ 
        \pkg{e\_total\_flux\_source-finder} & mJy & The error in the total flux detected by Selavy across all observing frequencies\\
        \pkg{major\_axis\_source-finder} & arcsec & Semimajor axis of the polarized component from Selavy \\
        \pkg{minor\_axis\_source-finder} & arcsec & Semiminor axis of the polarized component from Selavy \\
        \pkg{position\_angle\_sourcefinder} & deg & Position angle of the polarized component from Selavy, increasing east from north. \\
        \pkg{e\_major\_axis\_sourcefinder} & arcsec & Error in the semimajor axis of the polarized component from Selavy  \\
        \pkg{e\_minor\_axis\_sourcefinder} & arcsec & Error in the semiminor axis of the polarized component from Selavy\\
        \pkg{e\_position\_angle\_sourcefinder} & deg & Error in the position angle of the polarized component from Selavy\\
        \pkg{deconvolved\_major\_axis\_sourcefinder} & arcsec & Deconvolved semimajor axis of the polarized component from Selavy\\
        \pkg{deconvolved\_minor\_axis\_sourcefinder} & arcsec & Deconvolved semiminor axis of the polarized component from Selavy\\
        \pkg{deconvolved\_position\_angle\_sourcefinder} & deg & Deconvolved position angle of the polarized component from Selavy \\
        \pkg{e\_deconvolved\_major\_axis\_sourcefinder} & arcsec & Error in the deconvolved semimajor axis from Selavy\\
        \pkg{e\_deconvolved\_minor\_axis\_sourcefinder} & arcsec & Error in the deconvolved semiminor axis from Selavy\\
        \pkg{e\_deconvolved\_position\_angle\_sourcefinder} & deg & Error in the deconvolved position angle from Selavy\\
        \pkg{chi\_squared\_fit\_sourcefinder} & None & $\chi^2$ value of Selavy fit\\
        \pkg{rms\_fit\_gauss} & None & Root-mean-squared value of Selavy fit\\
        \pkg{spectral\_index\_sourcefinder} & None & Spectral index reported by Selavy \\
        \pkg{spectral\_curvature} & None & Spectral curvature reported by Selavy; set to $-99$ when no curvature is obtained\\
        \pkg{e\_spectral\_index\_sourcefinder} & None & Error in the spectral index reported by Selavy\\
        \pkg{spectral\_curvature\_err} & None & Error in the spectral curvature reported by Selavy\\
        \pkg{local\_I\_rms\_sourcefinder} & Jy beam$^{-1}$ & Local root-mean-sqaured noise reported by Selavy \\
        \pkg{has\_siblings\_sourcefinder} & None & Boolean flag. Is true if the component is one of many fitted to the same island\\
        \pkg{fit\_is\_estimate} & None & Boolean flag. True if the fit failed; the reported parameter values come from the initial estimate\\
        \pkg{spectral\_index\_from\_TT} & None & Boolean. True when spectral curvature and index calculated from Taylor-term images.\\
        \pkg{flag\_c4} & None &  Boolean flag to indicate that the fitted component is formally bad  \\
        \pkg{comment} & None & Any comments included \\
        \pkg{polyOrd} & None & Order of log-polynomial fits to Stokes I by \pkg{RM-Tools}\\
        \pkg{IfitStat} & None & Stokes $I$ fit status code reported by \pkg{RM-Tools}\\
        \pkg{IfitChiSqRd} & None & Reduced $\chi^2$ for a Stokes $I$ fit reported by \pkg{RM-Tools}\\
        \pkg{sigmaAddQ}& None & $\sigma_\mathrm{add}$ metric for Stokes $Q$\\
        \pkg{dsigmaAddMinusQ} & None & Lower-bound error in Stokes $Q$ $\sigma_{\mathrm{add}}$ metric\\
        \pkg{dsigmaAddPlusQ} & None & Upper-bound error in Stokes $Q$ $\sigma_{\mathrm{add}}$ metric\\
        \pkg{sigmaAddU} & None & $\sigma_\mathrm{add}$ metric for Stokes $U$\\
        \pkg{dsigmaAddMinusU} & None & Lower-bound error in Stokes $U$ $\sigma_{\mathrm{add}}$ metric\\
        \pkg{dsigmaAddPlusU} & None & Upper-bound error in Stokes $U$ $\sigma_{\mathrm{add}}$ metric\\
        \pkg{I\_curvature} & None &  Spectral curvature in Stokes $I$ model from \pkg{RM-Tools}\\
        \pkg{I\_curvature\_err} & None & Error in the spectral curvature in Stokes $I$ model from \pkg{RM-Tools}\\
        \pkg{spectral\_index} & None & Spectral index in Stokes $I$ model from \pkg{RM-Tools}\\
        \pkg{spectral\_index\_err} & None & Error in spectral index in Stokes $I$ model from \pkg{RM-Tools}\\
        \pkg{stokesI} & Jy beam$^{-1}$ & Stokes $I$ model value at polarization reference frequency \\
        \pkg{stokesI\_err} & Jy beam$^{-1}$ & Error in Stokes $I$ model value at polarization reference frequency\\
        \pkg{FDF\_noise\_empirical} & Jy beam$^{-1}$ RMSF$^{-1}$ & Empirically derived noised in the Faraday dispersion function from \pkg{RM-Tools}\\
        \pkg{rm} & rad m$^{-2}$ & Observed rotation measure, obtained from RM-synthesis with \pkg{RM-Tools}\\
        \pkg{rm\_err}  & rad m$^{-2}$  & Error in the observed rotation measure\\
        \pkg{polint} & Jy beam$^{-1}$ & Peak polarized intensity (at polarization reference frequency)\\
        \pkg{polint\_err} & Jy beam$^{-1}$ & Error in peak polarized intensity\\
        \pkg{SNR\_PI} & None & Peak signal-to-noise ratio in polarized intensity \\
        \pkg{stokesU} & Jy beam$^{-1}$ & Stokes $U$ value at polarization reference frequency\\
        \pkg{polangle} & deg & Polarization angle at polarization reference frequency\\
        \pkg{polangle\_err} & deg & Error in polarization angle at polarization reference frequency\\
        \pkg{derot\_polangle} & deg & Polarization angle at zero wavelength (assuming a Farday-thin model)\\
        \pkg{derot\_polangle\_err} & deg &  Error in zero-wavelength polarization angle\\
        \pkg{reffreq\_I} & Hz & Reference frequency for Stokes $I$ model\\
        \pkg{reffreq\_pol} & Hz & Reference frequency for polarization\\
        \pkg{rmsf\_fwhm} & rad m$^{-2}$ & Full-width at half maximum of the rotation measure spread function\\
        \pkg{noise\_chan} & Jy beam$^{-1}$ & Median noise per channel in Stokes $Q$ and $U$\\
        \pkg{FDF\_noise\_theoretical} & Jy beam$^{-1}$ RMSF$^{-1}$ & The theoretical noise value in the Faraday dispersion function from \pkg{RM-Tools} \\
        \pkg{minfreq} & Hz & Lowest frequency used in RM-synthesis\\
        \pkg{maxfreq} & Hz & Highest frequency used in RM-synthesis\\
        \pkg{Nchan} & None & Number of channels used in RM-synthesis \\
        \pkg{channelwidht} & Hz & Median width of channels\\
        \pkg{fracpol} & None & Peak fractional polarization\\
        \pkg{beam\_major} & deg & Semimajor axis size of the synthesized beam\\
        \pkg{beam\_minor} & deg & Semiminor axis size of the synthesized beam\\
        \pkg{beam\_pa} & deg & Position angle of the synthesized beam\\
        \pkg{rrm} & rad m$^{-2}$ & RRM obtained from using the ERGS\\
        \pkg{grm\_ergs} & rad m$^{-2}$ & Estimated GRM from using ERGS\\
        \pkg{grm\_hut} & rad m$^{-2}$ & GRM obtained from the H22 GRM map\\
        \pkg{rrm\_err} & rad m$^{-2}$ & Error in the RRM obtained from ERGS\\
        \pkg{qu\_model} & None  & Equation number of the best-fit $QU-$model from this work\\
        \pkg{qu\_model\_red\_chi2} & None & Reduced $\chi^2$ of the best-fit $QU-$model. \\
        \pkg{SigmaRM} & rad m$^{-2}$ & $\Sigma_{\mathrm{RM}}$ obtained from $QU-$fitting\\
        \pkg{SigmaRMerrm} & rad m$^{-2} $ & Lower-bound error on $\Sigma_{\mathrm{RM}}$ obtained from $QU-$fitting\\
        \pkg{SigmaRMerrp} & rad m$^{-2}$ & Upper-bound error on $\Sigma_{\mathrm{RM}}$ obtained from $QU-$fitting\\
        \pkg{complex\_flag} & None & Boolean flag for the Faraday complexity of the polarized component\\
        \pkg{z} & None & The redshift of the source\\
        \pkg{z\_err} & None & Error on the redshift \\
        \pkg{z\_source} & None & Source of the redshift (spectroscopic, photometric or `-' for no redshift)\\
        \pkg{z\_spec\_ref} & None &Bibcode of the reference, if the spectroscopic redshift was used \\
        \pkg{in\_clust} & None & Boolean flag for the source being embedded in A3581
\enddata

\end{deluxetable*}
\end{longrotatetable}

\FloatBarrier
\bibliography{main}{}
\bibliographystyle{aasjournal}



\end{document}